\documentclass{article}

\usepackage{arxiv}
\PassOptionsToPackage{numbers}{natbib}
\usepackage[utf8]{inputenc} 
\usepackage[T1]{fontenc}    
\usepackage{hyperref}
\usepackage{graphicx}
\usepackage{url}
\usepackage[utf8]{inputenc} 
\usepackage[T1]{fontenc}    
\usepackage[utf8]{inputenc} 
\usepackage[T1]{fontenc}    
\usepackage{hyperref}       
\usepackage{url}            
\usepackage{booktabs}       
\usepackage{amsmath,amssymb,amsfonts}
\usepackage{nicefrac}       
\usepackage{microtype}      
\usepackage{lipsum}		
\usepackage{graphicx}
\usepackage{natbib}
\usepackage{doi}
\usepackage{tabularx}
\usepackage{textcomp}
\usepackage{hyperref} 
\usepackage{booktabs}       
\usepackage{amsfonts}       
\usepackage{nicefrac}       
\usepackage{microtype}      
\usepackage{xcolor} 
\usepackage{graphicx}
\usepackage{colortbl}
\usepackage{color}
\usepackage{array}
\usepackage{amsmath}
\usepackage{algorithm}
\usepackage{algpseudocode}
\usepackage{threeparttable}
\usepackage{multirow}

\usepackage{wrapfig}
\usepackage{subfigure}
\usepackage{caption}
\usepackage{subcaption}
\usepackage{amsthm}
\usepackage{mathtools}
\usepackage{amssymb}
\usepackage{bm}

\usepackage[utf8]{inputenc}
\usepackage[T1]{fontenc}
\usepackage{newunicodechar}
\usepackage{amsfonts}

\usepackage{longtable}

\usepackage{xcolor} 

\newcommand{\smilingcat}{\includegraphics[height=1em]{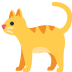}} 

\title{\smilingcat NECOMIMI: Neural-Cognitive Multimodal EEG-informed Image Generation with Diffusion Models}
\definecolor{myorange}{RGB}{255, 140, 0}

\date{} 					

\author{Chi-Sheng Chen \\
Neuro Industry, Inc.\\
San Francisco, CA 94114, USA \\
\texttt{m50816m50816@gmail.com} \\
}

\renewcommand{\shorttitle}{\textit{arXiv} Template}

\hypersetup{
pdftitle={A template for the arxiv style},
pdfsubject={cs.CV},
pdfauthor={Chi-Sheng ~Chen},
pdfkeywords={deep learning, multimodal learning, self-supervised learning, EEG, generative model},
}

\begin{document}
\maketitle

\begin{abstract}
\textcolor{myorange}{NECOMIMI} (\textcolor{myorange}{NE}ural-\textcolor{myorange}{CO}gnitive \textcolor{myorange}{M}ult\textcolor{myorange}{I}modal EEG-Infor\textcolor{myorange}{M}ed \textcolor{myorange}{I}mage Generation with Diffusion Models) introduces a novel framework for generating images directly from EEG signals using advanced diffusion models. Unlike previous works that focused solely on EEG-image classification through contrastive learning, NECOMIMI extends this task to image generation. The proposed NERV EEG encoder demonstrates state-of-the-art (SoTA) performance across multiple zero-shot classification tasks, including 2-way, 4-way, and 200-way, and achieves top results in our newly proposed Category-based Assessment Table (CAT) Score, which evaluates the quality of EEG-generated images based on semantic concepts. A key discovery of this work is that the model tends to generate abstract or generalized images, such as landscapes, rather than specific objects, highlighting the inherent challenges of translating noisy and low-resolution EEG data into detailed visual outputs. Additionally, we introduce the CAT Score as a new metric tailored for EEG-to-image evaluation and establish a benchmark on the ThingsEEG dataset. This study underscores the potential of EEG-to-image generation while revealing the complexities and challenges that remain in bridging neural activity with visual representation.
\end{abstract}


\begin{figure}[h]
    \includegraphics[width=1\linewidth]{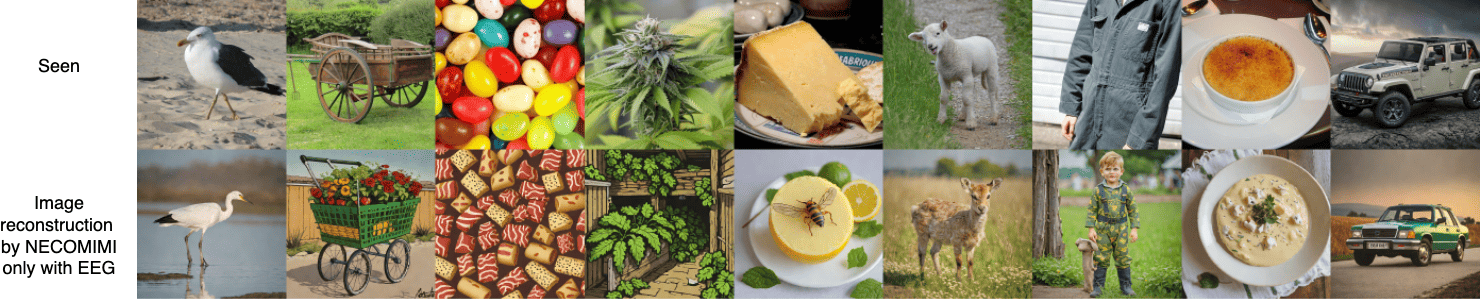}
    \caption{This image demonstrates the capability of the NECOMIMI model to reconstruct images purely from EEG data without using the "Seen" images (ground truth) as embeddings during the generation process. The two-stage NECOMIMI architecture effectively extracts semantic information from noisy EEG signals, showing that it can capture and represent the underlying concepts from brainwave activity. The bottom row of images, generated solely from EEG input, highlights the potential of NECOMIMI to approximate the content of the "Seen" images in the top row, even in the absence of any direct visual reference or embedding.}
    \label{fig:landing_demo}
\end{figure}

\section{Introduction}
Electroencephalography (EEG) is one of the most ancient techniques used to measure neuronal activity in the human brain \cite{mary_1959, millett_2001}. Its application has significant value in clinical practice, particularly in diagnosing epilepsy \cite{reif_strzelczyk_rosenow_2016}, depression \cite{li_chen_cheng_chen_chen_chen_bai_tsai_2023} and sleep disorders \cite{Hussain2022Quantitative}, as well as in assessing dysfunctions in sensory transmission pathways \cite{Thoma2003Lateralization} and more \cite{Perrottelli2021EEG-Based}. Historically, the analysis of EEG signals was limited to visual inspection of amplitude and frequency changes over time. However, with advancements in digital technology, the methodology has evolved significantly, shifting towards a more comprehensive analysis of the temporal and spatial characteristics of these signals \cite{frey_2016}. As a result of this evolution, EEG has gained recognition as a potent tool for capturing brain functions in real-time, particularly in the sub-second range. Despite its advantages, EEG has traditionally suffered from poor spatial resolution, making it challenging to pinpoint the precise brain areas responsible for the measured neuronal activity at the scalp \cite{Li2022Concurrent}. In recent years, there has been a surge of interest in utilizing EEG for more sophisticated applications, such as image recognition and reconstruction \cite{mai_zhang_fang_zhang_2023}. These advancements have led to significant improvements in the accuracy of image recognition tasks, underscoring the potential of EEG as a bridge between neural activity and visual representation \cite{spampinato_2016, brain2img}. The growing interest in using EEG for image recognition is rooted in its ability to capture the temporal dynamics of neuronal activity, though its spatial resolution remains a challenge. Innovative methodologies, including deep learning techniques and generative models like Generative Adversarial Networks (GANs) \cite{goodfellow_2014} and diffusion models \cite{ho_jain_abbeel_2020}, have enhanced the accuracy and effectiveness of EEG-based systems, allowing for the generation of photorealistic images based on neural signals \cite{brain2img, kumar_2017, singh_pandey_miyapuram_raman_2023}. Notably, studies have demonstrated the feasibility of decoding natural images from EEG signals, employing innovative frameworks that align EEG responses with paired image stimuli \cite{bai_wang_cao_ge_yuan_shan_2023}. However, most of the current works claiming to be EEG-to-image are essentially still image-to-image in nature, with EEG information primarily used to slightly guide the transformation of the input image by adding noise \cite{brain2img, palazzo_spampinato_kavasidis_giordano_shah, khare_choubey_amar_udutalapalli_2022, bai_wang_cao_ge_yuan_shan_2023}. In order to achieve a truly meaningful EEG-to-image generation, this work, named NECOMIMI (NEural-COgnitive MultImodal eeg-inforMed Image generation with diffusion models), introduces an innovative framework focused on EEG-based image generation, combining advanced diffusion model techniques.

This paper presents several key innovations as follows:

\begin{itemize}
    \item We propose a novel EEG encoder, NERV, which achieves state-of-the-art performance in multimodal contrastive learning tasks.
    \item Unlike previous work that primarily focused on image-to-image generation with EEG features as guidance, we introduce a comprehensive two-stage EEG-to-image multimodal generative framework. This not only extends prior contrastive learning between EEG and images but also applies it to image generation.
    \item To address the conceptual differences between EEG-to-image and traditional text-to-image tasks, we propose a new quantification method, the Category-based Assessment Table (CAT) Score, which evaluates image generation performance based on semantic concepts rather than image distribution.
    \item We establish a CAT score benchmark standard using Vision Language Model (VLM) on the ThingsEEG dataset.
    \item Additionally, we uncover some notable findings and phenomena regarding the EEG-to-image generation process.
\end{itemize}

\section{Related work}
\label{sec:headings}


\subsection{The potential of EEG data}
In a typical experiment studying brain responses related to visual processes, a person looks at a series of images while a brain scanner or recording device captures their brain signals for analysis. There are various non-invasive methods to capture these brain responses, like fMRI, EEG, and MEG, each with different sensitivity levels. However, we still don't fully understand what this data really means, and even more importantly, how to interpret it. In a pioneering study \cite{NISHIMOTO20111641}, the researchers tried to generate impressions of what the subjects saw using fMRI images, based on a large image dataset taken from YouTube. However, this method has challenges, like the complexity and high cost of using an fMRI scanner. To overcome these drawbacks, a lot of research has shifted to using electrophysiological responses, particularly EEG, which has lower spatial resolution than most other methods but much higher temporal resolution. EEG recordings are also cheaper and easier to conduct, but the data is often noisy and affected by external factors, making it harder to reconstruct the original stimulus. Most image recognition and/or generation from brain signals nowadays is done using fMRI data \cite{zhang_zhang_zhang_kweon_2023}, while EEG, being noisier, is used much less often.

\subsection{Using EEG information on image generation and reconstruction}

Building on this shift towards EEG, prior to efforts in generating images directly from brain data, the concept of using EEG signals for image classification was introduced by the study \cite{8099962}. This work first demonstrated the feasibility of decoding visual categories from EEG recordings using deep learning models, setting a foundation for leveraging neural signals in image-related tasks. However, the dataset they used was relatively small, which limited the generalization of their findings. Further advancements in generative models, specifically with the introduction of Variational Autoencoders (VAE) and Generative Adversarial Networks (GAN), opened new possibilities for image generation. The VAE model proposed by \cite{kingma_welling_2013, kingma_welling_2019} achieved data generation and reconstruction by learning the latent distribution of data. The GAN model introduced by \cite{goodfellow_2014} utilized adversarial training between a generator and a discriminator to produce highly realistic images. Building on these methods, Brain2Image \cite{brain2img} was the first to use VAE to guide image generation from EEG features. Following that, EEG-GAN \cite{palazzo_spampinato_kavasidis_giordano_shah} presented the first EEG-based image generation model, using LSTM \cite{lstm_1997} to extract EEG information and guide the GAN for image generation. After this, there were still many EEG-to-image works based on GAN that emerged, with most of them focusing on improving the GAN architecture and the way it interacts with the EEG encoder, like in ThoughtViz \cite{ThoughtViz_2018}, VG-GAN-VC \cite{jiao_you_yang_li_zhang_shen_2019}, BrainMedia \cite{BrainMedia_2020}, and EEG2IMAGE \cite{singh_pandey_miyapuram_raman_2023}, etc. However, in all these works, a common and challenging problem is figuring out how to effectively use EEG data to guide image generation and reconstruction. 
This challenge of training neural networks to align multimodal information wasn't effectively addressed until the emergence of CLIP \cite{clip_2021}, which provided a much better solution. Since then, some works have also applied this approach to EEG-based image generation.

\subsection{Contrastive learning-based works on EEG-image tasks}
To the best of our knowledge, EEGCLIP \cite{EEGCLIP_2024} was the first to use contrastive learning to align EEG and image data. However, in this work, this aspect was only an exploratory attempt and did not further utilize the framework for downstream tasks like zero-shot image recognition. The next challenge lies in designing a better EEG encoder for contrastive learning, based on the rich image embeddings extracted from a CLIP-based image pre-trained encoder. Some recent works have explored this direction, such as NICE \cite{nice_2023}, MUSE \cite{muse_2024}, ATM \cite{atm_2024}, and \cite{vesdn_2024}. Some researchers have even attempted quantum-classical hybrid computing and quantum EEG encoder \cite{qeegnet_2024} to perform quantum contrastive learning \cite{chen_mcl_2024}. Most current works focus on tackling zero-shot classification, where the model is tested on unseen both EEG data and images that it hasn't encountered during training. The goal is to compute similarity scores for image recognition, aiming to enhance the model's generalization performance on out-of-sample data. As contrastive learning architectures for EEG-based image recognition mature, and inspired by test-to-image frameworks in other generative fields, the invention of diffusion models has addressed the instability issues associated with previous GAN-based generation methods to some extent. While there are already EEG-based image reconstruction efforts using diffusion models, such as NeuroVision \cite{khare_choubey_amar_udutalapalli_2022}, DreamDiffusion \cite{bai_wang_cao_ge_yuan_shan_2023}, DM-RE2I \cite{DM-RE2I_2023}, BrainViz \cite{BrainViz_2023}, NeuroImagen \cite{NeuroImagen_2023}, and EEGVision \cite{guo_2024}, most of these works still largely rely on image-based features, with EEG data serving as supplementary information for the diffusion process. While these methods have made significant strides in computer vision, they primarily rely on images as input and are not designed to process non-visual signals like EEG directly. Currently, models designed specifically for direct generation tasks using pure EEG features or embeddings, where EEG functions similarly to a prompt command, are still quite rare. This work seeks to introduce a flexible, plug-and-play architecture: NECOMIMI, which not only expands upon previous recognition-focused approaches but also extends them into EEG-to-image generation tasks based on modern diffusion models.

\section{Methodology}
\label{gen_inst}

\subsection{Overview}

\begin{figure}
    \centering
    \includegraphics[width=1\linewidth]{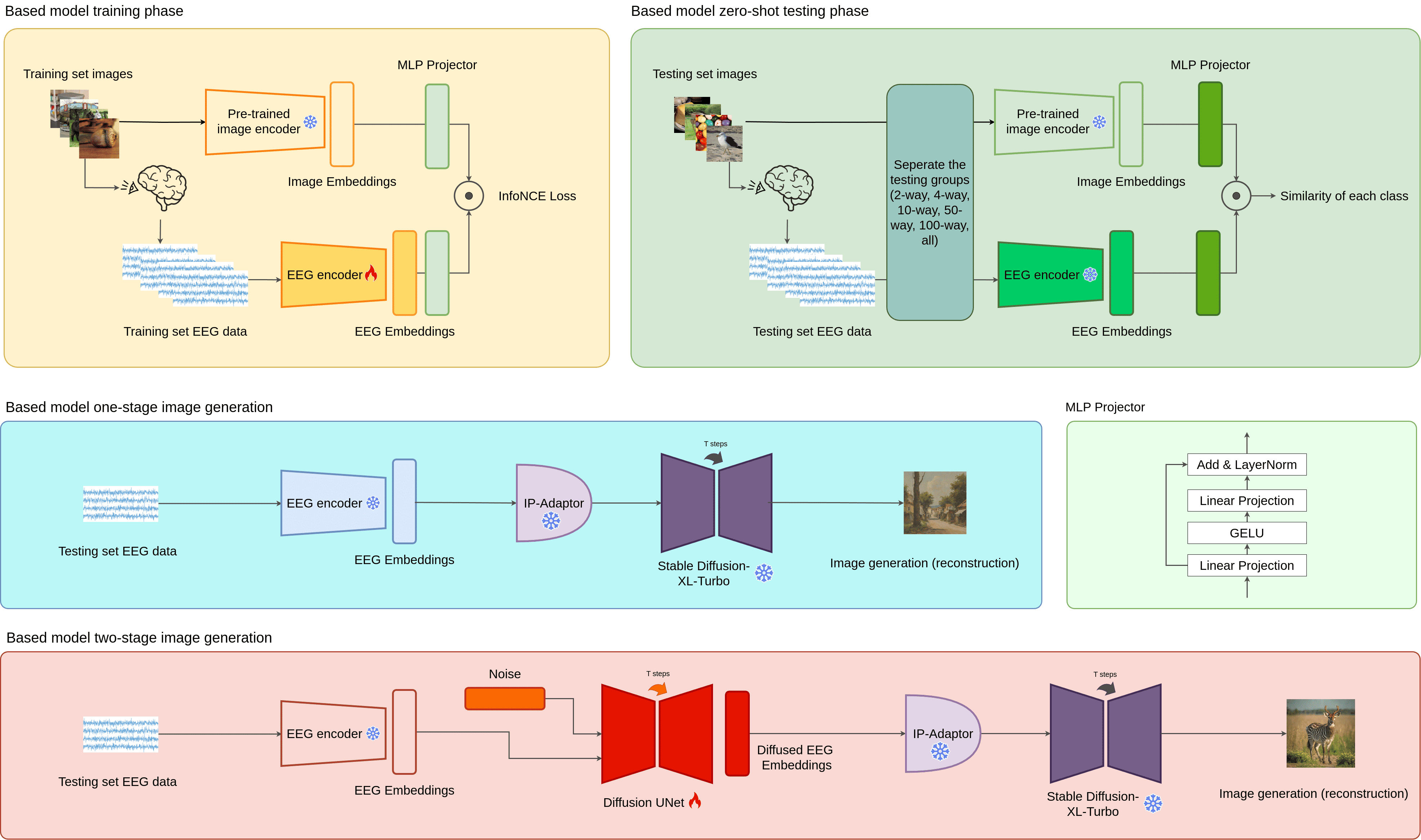}
    \caption{The figure illustrates the entire workflow of the EEG-based image generation model.}
    \label{fig:enter-label}
\end{figure}

This chapter provides a detailed overview of an advanced EEG-to-image generation model utilizing deep learning techniques and diffusion models. While the framework includes a one-stage image generation phase, we found that its performance was suboptimal. Consequently, the model is primarily designed as a two-stage process, which will be discussed in detail in later sections. The overall structure consists of four phases: the training phase, zero-shot testing, one-stage image generation, and two-stage image generation, each contributing to the transformation of raw EEG data into meaningful visual outputs.

\subsection{Training Phase}
In the initial training phase, both visual image $\in \mathbb{R}^{h \times w \times ch}$ and EEG data $\in \mathbb{R}^{e \times d}$ are processed in parallel to establish a shared embedding space, where $h$ is the height of the image, $w$ is the width of the image, $ch$ is the number of channels (e.g., RGB channels), $e$ is the number of electrodes (channels), and $d$ is the number of data points (time samples). Training set images are first passed through a pre-trained image encoder, which transforms the images into latent representations called image embeddings $I$. In this work, we use a pretrained Vision Transformer (ViT) \cite{vit_2020} from CLIP model \cite{clip_2021} as the image encoder, which outputs embeddings of size $\mathbb{R}^{1 \times 1024}$ for each image. Simultaneously, the EEG signals from the corresponding sessions are processed by a custom EEG encoder to produce EEG embeddings $E$. As for the EEG encoder, in this work, we extended several existing works like NICE \cite{nice_2023}, MUSE \cite{muse_2024}, Nervformer \cite{muse_2024} and ATM \cite{atm_2024} to enable EEG-to-image capabilities. Additionally, we proposed a new EEG encoder, NERV, which is specifically designed for noisy, multi-channel time series data like EEG, based on a multi-attention mechanism.

These embeddings are projected into a unified space via an MLP Projector, where they are trained using the InfoNCE loss. This contrastive learning loss function ensures that corresponding image and EEG embeddings are aligned in the latent space, enhancing the model's ability to understand and link neural patterns to visual stimuli. Standard contrastive learning employs the InfoNCE loss as defined by \cite{oord2018representation, he2020momentum, radford2021learning}:
\begin{equation}
\mathcal{L}_{InfoNCE} = -\mathbb{E}\left[\log \frac{\exp(S_{E, I}/\tau)}{\sum_{k=1}^N \exp(S_{E, I_k}/\tau)}\right]
\end{equation}

where the \(S_{E, I}\) represents the similarity score between the EEG embeddings $E$, and the paired image embeddings $I$, and the \(\tau\) is learned temperature parameter.

\subsection{Zero-shot Testing Phase}
Once trained, the model enters the zero-shot testing phase. This phase focuses on evaluating the model’s ability to generalize to unseen data. Here, the EEG signals and images from the test set are encoded using the pre-trained encoders, and their respective embeddings are projected through the MLP Projector. The testing groups are separated into multiple divisions—2-way, 4-way, 10-way, 50-way, 100-way and beyond—allowing for a structured comparison between the EEG and image embeddings. The final similarity scores between embeddings determine the model's classification accuracy, enabling the assessment of how well the model understands new EEG data without additional training.

\subsection{One-stage Image Generation}
In the one-stage image generation process, the EEG embeddings from the testing set are directly used as inputs to reconstruct images. By incorporating the IP-Adapter \cite{ipadaptor_2023}, which was originally designed to use images as prompts, due to its compact design, enhances image prompt flexibility within pre-trained text-to-image models. We adapt it in this work as a means to transform EEG embeddings into "feature prompts" for the image generation process.
The conditioned embeddings are then processed by the Stable Diffusion XL-Turbo model \cite{sdxl_2023, lcm_2024}, a faster version of Stable Diffusion XL designed for rapid image synthesis, which reconstructs the final images based on the input EEG data. This method offers a streamlined approach to EEG-based image generation, relying on a single transformation stage to produce meaningful visual outputs from neural signals. The start of the EEG-conditioned diffusion phase is critical for generating images based on EEG data. This phase uses a classifier-free guidance method, which pairs CLIP embeddings and EEG embeddings \((I, E)\). By applying advanced generative techniques, the diffusion process is adapted to use the EEG embedding \(E\) to model the distribution of the CLIP embeddings \(p(I|E)\). The CLIP embedding \(I\), generated during this stage, lays the foundation for the next phase of image generation. The architecture integrates a simplified U-Net model, represented as \(\epsilon_{\text{prior}}(I^t, t, E)\), where \(I^t\) is the noisy CLIP embedding at a specific diffusion step \(t\).

The classifier-free guidance method helps refine the diffusion model (DM) using a specific EEG condition \(E\). This approach synchronizes the outputs of both a conditional and an unconditional model. The final model equation is expressed as:

\begin{equation}
\epsilon_{\text{prior}}^{w}(I^t, t, E) = (1 + w)\epsilon_{\text{prior}}(I^t, t, E) - w\epsilon_{\text{prior}}(I^t, t),
\end{equation}

where \(w \geq 0\) controls the guidance scale. This technique allows for training both the conditional and unconditional models within the same network, periodically replacing the EEG embedding \(E\) with a null value to enhance training variation (about 10\% of the data points). The main goal is to improve the quality of generated images while maintaining diversity.

However, we were surprised to find that when using EEG embeddings directly as prompts for the diffusion model, the generated images mostly turned out to be landscapes, regardless of the category. We will discuss the detailed results in later sections. As a result, we attempted a 2-stage approach for image generation.

\subsection{Two-stage Image Generation}
The prior diffusion stage plays a crucial role in generating an intermediate representation \cite{632983}, such as a CLIP image embedding, from a text caption \cite{dalle_2022}. This representation is then used by the diffusion decoder to produce the final image. This two-stage process enhances image diversity, maintains photorealism, and allows for efficient and controlled image generation \cite{fmri_to_img_2023}. The two-stage image generation process introduces a more complex and refined method of synthesizing images from EEG data. In this approach, the EEG embeddings are first processed by a Diffusion U-Net, which applies additional transformations to enhance the representation of the neural data. After passing through the U-Net, the modified EEG embeddings are fed into the Stable Diffusion XL-Turbo model, with the assistance of the IP-Adaptor. This two-step transformation ensures a more nuanced generation process, potentially leading to higher-quality images by incorporating deeper layers of refinement. 
The first step of stage-1 is training the prior diffusion model. The main purpose of training is to let the model learn how to recover the original embedding from noisy embeddings. The specific steps are as follows: (a) Randomly replace conditional EEG embeddings $c_{\text{emb}}$ with None with a 10\% probability:
\begin{equation}
c_{\text{emb}} = \text{None}, \quad \text{if } \text{random}() < 0.1
\end{equation}
(b) Add random noise to the target embedding $h_{\text{emb}}$, perturb it using the scheduler at a timestep $t$, use the symbol $\mathcal{S}_{add\_noise}$ to represent the scheduler add noise function:
\begin{equation}
\hat{h}_{\text{emb}}(t) = \mathcal{S}_{add\_noise}(h_{\text{emb}}, \epsilon, t)
\end{equation}
where $\epsilon \sim \mathcal{N}(0, I)$ is the random noise, and $t$ is a randomly sampled timestep.
(c) The model receives the perturbed embedding $\hat{h}_{\text{emb}}(t)$ and conditional embedding $c_{\text{emb}}$, and predicts the noise. Use the symbol $\mathcal{D}_{\text{prior}}$ to represent the diffusion prior function:
\begin{equation}
\epsilon_{\text{pred}} = \mathcal{D}_{\text{prior}}(\hat{h}_{\text{emb}}(t), t, c_{\text{emb}})
\end{equation}
(d) Compute the loss using Mean Squared Error (MSE) between the predicted noise and the actual noise:
\begin{equation}
L = \frac{1}{N} \sum_{i=1}^{N} \left( \epsilon_{\text{pred}}^{(i)} - \epsilon^{(i)} \right)^2
\end{equation}
(e) Perform backpropagation on the loss $L$, and update the model parameters using the optimizer:
\begin{equation}
\theta \leftarrow \theta - \eta \nabla_{\theta} L
\end{equation}
where $\eta$ is the learning rate and $\theta$ represents the model's parameters.

The last step of stage-1 is generation process. The main purpose of the generation process is to gradually denoise and generate the final embedding based on the conditional EEG embedding $c_{\text{emb}}$, starting from random noise. The specific steps are as follows: (a) Generate a sequence of timesteps $t$, which will be used for the denoising process, define $\mathcal{T} = \{t_1, t_2, \dots, t_T\}$ to represent the set of time steps sampled from the total steps $T$:
\begin{equation}
\{t_1, t_2, \dots, t_T\} \sim \mathcal{T}(T)
\end{equation}
where $T$ is the total number of denoising steps.
(b) Initialize random noise embedding $h_T$, which serves as the starting point for the generation process:
\begin{equation}
h_T \sim \mathcal{N}(0, I)
\end{equation}
(c) Starting from timestep $T$, iteratively apply the model to predict noise and denoise the embedding until $t = 0$. Each step depends on the conditional embedding $c_{\text{emb}}$:

If using conditional embedding, perform both unconditional and conditional noise prediction at each step:
\begin{equation}
\epsilon_{\text{pred\_cond}} = \mathcal{D}_{\text{prior}}(h_t, t, c_{\text{emb}})
\end{equation}
\begin{equation}
\epsilon_{\text{pred\_uncond}} = \mathcal{D}_{\text{prior}}(h_t, t)
\end{equation}
Then combine the results using classifier-free guidance, define $\alpha_{\text{guide}}$ as the guidance scale:
\begin{equation}
\epsilon_{\text{pred}} = \epsilon_{\text{pred\_uncond}} + \alpha_{\text{guide}} \times (\epsilon_{\text{pred\_cond}} - \epsilon_{\text{pred\_uncond}})
\end{equation}

Finally, update the noisy embedding based on the predicted noise, use the symbol $\mathcal{S}_{step}$ to represent the scheduler step function:
\begin{equation}
h_{t-1} = \mathcal{S}_{step}(\epsilon_{\text{pred}}, t, h_t)
\end{equation}

(d) After the denoising process is complete, $h_{output}$ represents the final generated embedding of a EEG, which is the model's output:
\begin{equation}
h_{output} = h_{\text{generated}} \in \mathbb{R}^{1 \times 1024}
\end{equation}

The stage-2 is input the $h_{output}$ into the IP-adaptor as a prompt to generate the image by Stable Diffusion XL-Turbo model.

\subsection{NERV EEG Encoder}

\begin{figure}
    \centering
    \includegraphics[width=.7\linewidth]{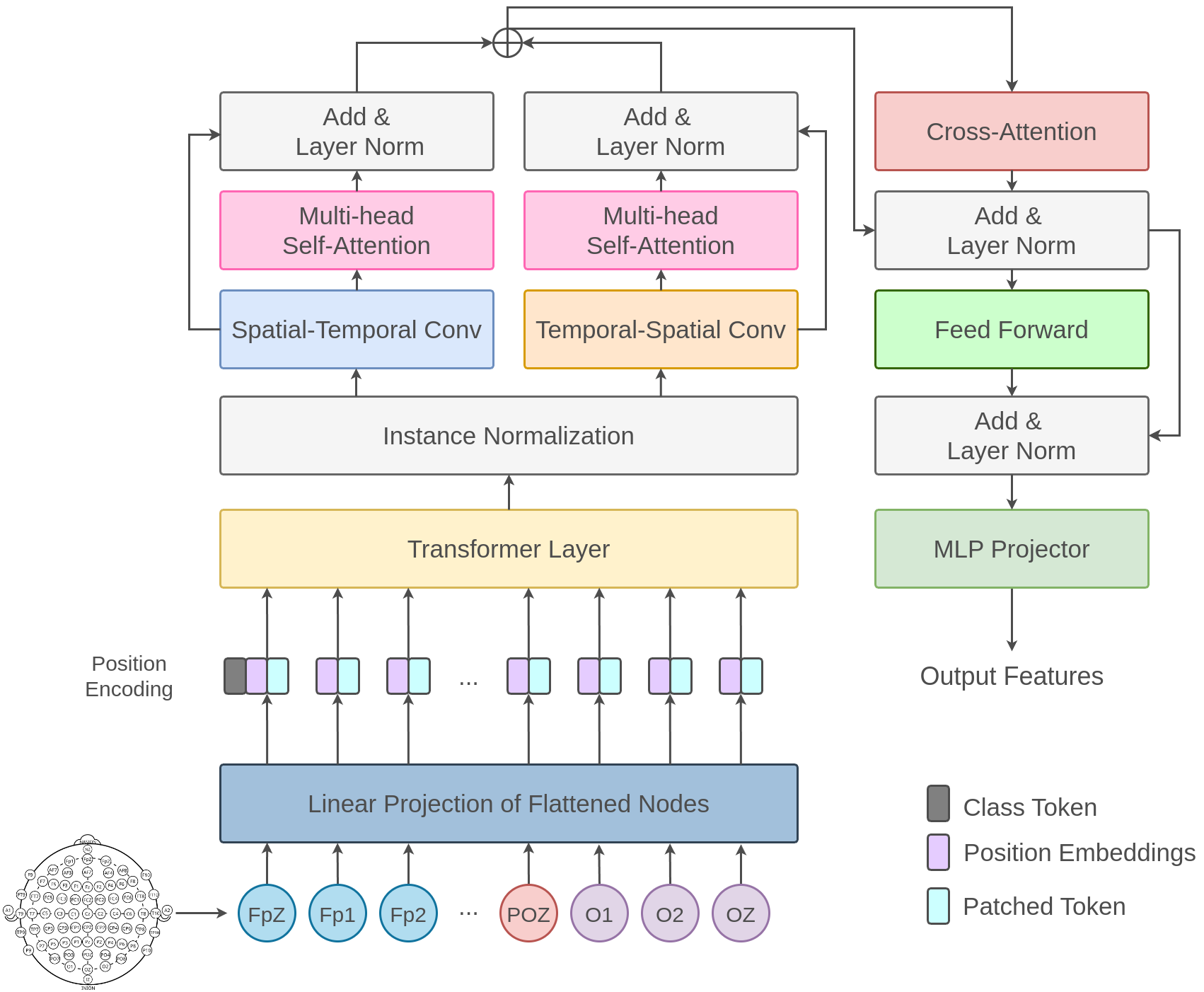}
    \caption{This diagram shows the overall structure and workflow of the NERV EEG encoder model.}
    \label{fig:nerv2}
\end{figure}

This diagram \ref{fig:nerv2} illustrates the structure of NERV, a neural network encoder designed for EEG signal processing. The workflow starts with a linear projection of the flattened EEG nodes, followed by position encoding to retain temporal information. EEG signals pass through a Transformer layer and undergo instance normalization. The model then applies both spatial-temporal convolution (blue) to extract spatial features followed by temporal features and temporal-spatial convolution (yellow) to extract temporal features first, then spatial features. Multi-head self-attention mechanisms are applied to both feature sets, followed by layer normalization and residual connections. The cross-attention block (red) fuses the temporal and spatial features, which are further processed by a feed-forward layer before final output. The class token, position embeddings, and patch tokens are all part of the input sequence processed through these steps, ultimately yielding the output features for EEG-based tasks.

\subsection{Category-based Assessment Table (CAT) Score}

Unlike traditional image-to-image or text-to-image models driven by image representations, EEG-to-image models face unique challenges. In the current NECOMIMI architecture, the model can only capture broad semantic information from EEG signals rather than fine-grained details. For example, suppose the ground truth EEG data was recorded while a subject was observing an aircraft carrier. When using Model A as the EEG encoder in NECOMIMI, the generated image is a jet, while using Model B results in an image of a sheep. To objectively assess performance, we need a standard that scores Model A higher than Model B in such cases.

Why not use existing evaluation metrics? Traditional metrics like Structural Similarity Index (SSIM) \cite{SSIM} measure structural similarity between the ground truth and generated image, while the Inception Score (IS) \cite{is_2016} and Fréchet Inception Distance (FID) \cite{NIPS2017_FID} focus on the accuracy of image categories and its distribution. However, EEG captures more abstract semantic information, and we cannot guarantee that the subject's thoughts during EEG recording perfectly align with the ground truth image. This makes traditional evaluation methods unfair for EEG-to-image tasks.

To address this, we propose the Category-Based Assessment Table (CAT) Score, a new metric specifically designed for EEG-to-image evaluation. In the ThingsEEG test dataset (which contains 200 categories with one image per category), each image is manually labeled with two tags for broad categories, one for a specific category, and one for background content, resulting in a total of five tags per image. We extracted the tags by ChatGPT-4o \cite{openai_2023}. The entire test dataset thus comprises 200 images × 5 tags = 1,000 points. Using manual annotation, we can determine whether the categories of generated images match these labels, providing a fair assessment for EEG-to-image models. For more details on the ThingsEEG categories, please refer to the appendix.

\section{Experiments}
\subsection{Datasets and Preprocessing}
The ThingsEEG dataset \cite{gifford2022large} consists of a large set of EEG recordings obtained through a rapid serial visual presentation (RSVP) paradigm. The responses were collected from 10 participants who viewed a total of 16,740 natural images from the THINGS database \cite{hebart2019things}. The dataset contains 1654 training categories, each with 10 images, and 200 test categories, each with a single image. The EEG data were recorded using 64-channel EASYCAP equipment, and preprocessing involved segmenting the data into trials from 0 to 1000 ms after the stimulus was shown, with baseline correction based on the pre-stimulus period. EEG responses for each image were averaged over multiple repetitions.

\subsection{Experiment Details}
Due to the significant impact that different versions of the CLIP package can have on the results of contrastive learning, this work ensures a fair comparison of various EEG encoders by rerunning all experiments using a unified CLIP-ViT environment, where available open-source code (e.g., \cite{nice_2023}\footnote{\url{https://github.com/eeyhsong/NICE-EEG}}, \cite{muse_2024}\footnote{\url{https://github.com/ChiShengChen/MUSE_EEG}}, \cite{atm_2024}\footnote{\url{https://github.com/dongyangli-del/EEG_Image_decode}}) was utilized. Another factor that can influence contrastive learning is batch size. Therefore, all experiments in this work were conducted with a batch size of 1024. The final results are averaged from the best outcomes of 5 random seed training sessions, each running for 200 epochs. We employ the AdamW optimizer, setting the learning rate to 0.0002 and parameters \(\beta_1\)=0.5 and \(\beta_2\)=0.999. The \(\tau\) in contrastive learning initialized with $log(1/0.07)$. The NERV model achieves the best results with 5 multi-heads, while the Transformer layer has 1 multi-head and the cross-attention layer has 8 multi-heads. The time step is 50 in diffusion model.  All experiments, including both EEG encoder training and prior diffusion model processing, were performed on a machine equipped with an A100 GPU. 

\subsection{Classification Results}
In Table \ref{tab:per_com_2_4}, the classification accuracy for both 2-way and 4-way zero-shot tasks is evaluated across ten subjects. Our new model NERV consistently achieves the best performance, particularly excelling in the 2-way classification task, where it maintains top accuracy across most subjects. It achieves an average accuracy of 94.8\% in the 2-way classification and 86.8\% in the 4-way classification, outperforming other methods like NICE \cite{nice_2023}, MUSE \cite{muse_2024}, and ATM-S \cite{atm_2024}. While NICE and MUSE perform strongly in some subjects, they often fall short of NERV's performance. NICE has an average of 91.3\% in the 2-way task and 81.3\% in the 4-way task, with MUSE trailing behind with averages of 92.2\% (2-way) and 82.8\% (4-way). ATM-S performs comparably to NICE and MUSE in some subjects but falls short on average with 86.5\% in the 4-way classification. In Table \ref{tab:per_com_200}, the results for the more challenging 200-way zero-shot classification task show that NERV also performs the best, especially in the top-1 accuracy. ATM-S and NERV perform similarly, but NERV shows stronger performance in most subjects. NERV achieves an average top-1 accuracy of 27.9\% and top-5 accuracy of 54.7\%, leading over all other methods. In contrast, Nervformer \cite{muse_2024} and BraVL \cite{du2023decoding} show weaker performance, especially in the top-1 accuracy, where they average 19.8\% and 5.8\%, respectively. For the results of other 10-way, 50-way, and 100-way zero-shot classifications, please refer to the appendix. In summary, NERV consistently outperforms its competitors in both tasks, demonstrating the strongest zero-shot classification capability, particularly when distinguishing between a large number of categories, making it the most effective model in these experiments.

\begin{table}[h]
  \caption{Overall accuracy (\%) of 2-way and 4-way zero-shot classification using CLIP-ViT as image encoder: top-1 and top-5. The parts in bold represent the best results, while the underlined parts are the second best.}
  \label{tab:per_com_2_4}
  \centering
  \Huge
\resizebox{\linewidth}{!}{
  \begin{tabular}{lcccccccccccccccccccccc}
    \toprule
    & \multicolumn{2}{c}{Subject 1} & \multicolumn{2}{c}{Subject 2} & \multicolumn{2}{c}{Subject 3} & \multicolumn{2}{c}{Subject 4} & \multicolumn{2}{c}{Subject 5} & \multicolumn{2}{c}{Subject 6} & \multicolumn{2}{c}{Subject 7} & \multicolumn{2}{c}{Subject 8} & \multicolumn{2}{c}{Subject 9} & \multicolumn{2}{c}{Subject 10} & \multicolumn{2}{c}{Ave} \\
    \cmidrule(r){2-3} \cmidrule(r){4-5} \cmidrule(r){6-7} \cmidrule(r){8-9} \cmidrule(r){10-11} \cmidrule(r){12-13} \cmidrule(r){14-15} \cmidrule(r){16-17} \cmidrule(r){18-19} \cmidrule(r){20-21} \cmidrule(r){22-23}
    Method & 2-way & 4-way & 2-way & 4-way & 2-way & 4-way & 2-way & 4-way & 2-way & 4-way & 2-way & 4-way & 2-way & 4-way & 2-way & 4-way & 2-way & 4-way & 2-way & 4-way & 2-way & 4-way \\
    \midrule
    \multicolumn{23}{c}{Subject dependent - train and test on one subject} \\
    \midrule
    Nervformer & 89.9 & 76.9 & 91.3 & 80.7 & 91.6 & 80.8 & 94.3 & 85.9 & 86.3 & 70.4 & 91.1 & 82.5 & 92.5 & 81.6 & 96.2 & 88.3 & 92.0 & 83.7 & 92.4 & 83.1 & 91.8 & 81.4\\
    NICE & 91.7 & 80.4 & 89.8 & 77.4 & 93.5 & 83.7 & 94.0 & 84.9 & 85.9 & 70.3 & 89.1 & 81.7 & 91.2 & 81.7 & 95.8 & 89.2 & 87.9 & 76.5 & 93.8 & 87.1 & 91.3 & 81.3 \\
    MUSE & 90.1 & 78.4 & 90.3 & 76.8 & 93.4 & 85.6 & 93.6 & 87.5 & 88.3 & 74.2 & 93.1 & 85.3 & 93.1 & 82.8 & 95.4 & 87.7 & 90.5 & 81.8 & 94.4 & 88.1 & 92.2 & 82.8\\
    ATM-S & 94.8 & 84.9 & 93.5 & 86.3 & 95.3 & 89.0 & 95.9 & 87.3 & 90.8 & 78.5 & 94.1 & 85.2 & 94.2 & 87.1 & 96.6 & 92.9 & 94.1 & 86.8 & 94.7 & 87.0 & 94.4 & 86.5\\
    NERV (ours) & 95.3 & 85.7 & 96.0 & 88.8 & 95.9 & 91.2 & 95.8 & 87.4 & 90.8 & 80.4 & 93.6 & 84.0 & 94.7 & 86.2 & 96.8 & 92.3 & 94.4 & 84.2 & 94.8 & 87.6 & \textbf{94.8} & \textbf{86.8}\\
    \bottomrule
  \end{tabular}}
\end{table}

\begin{table}[h]
  \caption{Overall accuracy (\%) of 200-way zero-shot classification using CLIP-ViT as image encoder: top-1 and top-5. The parts in bold represent the best results, while the underlined parts are the second best.}
  \label{tab:per_com_200}
  \centering
  \Huge
\resizebox{\linewidth}{!}{
  \begin{tabular}{lcccccccccccccccccccccc}
    \toprule
    & \multicolumn{2}{c}{Subject 1} & \multicolumn{2}{c}{Subject 2} & \multicolumn{2}{c}{Subject 3} & \multicolumn{2}{c}{Subject 4} & \multicolumn{2}{c}{Subject 5} & \multicolumn{2}{c}{Subject 6} & \multicolumn{2}{c}{Subject 7} & \multicolumn{2}{c}{Subject 8} & \multicolumn{2}{c}{Subject 9} & \multicolumn{2}{c}{Subject 10} & \multicolumn{2}{c}{Ave} \\
    \cmidrule(r){2-3} \cmidrule(r){4-5} \cmidrule(r){6-7} \cmidrule(r){8-9} \cmidrule(r){10-11} \cmidrule(r){12-13} \cmidrule(r){14-15} \cmidrule(r){16-17} \cmidrule(r){18-19} \cmidrule(r){20-21} \cmidrule(r){22-23}
    Method & top-1 & top-5 & top-1 & top-5 & top-1 & top-5 & top-1 & top-5 & top-1 & top-5 & top-1 & top-5 & top-1 & top-5 & top-1 & top-5 & top-1 & top-5 & top-1 & top-5 & top-1 & top-5 \\
    \midrule
    \multicolumn{23}{c}{Subject dependent - train and test on one subject} \\
    \midrule
    BraVL & 6.1 & 17.9 & 4.9 & 14.9 & 5.6 & 17.4 & 5.0 & 15.1 & 4.0 & 13.4 & 6.0 & 18.2 & 6.5 & 20.4 & 8.8 & 23.7 & 4.3 & 14.0 & 7.0 & 19.7 & 5.8 & 17.5 \\   
    Nervformer & 15.0 & 36.7 & 15.6 & 40.0 & 19.7 & 44.9 & 23.3 & 54.4 & 13.0 & 29.1 & 18.9 & 42.2 & 19.5 & 42.0 & 30.3 & 60.0 & 20.1 & 46.3 & 22.9 & 47.1 & 19.8 & 44.3\\
    NICE & 19.3 & 44.8 & 15.2 & 38.2 & 23.9 & 51.4 & 24.1 & 51.6 & 11.0 & 30.7 & 18.5 & 43.8 & 21.0 & 47.9 & 32.5 & 63.5 & 18.2 & 42.4 & 27.4 & 57.1 & 21.1 & 47.1 \\
    MUSE & 19.8 & 41.1 & 15.3 & 34.2 & 24.7 & 52.6 & 24.7 & 52.6 & 12.1 & 33.7 & 22.1 & 51.9 & 21.0 & 48.6 & 33.2 & 59.9 & 19.1 & 43.0 & 25.0 & 55.2 & 21.7 & 47.3\\
    ATM-S & 25.8 & 54.1 & 24.6 & 52.6 & 28.4 & 62.9 & 25.9 & 57.8 & 16.2 & 41.9 & 21.2 & 53.0 & 25.9 & 57.2 & 37.9 & 71.1 & 26.0 & 53.9 & 30.0 & 60.9 & \underline{26.2} & \textbf{56.5}\\
    NERV (ours) & 25.4 & 51.2 & 24.1 & 51.1 & 28.6 & 53.9 & 30.0 & 58.4 & 19.3 & 43.9 & 24.9 & 52.3 & 26.1 & 51.6 & 40.8 & 67.4 & 27.0 & 55.2 & 32.3 & 61.6 & \textbf{27.9} & \underline{54.7}\\
    \bottomrule
  \end{tabular}}
\end{table}

\subsection{Performance Comparison of different generative models}
Here, we introduce our newly proposed CAT Score method, which quantifies and evaluates the quality of EEG-generated images based on semantic concepts rather than pixel structure. Detailed CAT Score labels can be found in the appendix. To our surprise, while our proposed NERV method achieved SoTA on the CAT Score, no EEG encoder has surpassed a score of 500 in this evaluation out of a possible 1000 points. This highlights both the rigor of the CAT Score and the challenging nature of the pure EEG-to-Image task.

\begin{table}[h]
  \caption{Overall CAT score $\times 1000$ of NECOMIMI EEG-to-Image generation with several EEG encoders.}
  \label{tab:cat_com}
  \centering
  \Huge
\resizebox{\linewidth}{!}{
  \begin{tabular}{lccccccccccc}
    \toprule
    & Subject 1 & Subject 2 & Subject 3 & Subject 4 & Subject 5 & Subject 6 & Subject 7 & Subject 8 & Subject 9 & Subject 10 & Ave \\
    \midrule
    EEG Encoder &  &  &  &  &  & CAT Score &  &  &  &  &  \\
    \midrule
    Nervformer & 432 & 457 & 429 & 454 & 475 & 463 & 404 & 438 & 427 & 410 & 438.9 \\
    NICE & 426 & 456 & 445 & 447 & 411 & 454 & 438 & 443 & 426 & 429 & 437.5 \\
    MUSE & 438 & 456 & 434 & 416 & 426 & 463 & 443 & 437 & 410 & 468 &  439.1\\
    ATM-S & 413 & 419 & 411 & 464 & 427 & 469 & 442 & 472 & 431 & 445 &  \underline{439.3} \\
    NERV (ours) & 445 & 436 & 432 & 456 & 438 & 466 & 410 & 437 & 433 & 444 & \textbf{439.7} \\
    \bottomrule
  \end{tabular}}
\end{table}

\subsection{Findings in EEG-to-Image}
\begin{figure}
    \centering
    \includegraphics[width=1\linewidth]{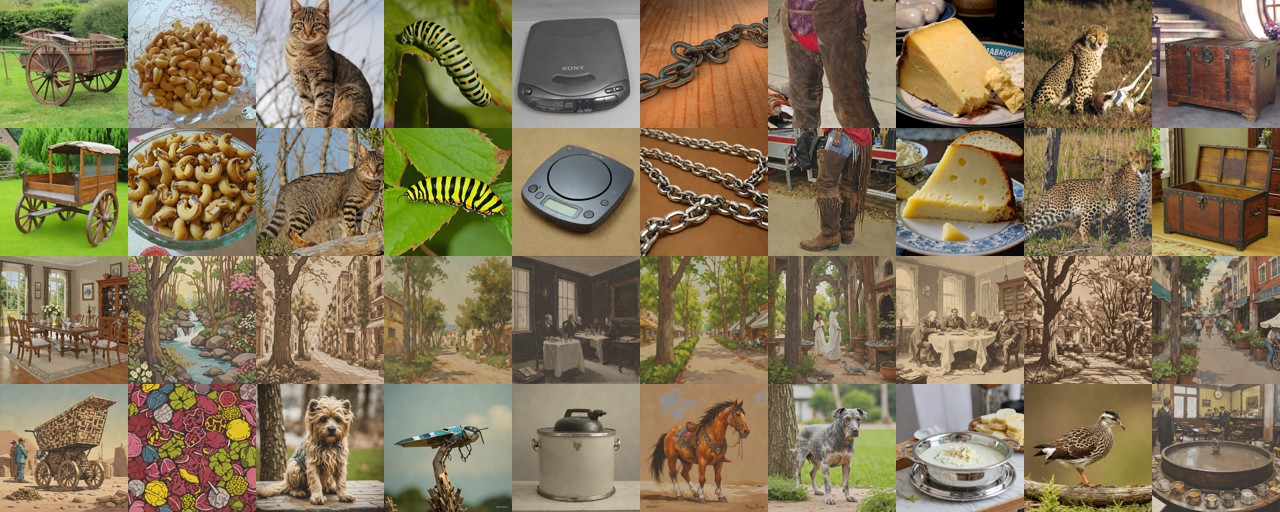}
    \caption{The image illustrates the progression of visual representations generated using different embedding techniques in a diffusion model: (a) Top row: The original images shown to subjects (ground truth). (b) Second row: Images generated by the CLIP-ViT embeddings of the original images. (c) Third row: Images generated by one-stage method using pure EEG embeddings with NERV EEG encoder. (d) Fourth row: Images generated by two-stage NECOMIMI method using pure EEG embeddings with NERV EEG encoder.}
    \label{fig:gen_img}
\end{figure}

We have observed some interesting findings from the pure EEG-to-Image process. As shown in the third row of Figure \ref{fig:gen_img}, the images generated by the diffusion model from embeddings compressed from EEG signals mainly consist of landscapes, which differ significantly from the original images (ground truth). Several factors may contribute to this phenomenon. For example, EEG signals are a high-noise, low-resolution form of data, capturing only certain aspects of brain activity. Moreover, we are currently unable to assess whether the brainwave data recorded from the subjects accurately captures the complete information of the original images, as the subjects might have been distracted and thinking about other things during the recording. This makes it difficult for the embeddings extracted from EEG signals to capture sufficient details, particularly when it comes to high-resolution object recognition (such as cats or specific items). As a result, the model tends to generate relatively vague or abstract images, like landscapes. Alternatively, the EEG signals may reflect higher-level abstract concepts or emotions associated with viewing the images rather than concrete objects or details. Since these abstract concepts are often related to the scene, background, or the brain’s broad perception of the environment, the model is more likely to generate abstract or general images, such as landscapes, instead of specific objects.

Additionally, the training of the model on EEG signals may still be insufficient. The diffusion model may not yet fully understand and generate images from EEG signals, especially when it lacks enough data or optimization to map EEG signals to specific visual information. As a result, the model might more easily generate the types of images it is "accustomed" to producing, such as landscapes, which may constitute a significant portion of the training data. The gap between the vision modality and the neural modality (EEG) is also substantial. EEG signals may not directly correspond to detailed objects in images, so the model tends to generate "safe options," like landscapes, which may have been more prevalent in the image generation samples during training. This leads to what can be described as "hallucinations." These factors collectively contribute to the significant differences between the images generated from EEG signals and the ground truth, particularly the failure in specific object recognition. This work can be considered a forward-looking exploration, as this field is just beginning to develop.

\section{Discussion and Conclusion}
The NECOMIMI framework expands previous works on EEG-Image contrastive learning classification by enabling image generation, filling a gap in prior research and opening new possibilities for EEG applications. We introduced the SoTA EEG encoder, NERV, which achieved top performance in 2-way, 4-way, and 200-way zero-shot classification tasks, as well as in the CAT Score evaluation, demonstrating its effectiveness in EEG-based generative tasks. A key finding is that the model often generates abstract images, like landscapes, rather than specific objects. This suggests that EEG data, being noisy and low-resolution, captures broad semantic concepts rather than detailed visuals. The gap between neural signals and visual stimuli remains a challenge for precise image generation. We also proposed the CAT Score, a new metric tailored for EEG-to-image generation, and established its benchmark on the ThingsEEG dataset. Surprisingly, we found that EEG encoder performance may not strongly correlate with the quality of generated images, providing new insights into the limitations and challenges of this task. In conclusion, NECOMIMI demonstrates the potential of EEG-to-image generation while highlighting the complexities of translating neural signals into accurate visual representations. Future research should focus on refining models to better capture detailed information from EEG signals.

\section*{Acknowledgment}

 I would like to thank Aidan Hung-Wen Tsai for brainstorming this fascinating title with me, which originated from a conversation in Silicon Valley about NeuroSky's cat-ear EEG product, Necomimi. I also appreciate Neuro Industry, Inc. for providing A100 computational resources on GCP. Special thanks to Dr. Guan-Ying Chen for proofreading and providing valuable feedback on revisions.

\bibliographystyle{unsrtnat}
\bibliography{references}  

\begin{thebibliography}{57}
\providecommand{\natexlab}[1]{#1}
\providecommand{\url}[1]{\texttt{#1}}
\expandafter\ifx\csname urlstyle\endcsname\relax
  \providecommand{\doi}[1]{doi: #1}\else
  \providecommand{\doi}{doi: \begingroup \urlstyle{rm}\Url}\fi

\bibitem[Mary(1959)]{mary_1959}
Mary.
\newblock The eeg in epilepsy a historical note.
\newblock \emph{Epilepsia}, 1\penalty0 (1-5):\penalty0 328–336, Jan 1959.
\newblock \doi{https://doi.org/10.1111/j.1528-1157.1959.tb04270.x}.
\newblock URL \url{https://onlinelibrary.wiley.com/doi/10.1111/j.1528-1157.1959.tb04270.x}.

\bibitem[Millett(2001)]{millett_2001}
David Millett.
\newblock Hans berger: From psychic energy to the eeg.
\newblock \emph{Perspectives in Biology and Medicine}, 44\penalty0 (4):\penalty0 522–542, Sep 2001.
\newblock \doi{https://doi.org/10.1353/pbm.2001.0070}.
\newblock URL \url{https://muse.jhu.edu/article/26086}.

\bibitem[Reif et~al.(2016)Reif, Strzelczyk, and Rosenow]{reif_strzelczyk_rosenow_2016}
Philipp~S Reif, Adam Strzelczyk, and Felix Rosenow.
\newblock The history of invasive eeg evaluation in epilepsy patients.
\newblock \emph{Seizure}, 41:\penalty0 191–195, Apr 2016.
\newblock \doi{https://doi.org/10.1016/j.seizure.2016.04.006}.
\newblock URL \url{https://www.seizure-journal.com/article/S1059-1311(16)30022-X/fulltext}.

\bibitem[Li et~al.(2023)Li, Chen, Cheng, Chen, Chen, Chen, Bai, and Tsai]{li_chen_cheng_chen_chen_chen_bai_tsai_2023}
Cheng-Ta Li, Chi-Sheng Chen, Chih-Ming Cheng, Chung-Ping Chen, Jen-Ping Chen, Mu-Hong Chen, Ya-Mei Bai, and Shih-Jen Tsai.
\newblock Prediction of antidepressant responses to non-invasive brain stimulation using frontal electroencephalogram signals: Cross-dataset comparisons and validation.
\newblock \emph{Journal of Affective Disorders}, 343:\penalty0 86–95, Dec 2023.
\newblock \doi{https://doi.org/10.1016/j.jad.2023.08.059}.
\newblock URL \url{https://www.sciencedirect.com/science/article/abs/pii/S0165032723010388}.

\bibitem[Hussain et~al.(2022)Hussain, Hossain, Jany, Hossain, Uddin, Kamal, Ku, and Kim]{Hussain2022Quantitative}
I.~Hussain, Md.~Azam Hossain, Rafsan Jany, Md.~Azam Hossain, M.~Uddin, A.~Kamal, Y.~Ku, and Jik-Soo Kim.
\newblock Quantitative evaluation of eeg-biomarkers for prediction of sleep stages.
\newblock \emph{Sensors (Basel, Switzerland)}, 22, 2022.
\newblock \doi{10.3390/s22083079}.

\bibitem[Thoma et~al.(2003)Thoma, Hanlon, Moses, Edgar, Huang, Weisend, Irwin, Sherwood, Paulson, Bustillo, Adler, Miller, and Cañive]{Thoma2003Lateralization}
R.~Thoma, F.~Hanlon, S.~Moses, J.~Christopher Edgar, Mingxiong Huang, M.~Weisend, J.~Irwin, A.~Sherwood, K.~Paulson, J.~Bustillo, L.~Adler, Gregory~A. Miller, and J.~Cañive.
\newblock Lateralization of auditory sensory gating and neuropsychological dysfunction in schizophrenia.
\newblock \emph{The American journal of psychiatry}, 160 9:\penalty0 1595--605, 2003.
\newblock \doi{10.1176/APPI.AJP.160.9.1595}.

\bibitem[Perrottelli et~al.(2021)Perrottelli, Giordano, Brando, Giuliani, and Mucci]{Perrottelli2021EEG-Based}
A.~Perrottelli, G.~Giordano, F.~Brando, L.~Giuliani, and A.~Mucci.
\newblock Eeg-based measures in at-risk mental state and early stages of schizophrenia: A systematic review.
\newblock \emph{Frontiers in Psychiatry}, 12, 2021.
\newblock \doi{10.3389/fpsyt.2021.653642}.

\bibitem[EK;Frey(2016)]{frey_2016}
Louis EK;Frey.
\newblock Electroencephalography (eeg): An introductory text and atlas of normal and abnormal findings in adults, children, and infants [internet], 2016.
\newblock URL \url{https://pubmed.ncbi.nlm.nih.gov/27748095/}.

\bibitem[Li et~al.(2022)Li, Yang, Fang, Hong, Reiss, and Zhang]{Li2022Concurrent}
Rihui Li, Dalin Yang, Feng Fang, K.~Hong, A.~Reiss, and Yingchun Zhang.
\newblock Concurrent fnirs and eeg for brain function investigation: A systematic, methodology-focused review.
\newblock \emph{Sensors (Basel, Switzerland)}, 22, 2022.
\newblock \doi{10.3390/s22155865}.

\bibitem[Mai et~al.(2023)Mai, Zhang, Fang, and Zhang]{mai_zhang_fang_zhang_2023}
Weijian Mai, Jian Zhang, Pengfei Fang, and Zhijun Zhang.
\newblock Brain-conditional multimodal synthesis: A survey and taxonomy, 2023.
\newblock URL \url{https://arxiv.org/abs/2401.00430}.

\bibitem[Spampinato et~al.(2016)Spampinato, Palazzo, Kavasidis, Giordano, Shah, and Souly]{spampinato_2016}
Concetto Spampinato, Simone Palazzo, Isaak Kavasidis, Daniela Giordano, Mubarak Shah, and Nasim Souly.
\newblock Deep learning human mind for automated visual classification, 2016.
\newblock URL \url{https://arxiv.org/abs/1609.00344}.

\bibitem[Kavasidis et~al.(2017)Kavasidis, Palazzo, Spampinato, Giordano, and Shah]{brain2img}
Isaak Kavasidis, Simone Palazzo, Concetto Spampinato, Daniela Giordano, and Mubarak Shah.
\newblock Brain2image: Converting brain signals into images.
\newblock In \emph{Proceedings of the 25th ACM International Conference on Multimedia}, MM '17, page 1809–1817, New York, NY, USA, 2017. Association for Computing Machinery.
\newblock ISBN 9781450349062.
\newblock \doi{10.1145/3123266.3127907}.
\newblock URL \url{https://doi.org/10.1145/3123266.3127907}.

\bibitem[Goodfellow et~al.(2014)Goodfellow, Pouget-Abadie, Mirza, Xu, Warde-Farley, Ozair, Courville, and Bengio]{goodfellow_2014}
Ian~J Goodfellow, Jean Pouget-Abadie, Mehdi Mirza, Bing Xu, David Warde-Farley, Sherjil Ozair, Aaron Courville, and Yoshua Bengio.
\newblock Generative adversarial networks, 2014.
\newblock URL \url{https://arxiv.org/abs/1406.2661}.

\bibitem[Ho et~al.(2020)Ho, Jain, and Abbeel]{ho_jain_abbeel_2020}
Jonathan Ho, Ajay Jain, and Pieter Abbeel.
\newblock Denoising diffusion probabilistic models, 2020.
\newblock URL \url{https://arxiv.org/abs/2006.11239}.

\bibitem[Kumar et~al.(2017)Kumar, Saini, Roy, Sahu, and Dogra]{kumar_2017}
Pradeep Kumar, Rajkumar Saini, Partha~Pratim Roy, Pawan~Kumar Sahu, and Debi~Prosad Dogra.
\newblock Envisioned speech recognition using eeg sensors.
\newblock \emph{Personal and Ubiquitous Computing}, 22\penalty0 (1):\penalty0 185–199, Sep 2017.
\newblock \doi{https://doi.org/10.1007/s00779-017-1083-4}.
\newblock URL \url{https://link.springer.com/article/10.1007/s00779-017-1083-4}.

\bibitem[Singh et~al.(2023)Singh, Pandey, Miyapuram, and Raman]{singh_pandey_miyapuram_raman_2023}
Prajwal Singh, Pankaj Pandey, Krishna Miyapuram, and Shanmuganathan Raman.
\newblock Eeg2image: Image reconstruction from eeg brain signals, 2023.
\newblock URL \url{https://arxiv.org/abs/2302.10121}.

\bibitem[Bai et~al.(2023)Bai, Wang, Cao, Ge, Yuan, and Shan]{bai_wang_cao_ge_yuan_shan_2023}
Yunpeng Bai, Xintao Wang, Yan-pei Cao, Yixiao Ge, Chun Yuan, and Ying Shan.
\newblock Dreamdiffusion: Generating high-quality images from brain eeg signals, 2023.
\newblock URL \url{https://arxiv.org/abs/2306.16934}.

\bibitem[Palazzo et~al.(2017)Palazzo, Spampinato, Kavasidis, Giordano, and Shah]{palazzo_spampinato_kavasidis_giordano_shah}
S.~Palazzo, C.~Spampinato, I.~Kavasidis, D.~Giordano, and M.~Shah.
\newblock Generative adversarial networks conditioned by brain signals.
\newblock In \emph{2017 IEEE International Conference on Computer Vision (ICCV)}, pages 3430--3438, 2017.
\newblock \doi{10.1109/ICCV.2017.369}.

\bibitem[Khare et~al.(2022)Khare, Choubey, Amar, and Udutalapalli]{khare_choubey_amar_udutalapalli_2022}
Sanchita Khare, Rajiv~Nayan Choubey, Loveleen Amar, and Venkanna Udutalapalli.
\newblock Neurovision: perceived image regeneration using cprogan.
\newblock \emph{Neural Computing and Applications}, 34\penalty0 (8):\penalty0 5979–5991, Jan 2022.
\newblock \doi{https://doi.org/10.1007/s00521-021-06774-1}.
\newblock URL \url{https://link.springer.com/article/10.1007/s00521-021-06774-1}.

\bibitem[Nishimoto et~al.(2011)Nishimoto, Vu, Naselaris, Benjamini, Yu, and Gallant]{NISHIMOTO20111641}
Shinji Nishimoto, An T. Vu, Thomas Naselaris, Yuval Benjamini, Bin Yu, and Jack L. Gallant.
\newblock Reconstructing visual experiences from brain activity evoked by natural movies.
\newblock \emph{Current Biology}, 21\penalty0 (19):\penalty0 1641--1646, 2011.
\newblock ISSN 0960-9822.
\newblock \doi{https://doi.org/10.1016/j.cub.2011.08.031}.
\newblock URL \url{https://www.sciencedirect.com/science/article/pii/S0960982211009377}.

\bibitem[Zhang et~al.(2023)Zhang, Zhang, Zhang, and Kweon]{zhang_zhang_zhang_kweon_2023}
Chenshuang Zhang, Chaoning Zhang, Mengchun Zhang, and In~So Kweon.
\newblock Text-to-image diffusion models in generative ai: A survey, 2023.
\newblock URL \url{https://arxiv.org/abs/2303.07909}.

\bibitem[Spampinato et~al.(2017)Spampinato, Palazzo, Kavasidis, Giordano, Souly, and Shah]{8099962}
C.~Spampinato, S.~Palazzo, I.~Kavasidis, D.~Giordano, N.~Souly, and M.~Shah.
\newblock Deep learning human mind for automated visual classification.
\newblock In \emph{2017 IEEE Conference on Computer Vision and Pattern Recognition (CVPR)}, pages 4503--4511, 2017.
\newblock \doi{10.1109/CVPR.2017.479}.

\bibitem[Kingma and Welling(2013)]{kingma_welling_2013}
Diederik~P Kingma and Max Welling.
\newblock Auto-encoding variational bayes, 2013.
\newblock URL \url{https://arxiv.org/abs/1312.6114}.

\bibitem[Kingma and Welling(2019)]{kingma_welling_2019}
Diederik~P Kingma and Max Welling.
\newblock An introduction to variational autoencoders.
\newblock \emph{Foundations and Trends® in Machine Learning}, 12\penalty0 (4):\penalty0 307–392, Jan 2019.
\newblock \doi{https://doi.org/10.1561/2200000056}.
\newblock URL \url{https://arxiv.org/abs/1906.02691}.

\bibitem[Hochreiter and Schmidhuber(1997)]{lstm_1997}
Sepp Hochreiter and J\"{u}rgen Schmidhuber.
\newblock Long short-term memory.
\newblock \emph{Neural Comput.}, 9\penalty0 (8):\penalty0 1735–1780, nov 1997.
\newblock ISSN 0899-7667.
\newblock \doi{10.1162/neco.1997.9.8.1735}.
\newblock URL \url{https://doi.org/10.1162/neco.1997.9.8.1735}.

\bibitem[Tirupattur et~al.(2018)Tirupattur, Rawat, Spampinato, and Shah]{ThoughtViz_2018}
Praveen Tirupattur, Yogesh~Singh Rawat, Concetto Spampinato, and Mubarak Shah.
\newblock Thoughtviz: Visualizing human thoughts using generative adversarial network.
\newblock In \emph{Proceedings of the 26th ACM International Conference on Multimedia}, MM '18, page 950–958, New York, NY, USA, 2018. Association for Computing Machinery.
\newblock ISBN 9781450356657.
\newblock \doi{10.1145/3240508.3240641}.
\newblock URL \url{https://doi.org/10.1145/3240508.3240641}.

\bibitem[Jiao et~al.(2019)Jiao, You, Yang, Li, Zhang, and Shen]{jiao_you_yang_li_zhang_shen_2019}
Zhicheng Jiao, Haoxuan You, Fan Yang, Xin Li, Han Zhang, and Dinggang Shen.
\newblock Decoding eeg by visual-guided deep neural networks.
\newblock \emph{Ijcai.org}, page 1387–1393, 2019.
\newblock URL \url{https://www.ijcai.org/proceedings/2019/192}.

\bibitem[Fares et~al.(2020)Fares, Zhong, and Jiang]{BrainMedia_2020}
Ahmed Fares, Sheng-hua Zhong, and Jianmin Jiang.
\newblock Brain-media: A dual conditioned and lateralization supported gan (dcls-gan) towards visualization of image-evoked brain activities.
\newblock In \emph{Proceedings of the 28th ACM International Conference on Multimedia}, MM '20, page 1764–1772, New York, NY, USA, 2020. Association for Computing Machinery.
\newblock ISBN 9781450379885.
\newblock \doi{10.1145/3394171.3413858}.
\newblock URL \url{https://doi.org/10.1145/3394171.3413858}.

\bibitem[Radford et~al.(2021{\natexlab{a}})Radford, Kim, Hallacy, Ramesh, Goh, Agarwal, Sastry, Askell, Mishkin, Clark, Krueger, and Sutskever]{clip_2021}
Alec Radford, Jong~Wook Kim, Chris Hallacy, Aditya Ramesh, Gabriel Goh, Sandhini Agarwal, Girish Sastry, Amanda Askell, Pamela Mishkin, Jack Clark, Gretchen Krueger, and Ilya Sutskever.
\newblock Learning transferable visual models from natural language supervision, 2021{\natexlab{a}}.
\newblock URL \url{https://arxiv.org/abs/2103.00020}.

\bibitem[Singh et~al.(2024)Singh, Dalal, Vashishtha, Miyapuram, and Raman]{EEGCLIP_2024}
P.~Singh, D.~Dalal, G.~Vashishtha, K.~Miyapuram, and S.~Raman.
\newblock Learning robust deep visual representations from eeg brain recordings.
\newblock In \emph{2024 IEEE/CVF Winter Conference on Applications of Computer Vision (WACV)}, pages 7538--7547, Los Alamitos, CA, USA, jan 2024. IEEE Computer Society.
\newblock \doi{10.1109/WACV57701.2024.00738}.
\newblock URL \url{https://doi.ieeecomputersociety.org/10.1109/WACV57701.2024.00738}.

\bibitem[Song et~al.(2024)Song, Liu, Li, Shi, Wang, and Gao]{nice_2023}
Yonghao Song, Bingchuan Liu, Xiang Li, Nanlin Shi, Yijun Wang, and Xiaorong Gao.
\newblock Decoding natural images from eeg for object recognition, 2024.
\newblock URL \url{https://arxiv.org/abs/2308.13234}.

\bibitem[Chen and Wei(2024)]{muse_2024}
Chi-Sheng Chen and Chun-Shu Wei.
\newblock Mind’s eye: Image recognition by eeg via multimodal similarity-keeping contrastive learning, 2024.
\newblock URL \url{https://arxiv.org/abs/2406.16910}.

\bibitem[Li et~al.(2024)Li, Wei, Li, Zou, and Liu]{atm_2024}
Dongyang Li, Chen Wei, Shiying Li, Jiachen Zou, and Quanying Liu.
\newblock Visual decoding and reconstruction via eeg embeddings with guided diffusion, 2024.
\newblock URL \url{https://arxiv.org/abs/2403.07721}.

\bibitem[Chen et~al.(2024{\natexlab{a}})Chen, He, Liu, and Yang]{vesdn_2024}
Hongzhou Chen, Lianghua He, Yihang Liu, and Longzhen Yang.
\newblock Visual neural decoding via improved visual-eeg semantic consistency, 2024{\natexlab{a}}.
\newblock URL \url{https://arxiv.org/abs/2408.06788}.

\bibitem[Chen et~al.(2024{\natexlab{b}})Chen, Chen, Tsai, and Wei]{qeegnet_2024}
Chi-Sheng Chen, Samuel Yen-Chi Chen, Aidan Hung-Wen Tsai, and Chun-Shu Wei.
\newblock Qeegnet: Quantum machine learning for enhanced electroencephalography encoding, 2024{\natexlab{b}}.
\newblock URL \url{https://arxiv.org/abs/2407.19214}.

\bibitem[Chen et~al.(2024{\natexlab{c}})Chen, Tsai, and Huang]{chen_mcl_2024}
Chi-Sheng Chen, Aidan Hung-Wen Tsai, and Sheng-Chieh Huang.
\newblock Quantum multimodal contrastive learning framework, 2024{\natexlab{c}}.
\newblock URL \url{https://arxiv.org/abs/2408.13919}.

\bibitem[Zeng et~al.(2023)Zeng, Xia, Qian, Hattori, Wang, and Kong]{DM-RE2I_2023}
Hong Zeng, Nianzhang Xia, Dongguan Qian, Motonobu Hattori, Chu Wang, and Wanzeng Kong.
\newblock Dm-re2i: A framework based on diffusion model for the reconstruction from eeg to image.
\newblock \emph{Biomedical Signal Processing and Control}, 86:\penalty0 105125–105125, Sep 2023.
\newblock \doi{https://doi.org/10.1016/j.bspc.2023.105125}.
\newblock URL \url{https://www.sciencedirect.com/science/article/abs/pii/S174680942300558X?via%3Dihub}.

\bibitem[Fu et~al.(2023)Fu, Shen, Chin, and Wang]{BrainViz_2023}
Honghao Fu, Zhiqi Shen, Jing~Jih Chin, and Hao Wang.
\newblock Brainvis: Exploring the bridge between brain and visual signals via image reconstruction, 2023.
\newblock URL \url{https://arxiv.org/abs/2312.14871}.

\bibitem[Lan et~al.(2023)Lan, Ren, Wang, Zheng, Li, Lu, and Qiu]{NeuroImagen_2023}
Yu-Ting Lan, Kan Ren, Yansen Wang, Wei-Long Zheng, Dongsheng Li, Bao-Liang Lu, and Lili Qiu.
\newblock Seeing through the brain: Image reconstruction of visual perception from human brain signals, 2023.
\newblock URL \url{https://arxiv.org/abs/2308.02510}.

\bibitem[Guo(2024)]{guo_2024}
Huangtao Guo.
\newblock Eegvision: Reconstructing vision from human brain signals.
\newblock \emph{Applied Mathematics and Nonlinear Sciences}, 9\penalty0 (1), Jan 2024.
\newblock \doi{https://doi.org/10.2478/amns-2024-1856}.
\newblock URL \url{https://sciendo.com/article/10.2478/amns-2024-1856}.

\bibitem[Dosovitskiy et~al.(2020)Dosovitskiy, Beyer, Kolesnikov, Weissenborn, Zhai, Unterthiner, Dehghani, Minderer, Heigold, Gelly, Uszkoreit, and Houlsby]{vit_2020}
Alexey Dosovitskiy, Lucas Beyer, Alexander Kolesnikov, Dirk Weissenborn, Xiaohua Zhai, Thomas Unterthiner, Mostafa Dehghani, Matthias Minderer, Georg Heigold, Sylvain Gelly, Jakob Uszkoreit, and Neil Houlsby.
\newblock An image is worth 16x16 words: Transformers for image recognition at scale, 2020.
\newblock URL \url{https://arxiv.org/abs/2010.11929}.

\bibitem[Oord et~al.(2018)Oord, Li, and Vinyals]{oord2018representation}
Aaron van~den Oord, Yazhe Li, and Oriol Vinyals.
\newblock Representation learning with contrastive predictive coding.
\newblock \emph{arXiv preprint arXiv:1807.03748}, 2018.

\bibitem[He et~al.(2020)He, Fan, Wu, Xie, and Girshick]{he2020momentum}
Kaiming He, Haoqi Fan, Yuxin Wu, Saining Xie, and Ross Girshick.
\newblock Momentum contrast for unsupervised visual representation learning.
\newblock In \emph{Proceedings of the IEEE/CVF conference on computer vision and pattern recognition}, pages 9729--9738, 2020.

\bibitem[Radford et~al.(2021{\natexlab{b}})Radford, Kim, Hallacy, Ramesh, Goh, Agarwal, Sastry, Askell, Mishkin, Clark, et~al.]{radford2021learning}
Alec Radford, Jong~Wook Kim, Chris Hallacy, Aditya Ramesh, Gabriel Goh, Sandhini Agarwal, Girish Sastry, Amanda Askell, Pamela Mishkin, Jack Clark, et~al.
\newblock Learning transferable visual models from natural language supervision.
\newblock In \emph{International conference on machine learning}, pages 8748--8763. PMLR, 2021{\natexlab{b}}.

\bibitem[Ye et~al.(2023)Ye, Zhang, Liu, Han, and Yang]{ipadaptor_2023}
Hu~Ye, Jun Zhang, Sibo Liu, Xiao Han, and Wei Yang.
\newblock Ip-adapter: Text compatible image prompt adapter for text-to-image diffusion models, 2023.
\newblock URL \url{https://arxiv.org/abs/2308.06721}.

\bibitem[Podell et~al.(2023)Podell, English, Lacey, Blattmann, Dockhorn, Müller, Penna, and Rombach]{sdxl_2023}
Dustin Podell, Zion English, Kyle Lacey, Andreas Blattmann, Tim Dockhorn, Jonas Müller, Joe Penna, and Robin Rombach.
\newblock Sdxl: Improving latent diffusion models for high-resolution image synthesis, 2023.
\newblock URL \url{https://arxiv.org/abs/2307.01952}.

\bibitem[Luo et~al.(2024)Luo, Tan, Patil, Gu, Platen, Passos, Huang, Li, and Zhao]{lcm_2024}
Simian Luo, Yiqin Tan, Suraj Patil, Daniel Gu, von Platen, Apolinário Passos, Longbo Huang, Jian Li, and Hang Zhao.
\newblock Lcm-lora: A universal stable-diffusion acceleration module, 2024.
\newblock URL \url{https://arxiv.org/abs/2311.05556}.

\bibitem[Zhu and Mumford(1997)]{632983}
Song~Chun Zhu and D.~Mumford.
\newblock Prior learning and gibbs reaction-diffusion.
\newblock \emph{IEEE Transactions on Pattern Analysis and Machine Intelligence}, 19\penalty0 (11):\penalty0 1236--1250, 1997.
\newblock \doi{10.1109/34.632983}.

\bibitem[Ramesh et~al.(2022)Ramesh, Dhariwal, Nichol, Chu, and Chen]{dalle_2022}
Aditya Ramesh, Prafulla Dhariwal, Alex Nichol, Casey Chu, and Mark Chen.
\newblock Hierarchical text-conditional image generation with clip latents, 2022.
\newblock URL \url{https://arxiv.org/abs/2204.06125}.

\bibitem[Scotti et~al.(2023)Scotti, Banerjee, Goode, Shabalin, Nguyen, Cohen, Dempster, Verlinde, Yundler, Weisberg, Norman, and Abraham]{fmri_to_img_2023}
Paul~S Scotti, Atmadeep Banerjee, Jimmie Goode, Stepan Shabalin, Alex Nguyen, Ethan Cohen, Aidan~J Dempster, Nathalie Verlinde, Elad Yundler, David Weisberg, Kenneth~A Norman, and Tanishq~Mathew Abraham.
\newblock Reconstructing the mind’s eye: fmri-to-image with contrastive learning and diffusion priors, 2023.
\newblock URL \url{https://arxiv.org/abs/2305.18274}.

\bibitem[Wang et~al.(2004)Wang, Bovik, Sheikh, and Simoncelli]{SSIM}
Zhou Wang, A.C. Bovik, H.R. Sheikh, and E.P. Simoncelli.
\newblock Image quality assessment: from error visibility to structural similarity.
\newblock \emph{IEEE Transactions on Image Processing}, 13\penalty0 (4):\penalty0 600--612, 2004.
\newblock \doi{10.1109/TIP.2003.819861}.

\bibitem[Salimans et~al.(2016)Salimans, Goodfellow, Zaremba, Cheung, Radford, and Chen]{is_2016}
Tim Salimans, Ian Goodfellow, Wojciech Zaremba, Vicki Cheung, Alec Radford, and Xi~Chen.
\newblock Improved techniques for training gans, 2016.
\newblock URL \url{https://arxiv.org/abs/1606.03498}.

\bibitem[Heusel et~al.(2017)Heusel, Ramsauer, Unterthiner, Nessler, and Hochreiter]{NIPS2017_FID}
Martin Heusel, Hubert Ramsauer, Thomas Unterthiner, Bernhard Nessler, and Sepp Hochreiter.
\newblock Gans trained by a two time-scale update rule converge to a local nash equilibrium.
\newblock In I.~Guyon, U.~Von Luxburg, S.~Bengio, H.~Wallach, R.~Fergus, S.~Vishwanathan, and R.~Garnett, editors, \emph{Advances in Neural Information Processing Systems}, volume~30. Curran Associates, Inc., 2017.
\newblock URL \url{https://proceedings.neurips.cc/paper_files/paper/2017/file/8a1d694707eb0fefe65871369074926d-Paper.pdf}.

\bibitem[OpenAI et~al.(2023)OpenAI, Achiam, Adler, Agarwal, Ahmad, Akkaya, Aleman, Almeida, Altenschmidt, Altman, Anadkat, Avila, Babuschkin, Balaji, Balcom, Baltescu, Bao, Bavarian, Belgum, and Bello]{openai_2023}
OpenAI, Josh Achiam, Steven Adler, Sandhini Agarwal, Lama Ahmad, Ilge Akkaya, Florencia~Leoni Aleman, Diogo Almeida, Janko Altenschmidt, Sam Altman, Shyamal Anadkat, Red Avila, Igor Babuschkin, Suchir Balaji, Valerie Balcom, Paul Baltescu, Haiming Bao, Mohammad Bavarian, Jeff Belgum, and Irwan Bello.
\newblock Gpt-4 technical report, 2023.
\newblock URL \url{https://arxiv.org/abs/2303.08774}.

\bibitem[Gifford et~al.(2022)Gifford, Dwivedi, Roig, and Cichy]{gifford2022large}
Alessandro~T Gifford, Kshitij Dwivedi, Gemma Roig, and Radoslaw~M Cichy.
\newblock A large and rich eeg dataset for modeling human visual object recognition.
\newblock \emph{NeuroImage}, 264:\penalty0 119754, 2022.

\bibitem[Hebart et~al.(2019)Hebart, Dickter, Kidder, Kwok, Corriveau, Van~Wicklin, and Baker]{hebart2019things}
Martin~N Hebart, Adam~H Dickter, Alexis Kidder, Wan~Y Kwok, Anna Corriveau, Caitlin Van~Wicklin, and Chris~I Baker.
\newblock Things: A database of 1,854 object concepts and more than 26,000 naturalistic object images.
\newblock \emph{PloS one}, 14\penalty0 (10):\penalty0 e0223792, 2019.

\bibitem[Du et~al.(2023)Du, Fu, Li, and He]{du2023decoding}
Changde Du, Kaicheng Fu, Jinpeng Li, and Huiguang He.
\newblock Decoding visual neural representations by multimodal learning of brain-visual-linguistic features.
\newblock \emph{IEEE Transactions on Pattern Analysis and Machine Intelligence}, 2023.

\end{thebibliography}






\newpage
\appendix
\section{Appendix}
\subsection{More EEG encoder classification performance comparison }

\begin{table}[h]
  \caption{Overall accuracy (\%) of 10-way zero-shot classification using CLIP-ViT as image encoder: top-1 and top-5.}
  \label{tab:per_com_10}
  \centering
  \Huge
\resizebox{\linewidth}{!}{
  \begin{tabular}{lccccccccccc}
    \toprule
    & Subject 1 & Subject 2 & Subject 3 & Subject 4 & Subject 5 & Subject 6 & Subject 7 & Subject 8 & Subject 9 & Subject 10 & Ave \\
    \midrule
    Method & 10-way & 10-way & 10-way & 10-way & 10-way & 10-way & 10-way & 10-way & 10-way & 10-way & 10-way \\
    \midrule
    \multicolumn{12}{c}{Subject dependent - train and test on one subject} \\
    \midrule
    Nervformer & 59.4 & 62.0 & 65.4 & 72.0 & 50.7 & 63.4 & 63.7 & 78.3 & 67.0 & 68.8 & 65.1 \\
    NICE & 64.1 & 57.6 & 70.2 & 72.6 & 51.8 & 63.0 & 63.8 & 79.1 & 59.6 & 73.9 & 65.6 \\
    MUSE & 61.0 & 56.1 & 70.8 & 71.3 & 55.1 & 70.1 & 66.2 & 76.9 & 62.8 & 73.2 & 66.4 \\
    ATM-S & 72.5 & 70.4 & 76.3 & 74.1 & 64.6 & 72.2 & 73.6 & 83.2 & 70.6 & 75.8 & 73.3 \\
    NERV (ours) & 72.2 & 74.3 & 75.9 & 76.7 & 62.5 & 71.8 & 70.4 & 81.8 & 70.9 & 73.8 & 73.0 \\
    \bottomrule
  \end{tabular}}
\end{table}

\begin{table}[h]
  \caption{Overall accuracy (\%) of 50-way zero-shot classification using CLIP-ViT as image encoder: top-1 and top-5.}
  \label{tab:per_com_50}
  \centering
  \Huge
\resizebox{\linewidth}{!}{
  \begin{tabular}{lcccccccccccccccccccccc}
    \toprule
    & \multicolumn{2}{c}{Subject 1} & \multicolumn{2}{c}{Subject 2} & \multicolumn{2}{c}{Subject 3} & \multicolumn{2}{c}{Subject 4} & \multicolumn{2}{c}{Subject 5} & \multicolumn{2}{c}{Subject 6} & \multicolumn{2}{c}{Subject 7} & \multicolumn{2}{c}{Subject 8} & \multicolumn{2}{c}{Subject 9} & \multicolumn{2}{c}{Subject 10} & \multicolumn{2}{c}{Ave} \\
    \cmidrule(r){2-3} \cmidrule(r){4-5} \cmidrule(r){6-7} \cmidrule(r){8-9} \cmidrule(r){10-11} \cmidrule(r){12-13} \cmidrule(r){14-15} \cmidrule(r){16-17} \cmidrule(r){18-19} \cmidrule(r){20-21} \cmidrule(r){22-23}
    Method & top-1 & top-5 & top-1 & top-5 & top-1 & top-5 & top-1 & top-5 & top-1 & top-5 & top-1 & top-5 & top-1 & top-5 & top-1 & top-5 & top-1 & top-5 & top-1 & top-5 & top-1 & top-5 \\
    \midrule
    \multicolumn{23}{c}{Subject dependent - train and test on one subject} \\
    \midrule
    Nervformer & 28.4 & 66.0 & 32.0 & 71.8 & 37.4 & 73.9 & 44.8 & 81.6 & 24.6 & 57.1 & 33.8 & 74.4 & 33.6 & 69.2 & 49.9 & 87.2 & 36.8 & 75.6 & 38.8 & 76.6 & 36.0 & 73.3\\
    NICE & 36.0 & 72.2 & 30.2 & 66.8 & 43.0 & 77.8 & 44.0 & 80.3 & 24.8 & 58.2 & 35.6 & 70.4 & 36.9 & 72.5 & 53.3 & 86.0 & 34.4 & 65.4 & 45.8 & 82.8 & 38.4 & 73.2 \\
    MUSE & 33.9 & 70.9 & 29.9 & 65.7 & 43.6 & 79.4 & 42.8 & 79.8 & 26.1 & 63.4 & 39.8 & 79.4 & 39.8 & 73.3 & 49.8 & 84.2 & 34.4 & 72.7 & 44.5 & 81.1 & 38.5 & 74.9\\
    ATM-S & 45.3 & 78.7 & 44.5 & 80.5 & 49.8 & 85.0 & 46.2 & 83.2 & 33.3 & 69.2 & 42.8 & 81.1 & 47.5 & 80.8 & 59.7 & 91.0 & 45.8 & 79.3 & 50.6 & 82.4 & 46.6 & 81.1\\
    NERV (ours) & 41.1 & 74.8 & 43.2 & 80.5 & 47.9 & 82.8 & 48.1 & 83.5 & 36.4 & 70.7 & 43.0 & 77.6 & 43.5 & 77.3 & 59.2 & 88.4 & 46.1 & 79.4 & 51.0 & 81.7 & 46.0 & 79.7\\
    \bottomrule
  \end{tabular}}
\end{table}

\begin{table}[h]
  \caption{Overall accuracy (\%) of 100-way zero-shot classification using CLIP-ViT as image encoder: top-1 and top-5.}
  \label{tab:per_com_100}
  \centering
  \Huge
\resizebox{\linewidth}{!}{
  \begin{tabular}{lcccccccccccccccccccccc}
    \toprule
    & \multicolumn{2}{c}{Subject 1} & \multicolumn{2}{c}{Subject 2} & \multicolumn{2}{c}{Subject 3} & \multicolumn{2}{c}{Subject 4} & \multicolumn{2}{c}{Subject 5} & \multicolumn{2}{c}{Subject 6} & \multicolumn{2}{c}{Subject 7} & \multicolumn{2}{c}{Subject 8} & \multicolumn{2}{c}{Subject 9} & \multicolumn{2}{c}{Subject 10} & \multicolumn{2}{c}{Ave} \\
    \cmidrule(r){2-3} \cmidrule(r){4-5} \cmidrule(r){6-7} \cmidrule(r){8-9} \cmidrule(r){10-11} \cmidrule(r){12-13} \cmidrule(r){14-15} \cmidrule(r){16-17} \cmidrule(r){18-19} \cmidrule(r){20-21} \cmidrule(r){22-23}
    Method & top-1 & top-5 & top-1 & top-5 & top-1 & top-5 & top-1 & top-5 & top-1 & top-5 & top-1 & top-5 & top-1 & top-5 & top-1 & top-5 & top-1 & top-5 & top-1 & top-5 & top-1 & top-5 \\
    \midrule
    \multicolumn{23}{c}{Subject dependent - train and test on one subject} \\
    \midrule
    Nervformer & 21.0 & 50.8 & 21.6 & 55.1 & 27.6 & 58.5 & 33.0 & 67.8 & 17.0 & 43.4 & 24.7 & 56.2 & 24.5 & 54.8 & 39.8 & 75.6 & 26.8 & 62.3 & 30.2 & 63.6 & 26.6 & 58.8\\
    NICE & 28.0 & 60.5 & 21.8 & 53.2 & 33.1 & 64.2 & 32.2 & 65.9 & 16.8 & 43.9 & 26.0 & 57.6 & 28.0 & 59.0 & 40.7 & 76.0 & 24.5 & 54.5 & 37.2 & 71.0 & 28.8 & 60.6 \\
    MUSE & 25.4 & 56.7 & 21.2 & 49.8 & 33.9 & 67.6 & 32.2 & 65.7 & 18.0 & 49.6 & 30.4 & 67.2 & 29.5 & 60.8 & 39.0 & 73.3 & 26.1 & 58.7 & 33.6 & 67.0 & 28.9 & 61.6\\
    ATM-S & 34.9 & 67.7 & 33.1 & 66.9 & 38.1 & 74.3 & 36.0 & 70.2 & 24.6 & 55.6 & 28.4 & 67.4 & 35.1 & 67.9 & 48.3 & 82.1 & 33.2 & 68.6 & 39.1 & 73.0 & 35.1 & 69.4\\
    NERV (ours) & 31.1 & 64.4 & 33.1 & 66.9 & 36.6 & 74.1 & 39.0 & 70.2 & 26.1 & 57.1 & 32.9 & 65.2 & 34.2 & 66.0 & 50.4 & 78.0 & 35.5 & 67.7 & 41.1 & 72.5 & 36.0 & 68.2\\
    \bottomrule
  \end{tabular}}
\end{table}

\subsection{Details of Category-based Assessment Table (CAT) Score}
All the category-based labels are generated by ChatGPT-4o \footnote{\url{https://chatgpt.com}}, the prompt we used is "Please provide me with 5 one-word descriptions of the image, ranging from high level to low level.".

\begin{longtable}{|c|c|c|}
\hline
\textbf{Image Label} & \textbf{Test Image in ThingsEEG} & \textbf{Category-based label} \\
\hline
\endfirsthead

\hline
\textbf{Image Label} & \textbf{Test Image in ThingsEEG} & \textbf{Category-based label} \\
\hline
\endhead

\hline \multicolumn{3}{|r|}{\textit{Continued on next page}} \\ \hline
\endfoot

\hline
\endlastfoot
\hline
00001\_aircraft\_carrier & \includegraphics[width=.2\textwidth]{ 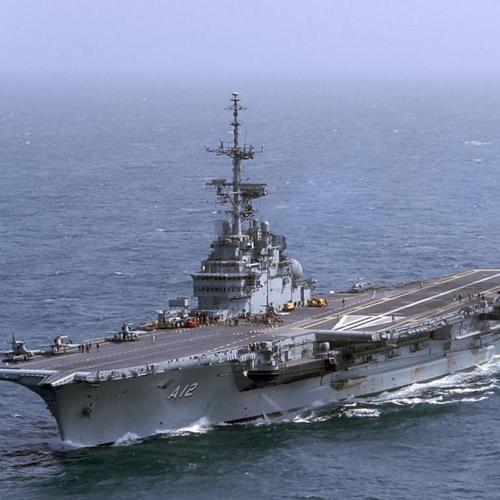} & 
\begin{tabular}{ccc}
    Ship & Carrier & Deck \\
    Island & Antenna & \\
\end{tabular} \\
\hline
\hline
00002\_antelope & \includegraphics[width=.2\textwidth]{ 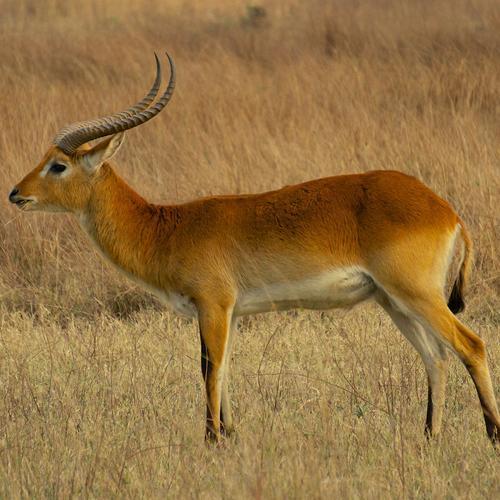} & 
\begin{tabular}{ccc}
     Animal &  Antelope &  Fur\\
     Grassland &  Horns & \\
\end{tabular} \\
\hline
\hline
00003\_backscratcher & \includegraphics[width=.2\textwidth]{ 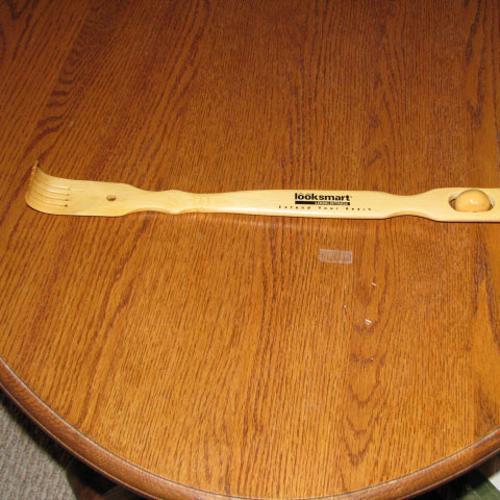} & 
\begin{tabular}{ccc}
     Object &  Tool &  Backscratcher\\
     Wood &  Handle & \\
\end{tabular} \\
\hline
\hline
00004\_balance\_beam & \includegraphics[width=.2\textwidth]{ 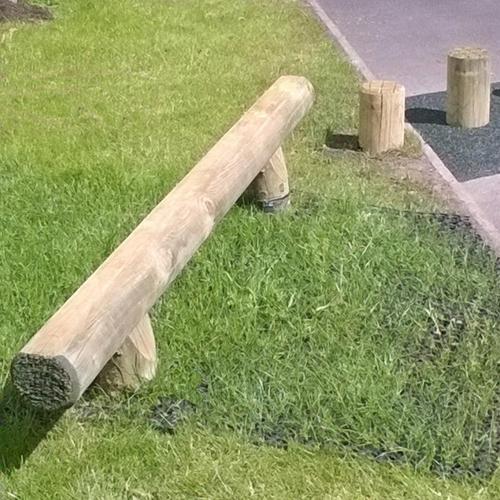} & 
\begin{tabular}{ccc}
     Structure &  Beam &  Wood\\
     Grass &  Support & \\
\end{tabular} \\
\hline
\hline
00005\_banana & \includegraphics[width=.2\textwidth]{ 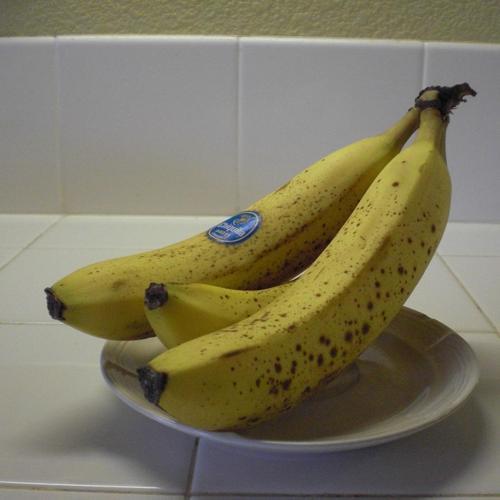} & 
\begin{tabular}{ccc}
     Fruit &  Banana &  Yellow \\
     Spotted &  Plate & \\
\end{tabular} \\
\hline
\hline
00006\_baseball\_bat & \includegraphics[width=.2\textwidth]{ 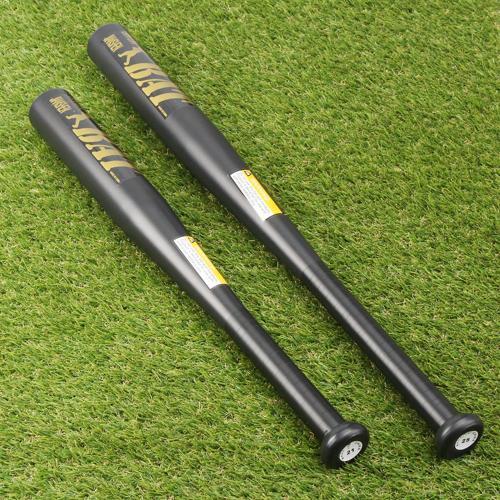} & 
\begin{tabular}{ccc}
     Sports & Bats & Baseball \\
     Black &  Grass & \\
\end{tabular} \\
\hline
\hline
00007\_basil & \includegraphics[width=.2\textwidth]{ 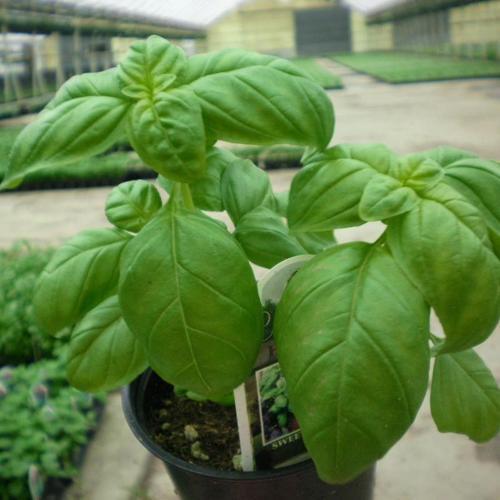} & 
\begin{tabular}{ccc}
    Plant & Herb & Basil \\
    Green & Leaves & \\
\end{tabular} \\
\hline
\hline
00008\_basketball & \includegraphics[width=.2\textwidth]{ 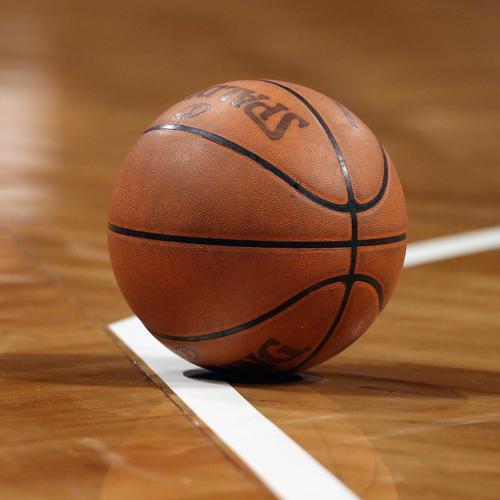} & 
\begin{tabular}{ccc}
    Sport & Basketball & Ball \\
    Orange & Court & \\
\end{tabular} \\
\hline
\hline
00009\_bassoon & \includegraphics[width=.2\textwidth]{ 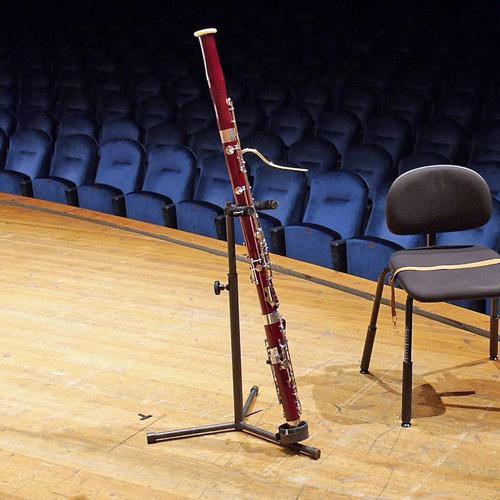} & 
\begin{tabular}{ccc}
    Instrument & Bassoon & Woodwind \\
    Stage & Chair & \\
\end{tabular} \\
\hline
\hline
00010\_baton4 & \includegraphics[width=.2\textwidth]{ 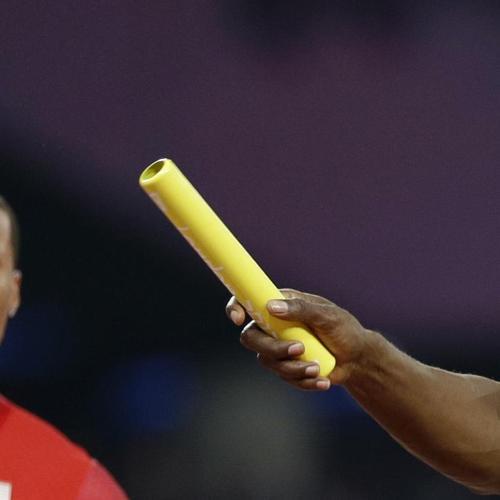} & 
\begin{tabular}{ccc}
    Race & Relay & Baton \\
    Yellow & Hand & \\
\end{tabular} \\
\hline
\hline
00011\_batter & \includegraphics[width=.2\textwidth]{ 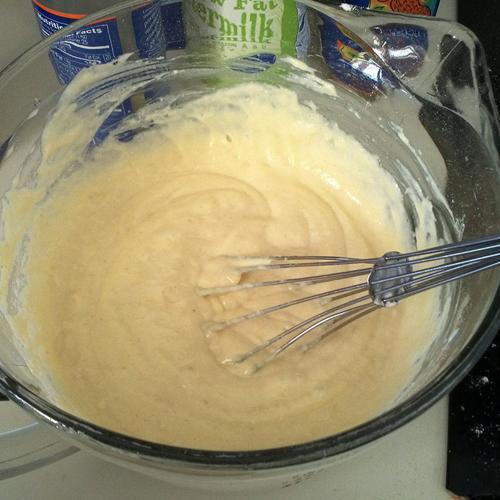} & 
\begin{tabular}{ccc}
    Cooking & Batter & Mixing \\
    Whisk & Bowl & \\
\end{tabular} \\
\hline
\hline
00012\_beaver & \includegraphics[width=.2\textwidth]{ 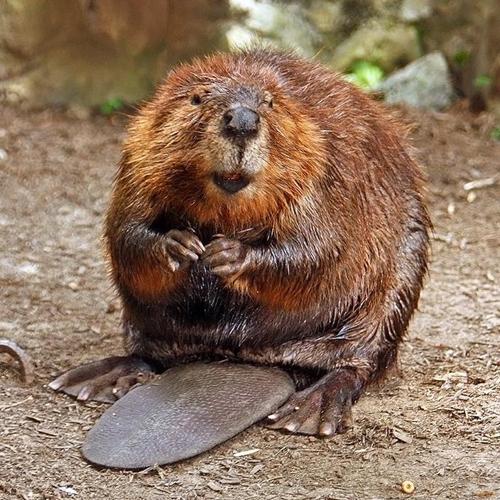} & 
\begin{tabular}{ccc}
     Animal & Beaver & Fur \\
     Tail & Paws & \\
\end{tabular} \\
\hline
\hline
00013\_bench & \includegraphics[width=.2\textwidth]{ 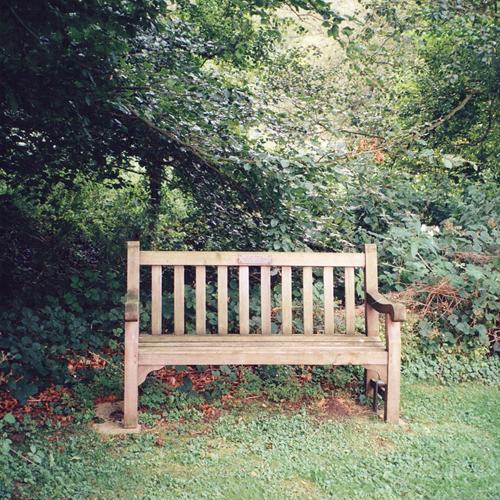} & 
\begin{tabular}{ccc}
     Outdoor & Bench & Wooden \\
     Garden & Trees & \\
\end{tabular} \\
\hline
\hline
00014\_bike & \includegraphics[width=.2\textwidth]{ 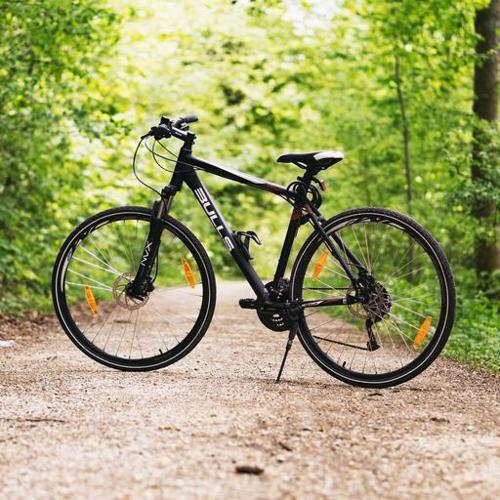} & 
\begin{tabular}{ccc}
     Bicycle & Road & Wheels \\
     Frame & Path & \\
\end{tabular} \\
\hline
\hline
00015\_birthday\_cake & \includegraphics[width=.2\textwidth]{ 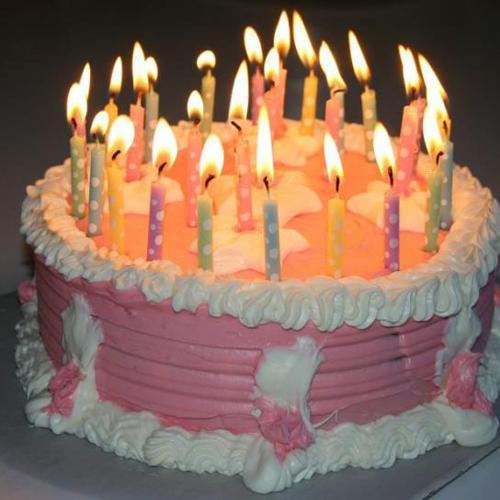} & 
\begin{tabular}{ccc}
     Cake & Candles & Flames \\
     Pink & Frosting & \\
\end{tabular} \\
\hline
\hline
00016\_blowtorch & \includegraphics[width=.2\textwidth]{ 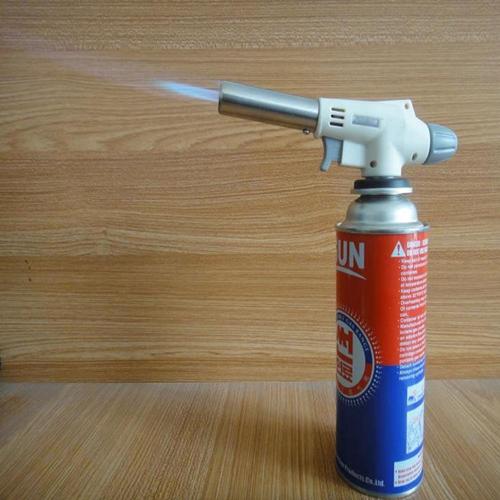} & 
\begin{tabular}{ccc}
     Tool & Blowtorch &  Flame \\
     Canister &  Gas & \\
\end{tabular} \\
\hline
\hline
00017\_boat & \includegraphics[width=.2\textwidth]{ 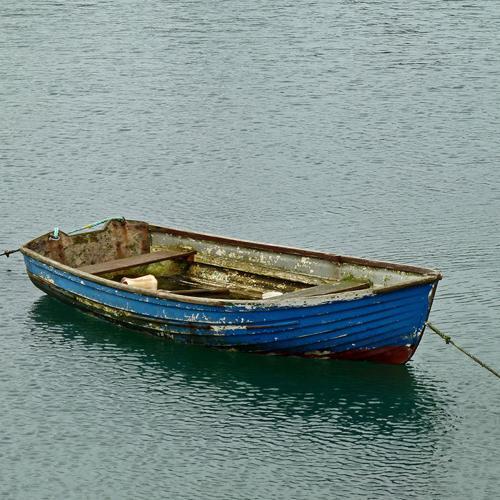} & 
\begin{tabular}{ccc}
    Boat & Water & Blue \\
    Old & Rowing & \\
\end{tabular} \\
\hline
\hline
00018\_bok\_choy & \includegraphics[width=.2\textwidth]{ 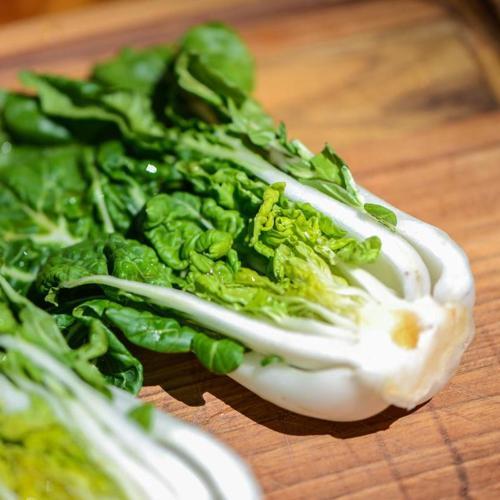} & 
\begin{tabular}{ccc}
     Vegetable & BokChoy & Green \\
     Leafy & Stems & \\
\end{tabular} \\
\hline
\hline
00019\_bonnet & \includegraphics[width=.2\textwidth]{ 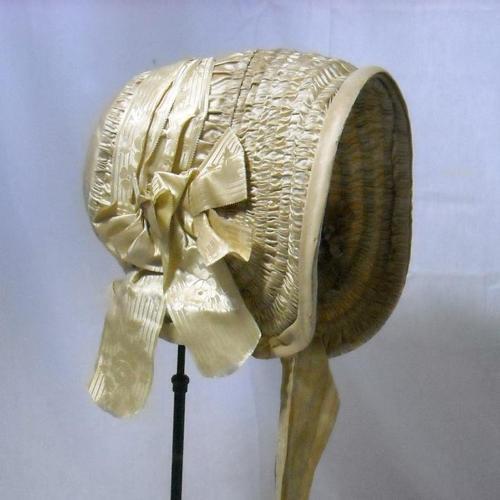} & 
\begin{tabular}{ccc}
    Hat & Bonnet & Ribbon \\
    Fabric & Vintage & \\
\end{tabular} \\
\hline
\hline
00020\_bottle\_opener & \includegraphics[width=.2\textwidth]{ 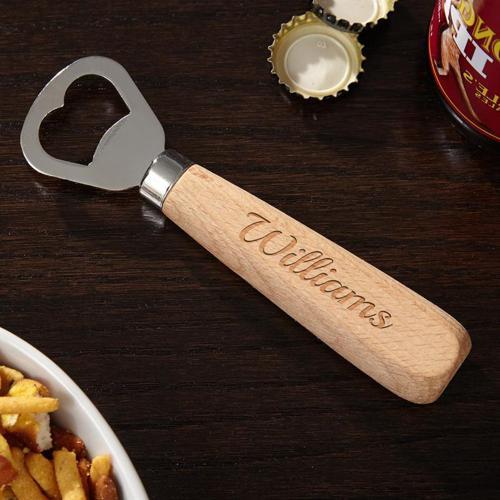} & 
\begin{tabular}{ccc}
   Tool  & Opener & Wooden \\
   Bottlecap  & Engraving & \\
\end{tabular} \\
\hline
\hline
00021\_brace & \includegraphics[width=.2\textwidth]{ 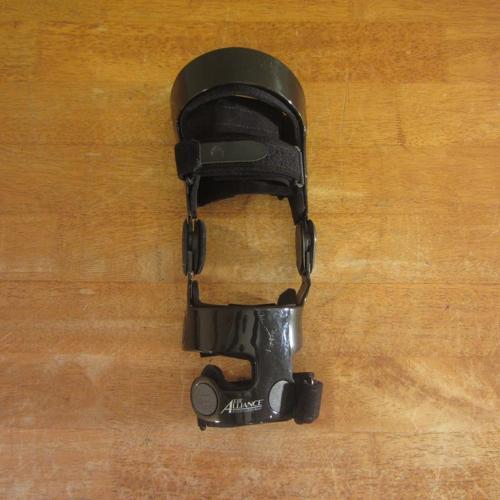} & 
\begin{tabular}{ccc}
   Support  & Brace & Joint \\
   Black  & Strap & \\
\end{tabular} \\
\hline
\hline
00022\_bread & \includegraphics[width=.2\textwidth]{ 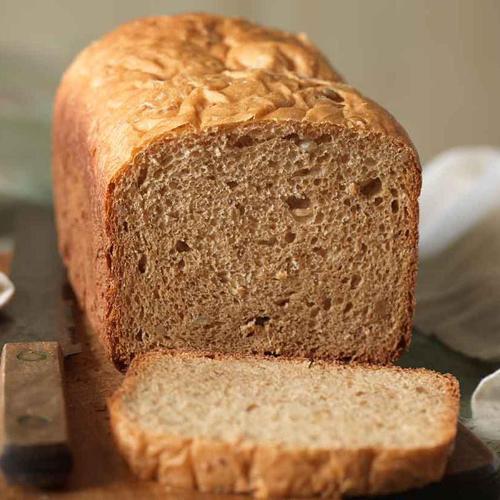} & 
\begin{tabular}{ccc}
     Food & Bread & Loaf \\
     Slice & Crust & \\
\end{tabular} \\
\hline
\hline
00023\_breadbox & \includegraphics[width=.2\textwidth]{ 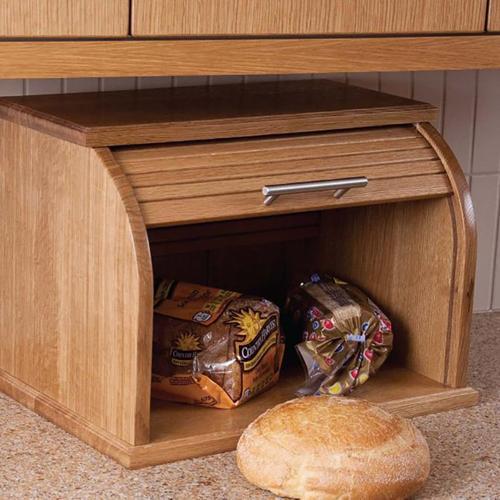} & 
\begin{tabular}{ccc}
     Storage & Breadbox & Wooden \\
     Bread &  Countertop & \\
\end{tabular} \\
\hline
\hline
00024\_bug & \includegraphics[width=.2\textwidth]{ 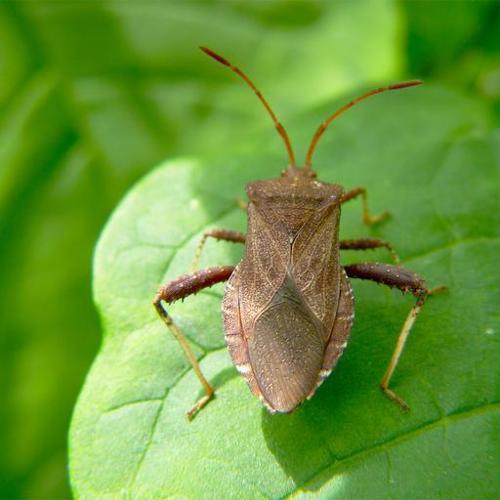} & 
\begin{tabular}{ccc}
     Insect & Bug & Leaf \\
     Brown & Antennae & \\
\end{tabular} \\
\hline
\hline
00025\_buggy & \includegraphics[width=.2\textwidth]{ 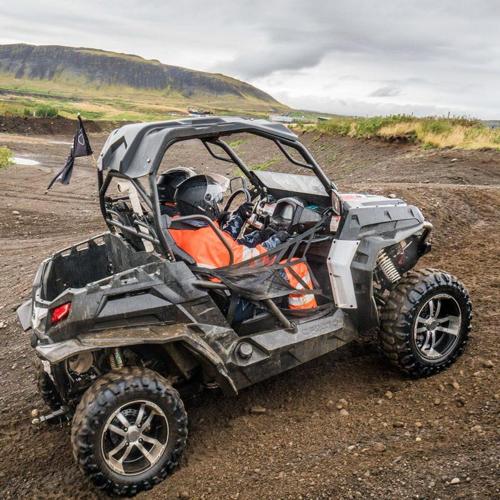} & 
\begin{tabular}{ccc}
     Vehicle & Buggy & Off-road \\
     Wheels & Helmet & \\
\end{tabular} \\
\hline
\hline
00026\_bullet & \includegraphics[width=.2\textwidth]{ 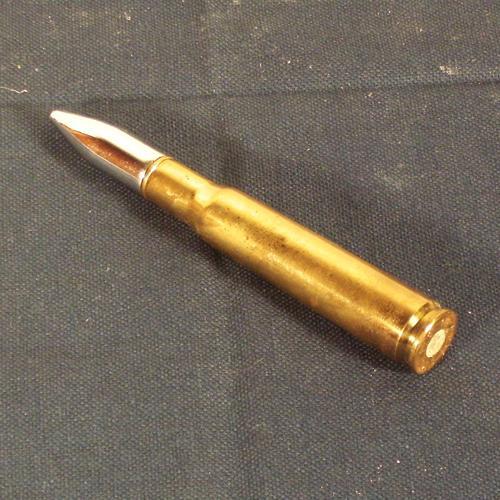} & 
\begin{tabular}{ccc}
    Ammunition & Bullet & Brass \\
    Cartridge & Metal & \\
\end{tabular} \\
\hline
\hline
00027\_bun & \includegraphics[width=.2\textwidth]{ 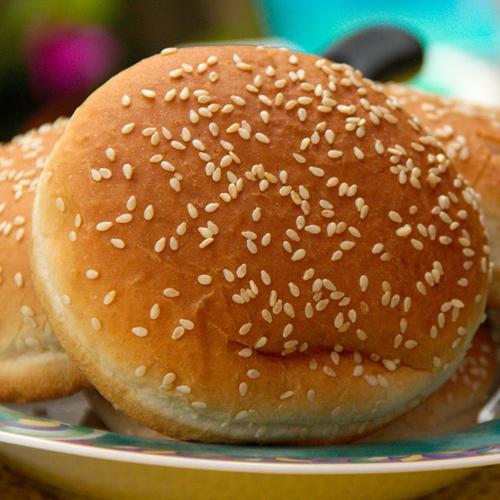} & 
\begin{tabular}{ccc}
   Food & Bun & Sesame \\
   Bread  & Round & \\
\end{tabular} \\
\hline
\hline
00028\_bush & \includegraphics[width=.2\textwidth]{ 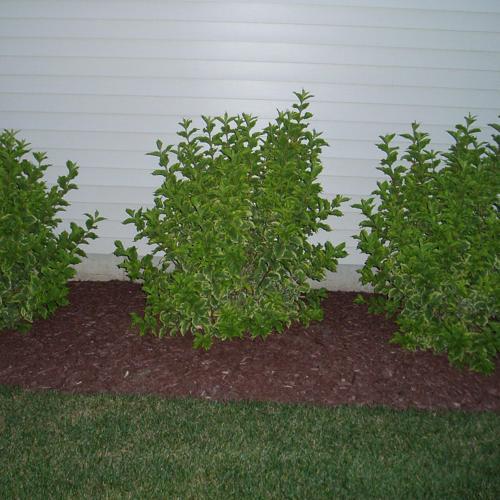} & 
\begin{tabular}{ccc}
    Plants & Bushes & Green \\
    Mulch & Shrub & \\
\end{tabular} \\
\hline
\hline
00029\_calamari & \includegraphics[width=.2\textwidth]{ 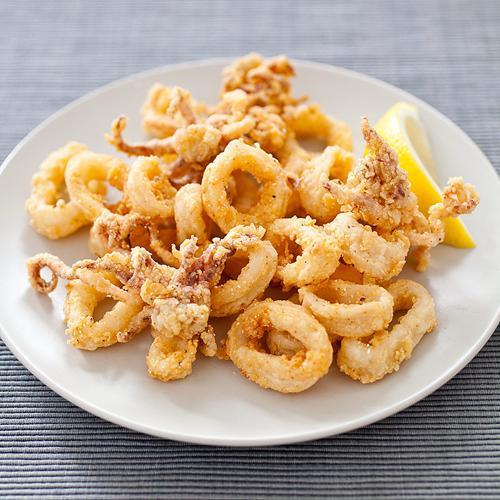} & 
\begin{tabular}{ccc}
    Food & Calamari & Fried \\
    Plate & Lemon & \\
\end{tabular} \\
\hline
\hline
00030\_candlestick & \includegraphics[width=.2\textwidth]{ 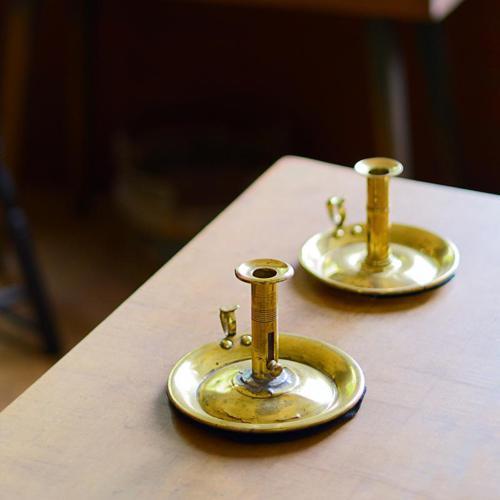} & 
\begin{tabular}{ccc}
    Candlesticks & Brass & Holders \\
    Antique & Table & \\
\end{tabular} \\
\hline
\hline
00031\_cart & \includegraphics[width=.2\textwidth]{ 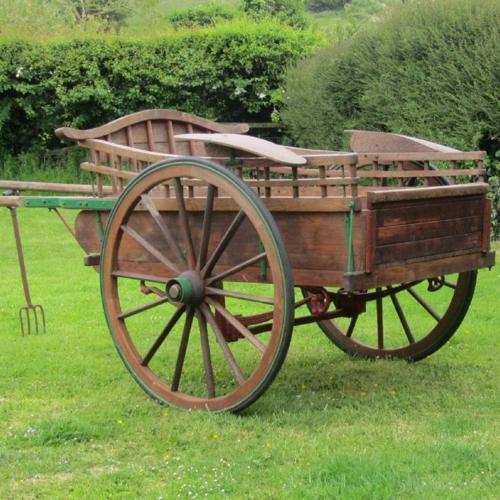} & 
\begin{tabular}{ccc}
   Cart  & Wheels & Wooden \\
   Farm  & Grass  & \\
\end{tabular} \\
\hline
\hline
00032\_cashew & \includegraphics[width=.2\textwidth]{ 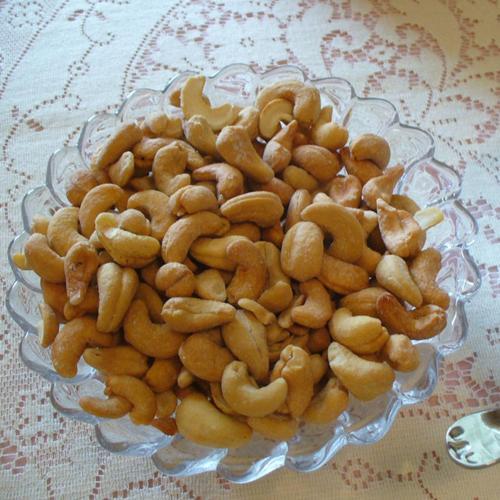} & 
\begin{tabular}{ccc}
    Nuts & Cashews & Bowl \\
    Snack & Glass & \\
\end{tabular} \\
\hline
\hline
00033\_cat & \includegraphics[width=.2\textwidth]{ 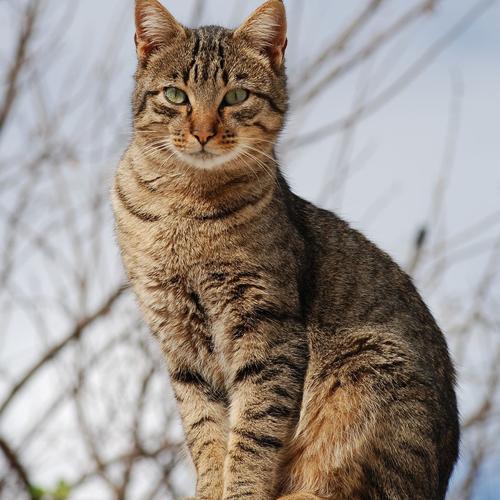} & 
\begin{tabular}{ccc}
    Animal & Cat & Tabby \\
    Fur & Whiskers & \\
\end{tabular} \\
\hline
\hline
00034\_caterpillar & \includegraphics[width=.2\textwidth]{ 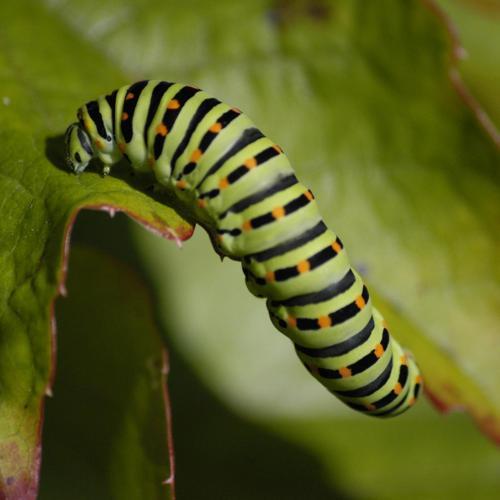} & 
\begin{tabular}{ccc}
    Insect & Caterpillar & Striped \\
    Green & Leaf & \\
\end{tabular} \\
\hline
\hline
00035\_cd\_player & \includegraphics[width=.2\textwidth]{ 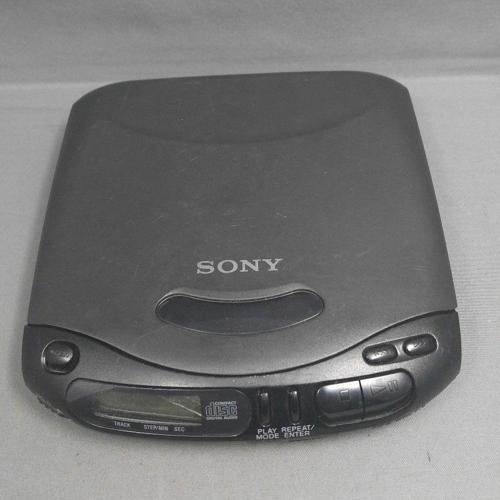} & 
\begin{tabular}{ccc}
    Device & CDPlayer & Portable \\
    Gray & Buttons & \\
\end{tabular} \\
\hline
\hline
00036\_chain & \includegraphics[width=.2\textwidth]{ 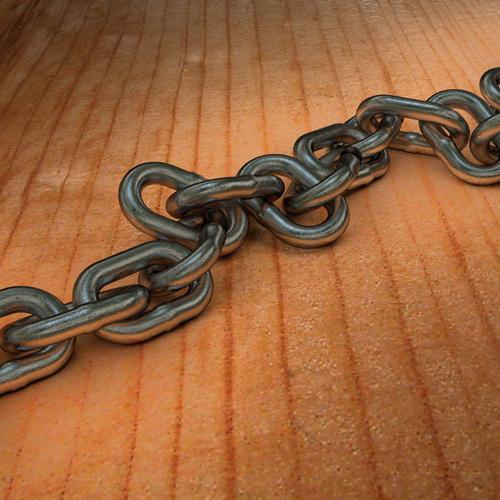} & 
\begin{tabular}{ccc}
    Metal & Chain & Links \\
    Rusty & Wood & \\
\end{tabular} \\
\hline
\hline
00037\_chaps & \includegraphics[width=.2\textwidth]{ 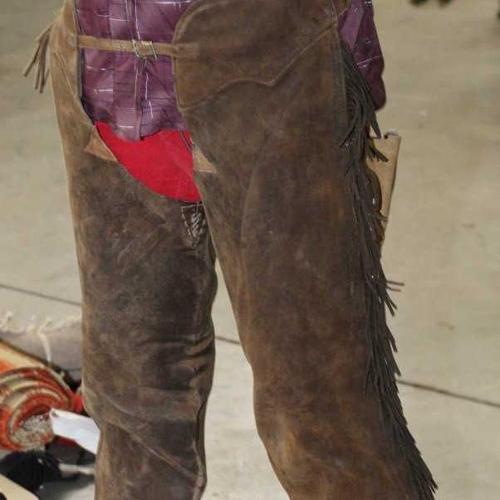} & 
\begin{tabular}{ccc}
    Clothing & Chaps & Leather \\
    Fringe & Brown & \\
\end{tabular} \\
\hline
\hline
00038\_cheese & \includegraphics[width=.2\textwidth]{ 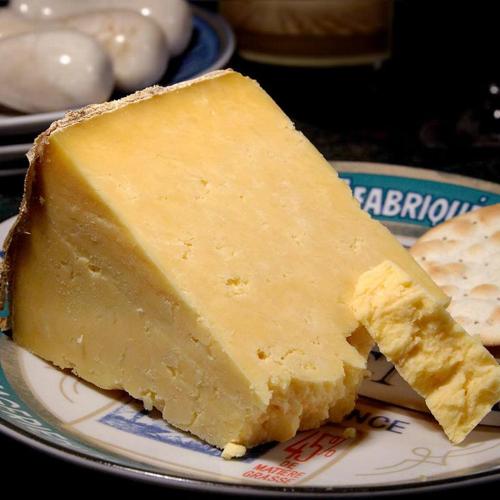} & 
\begin{tabular}{ccc}
   Food  & Cheese & Wedge \\
   Yellow  & Cracker & \\
\end{tabular} \\
\hline
\hline
00039\_cheetah & \includegraphics[width=.2\textwidth]{ 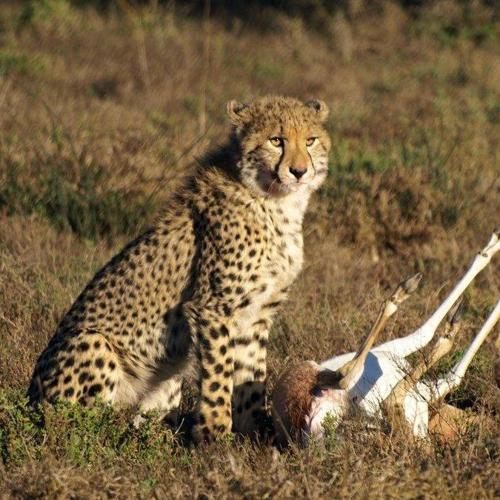} & 
\begin{tabular}{ccc}
   Animal  & Cheetah & Spotted \\
   Hunt  & Grassland & \\
\end{tabular} \\
\hline
\hline
00040\_chest2 & \includegraphics[width=.2\textwidth]{ 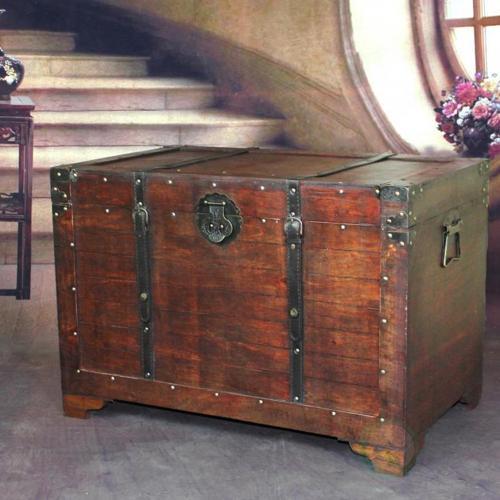} & 
\begin{tabular}{ccc}
   Furniture  & Chest & Wooden \\
   Vintage  & Lock & \\
\end{tabular} \\
\hline
\hline
00041\_chime & \includegraphics[width=.2\textwidth]{ 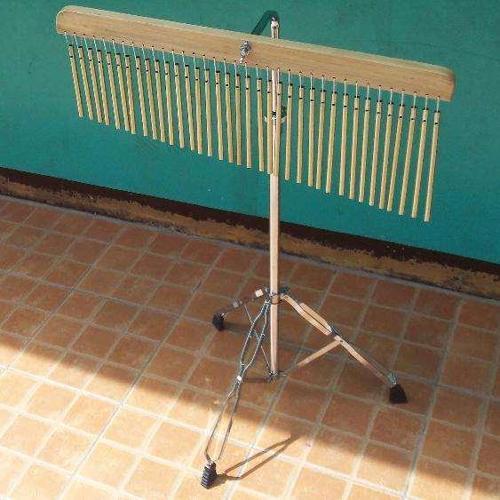} & 
\begin{tabular}{ccc}
    Instrument & Chime & Percussion \\
    Metal & Stand & \\
\end{tabular} \\
\hline
\hline
00042\_chopsticks & \includegraphics[width=.2\textwidth]{ 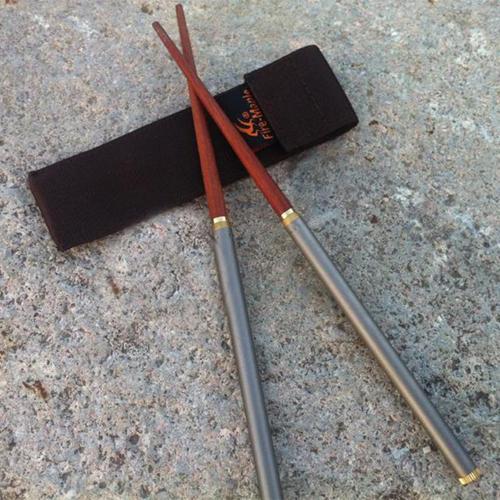} & 
\begin{tabular}{ccc}
   Utensils  & Chopsticks & Wooden \\
   Metal  & Case & \\
\end{tabular} \\
\hline
\hline
00043\_cleat & \includegraphics[width=.2\textwidth]{ 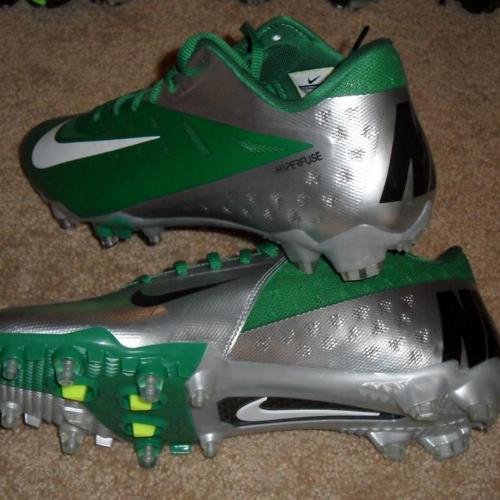} & 
\begin{tabular}{ccc}
    Footwear & Cleats & Shoe \\
    Green & Studs & \\
\end{tabular} \\
\hline
\hline
00044\_cleaver & \includegraphics[width=.2\textwidth]{ 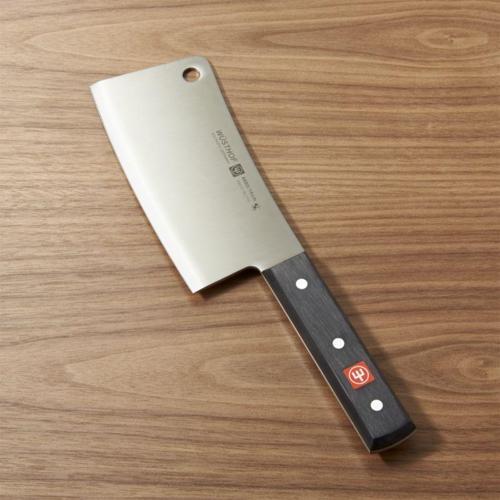} & 
\begin{tabular}{ccc}
    Tool & Cleaver & Blade \\
    Handle & Steel  & \\
\end{tabular} \\
\hline
\hline
00045\_coat & \includegraphics[width=.2\textwidth]{ 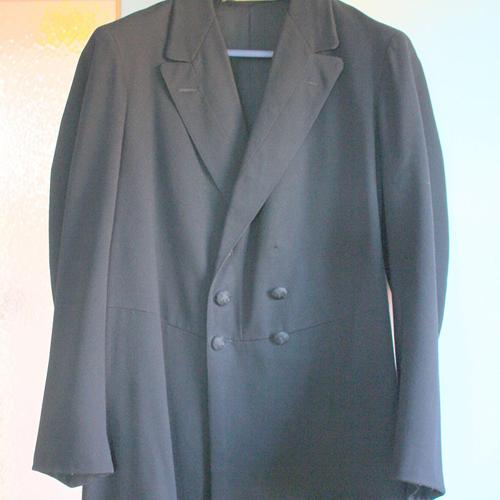} & 
\begin{tabular}{ccc}
   Clothing  & Coat & Black \\
   Double-breasted  & Hanger & \\
\end{tabular} \\
\hline
\hline
00046\_cobra & \includegraphics[width=.2\textwidth]{ 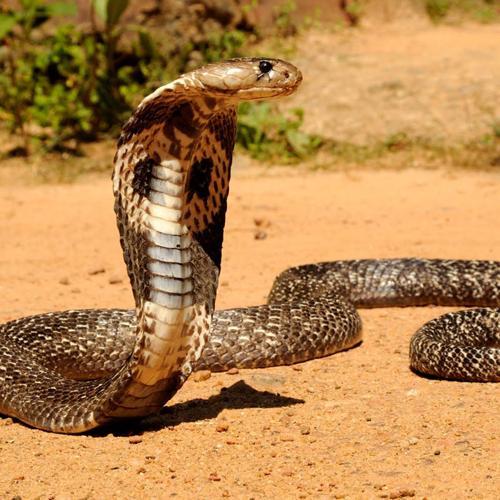} & 
\begin{tabular}{ccc}
    Animal & Cobra & Snake \\
    Hood & Sand & \\
\end{tabular} \\
\hline
\hline
00047\_coconut & \includegraphics[width=.2\textwidth]{ 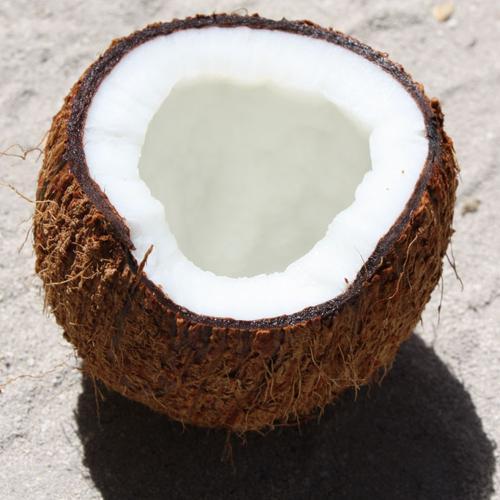} & 
\begin{tabular}{ccc}
   Fruit  & Coconut & Shell \\
   White  & Husk & \\
\end{tabular} \\
\hline
\hline
00048\_coffee\_bean & \includegraphics[width=.2\textwidth]{ 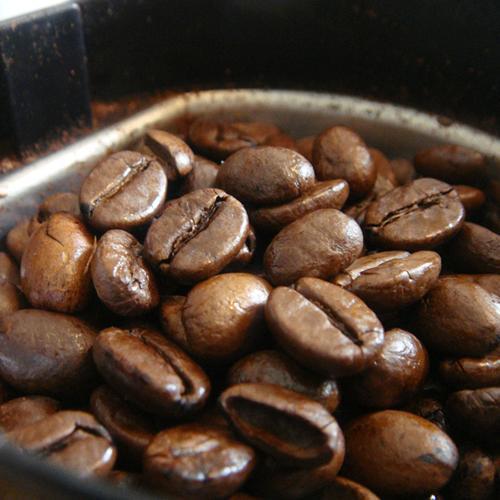} & 
\begin{tabular}{ccc}
   Coffee  & Beans & Roasted \\
   Brown  & Grinder & \\
\end{tabular} \\
\hline
\hline
00049\_coffeemaker & \includegraphics[width=.2\textwidth]{ 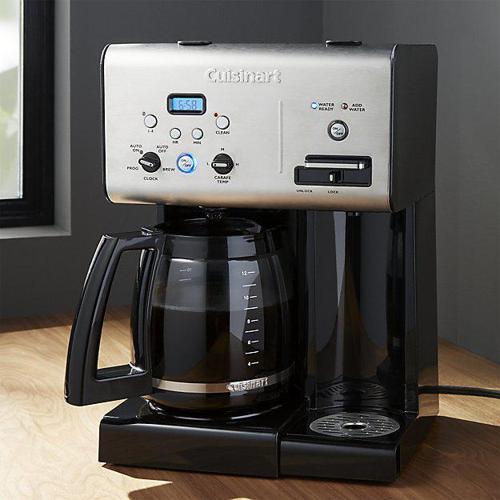} & 
\begin{tabular}{ccc}
   Appliance  & Coffeemaker & Machine \\
   Carafe  & Buttons & \\
\end{tabular} \\
\hline
\hline
00050\_cookie & \includegraphics[width=.2\textwidth]{ 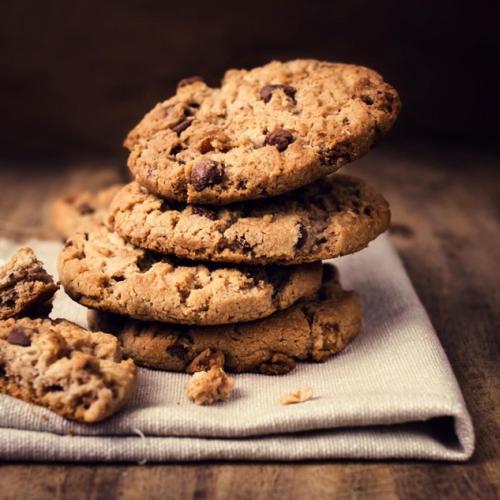} & 
\begin{tabular}{ccc}
   Cookies  & Snack & Chocolate \\
   Stack  & Crumb & \\
\end{tabular} \\
\hline
\hline
00051\_cordon\_bleu & \includegraphics[width=.2\textwidth]{ 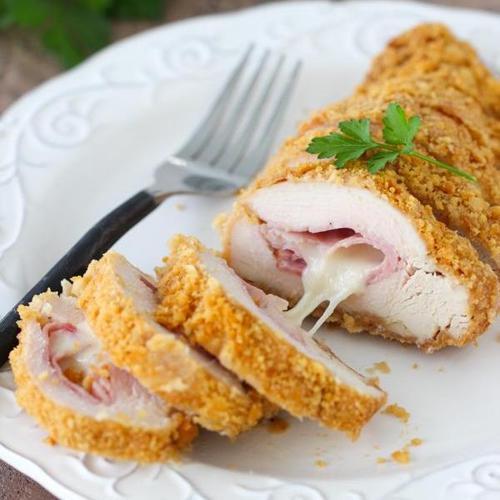} & 
\begin{tabular}{ccc}
    Food & Chicken & CordonBleu \\
    Breaded & Stuffed & \\
\end{tabular} \\
\hline
\hline
00052\_coverall & \includegraphics[width=.2\textwidth]{ 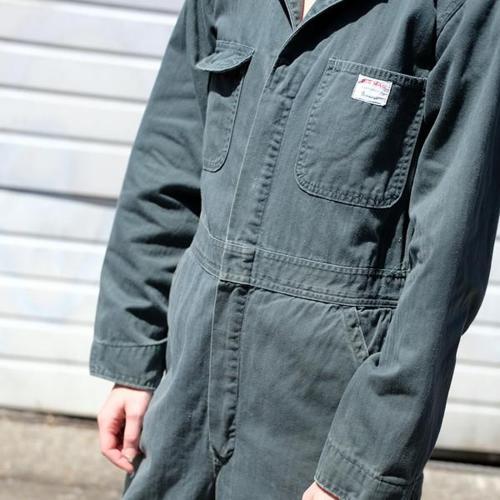} & 
\begin{tabular}{ccc}
   Clothing & Coverall & Workwear \\
   Pockets  & Green & \\
\end{tabular} \\
\hline
\hline
00053\_crab & \includegraphics[width=.2\textwidth]{ 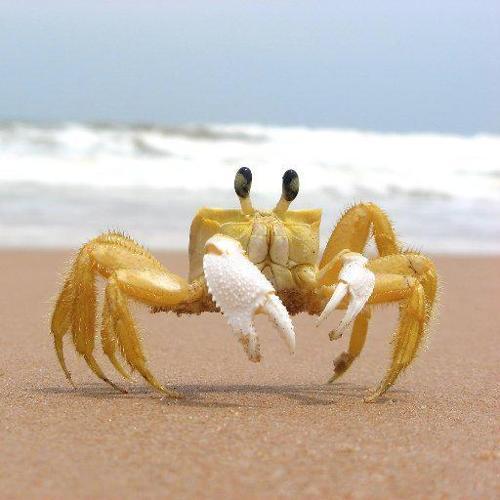} & 
\begin{tabular}{ccc}
    Animal & Crab & Beach \\
    Claws & Sand & \\
\end{tabular} \\
\hline
\hline
00054\_creme\_brulee & \includegraphics[width=.2\textwidth]{ 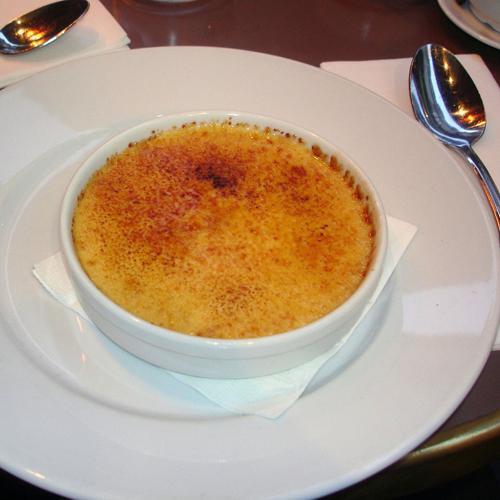} & 
\begin{tabular}{ccc}
   Dessert  & CrèmeBrûlée & Caramelized \\
   Custard  & Spoon & \\
\end{tabular} \\
\hline
\hline
00055\_crepe & \includegraphics[width=.2\textwidth]{ 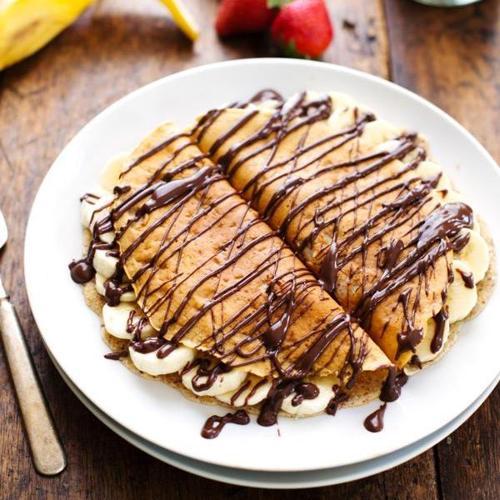} & 
\begin{tabular}{ccc}
   Dessert  & Crepe & Chocolate \\
   Banana  & Plate & \\
\end{tabular} \\
\hline
\hline
00056\_crib & \includegraphics[width=.2\textwidth]{ 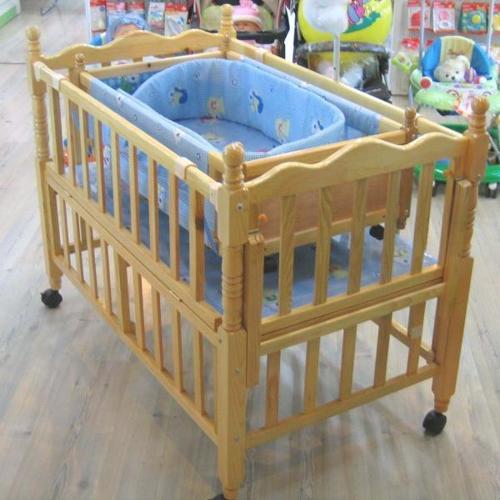} & 
\begin{tabular}{ccc}
   Furniture  & Crib & Wooden \\
   Baby  & Bedding & \\
\end{tabular} \\
\hline
\hline
00057\_croissant & \includegraphics[width=.2\textwidth]{ 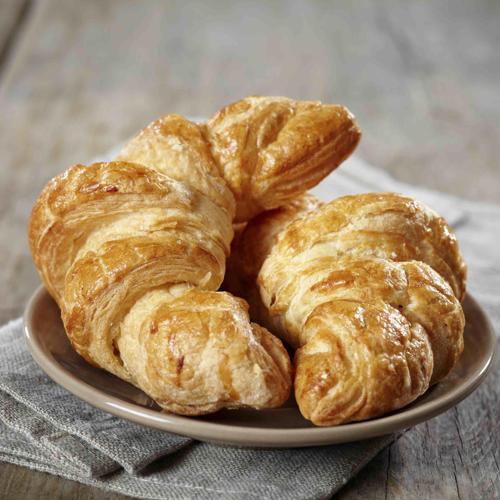} & 
\begin{tabular}{ccc}
     Pastry & Croissant & Flaky \\
    Golden & Plate & \\
\end{tabular} \\
\hline
\hline
00058\_crow & \includegraphics[width=.2\textwidth]{ 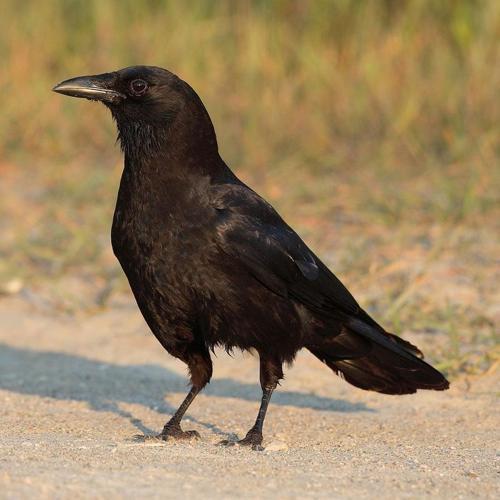} & 
\begin{tabular}{ccc}
    Bird & Crow & Black \\
    Feathers & Beak & \\
\end{tabular} \\
\hline
\hline
00059\_cruise\_ship & \includegraphics[width=.2\textwidth]{ 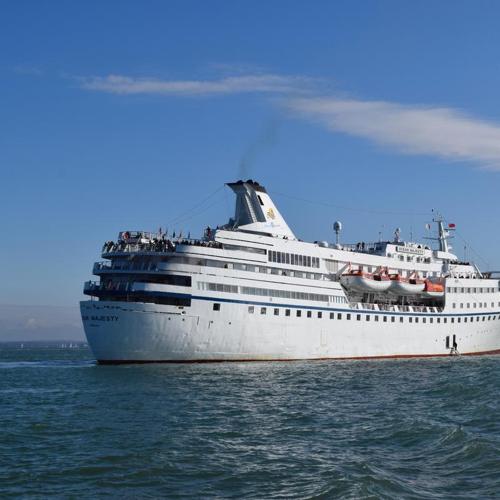} & 
\begin{tabular}{ccc}
     Vessel & Cruise & Ship \\
     Ocean & Deck & \\
\end{tabular} \\
\hline
\hline
00060\_crumb & \includegraphics[width=.2\textwidth]{ 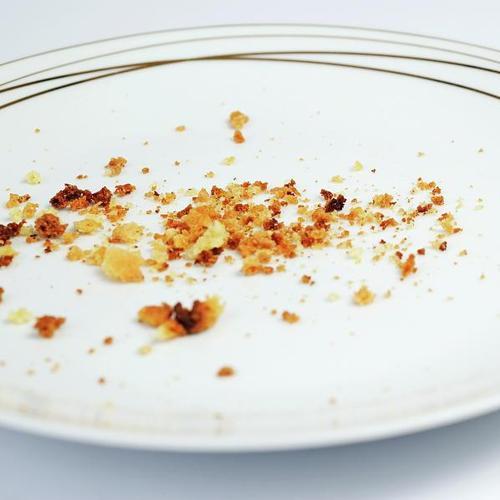} & 
\begin{tabular}{ccc}
    Crumbs & Plate & Food \\
    Leftovers & White & \\
\end{tabular} \\
\hline
\hline
00061\_cupcake & \includegraphics[width=.2\textwidth]{ 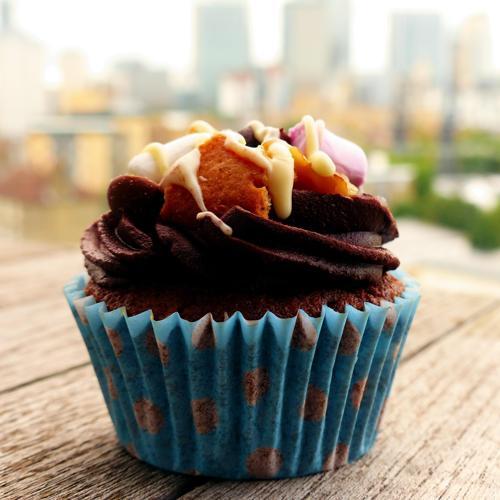} & 
\begin{tabular}{ccc}
    Cupcake & Dessert & Chocolate \\
    Icing & Wrapper & \\
\end{tabular} \\
\hline
\hline
00062\_dagger & \includegraphics[width=.2\textwidth]{ 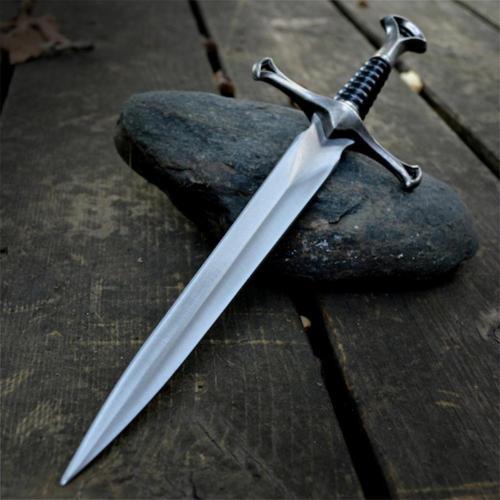} & 
\begin{tabular}{ccc}
    Weapon & Dagger & Blade \\
    Handle & Rock & \\
\end{tabular} \\
\hline
\hline
00063\_dalmatian & \includegraphics[width=.2\textwidth]{ 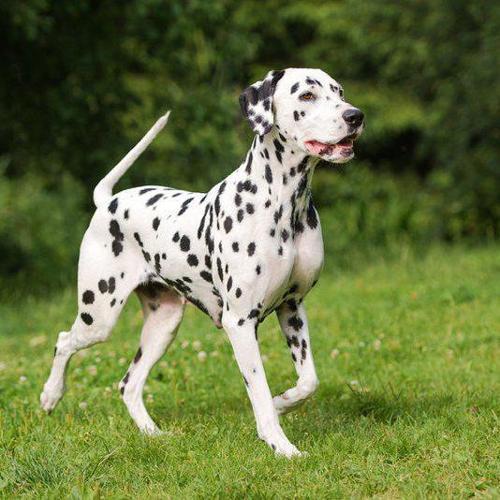} & 
\begin{tabular}{ccc}
    Dog & Dalmatian & Spotted \\
    White & Grass & \\
\end{tabular} \\
\hline
\hline
00064\_dessert & \includegraphics[width=.2\textwidth]{ 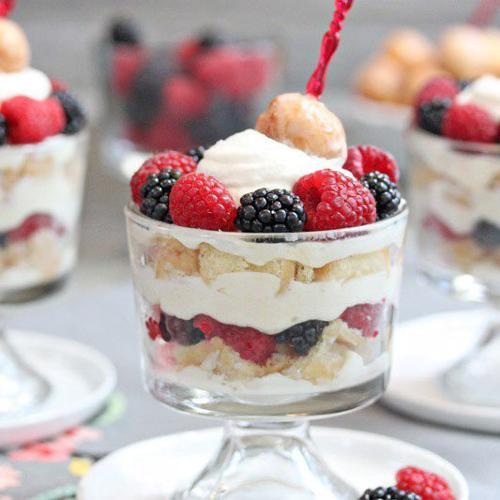} & 
\begin{tabular}{ccc}
    Dessert & Berries & Cream \\
    Trifle & Glass & \\
\end{tabular} \\
\hline
\hline
00065\_dragonfly & \includegraphics[width=.2\textwidth]{ 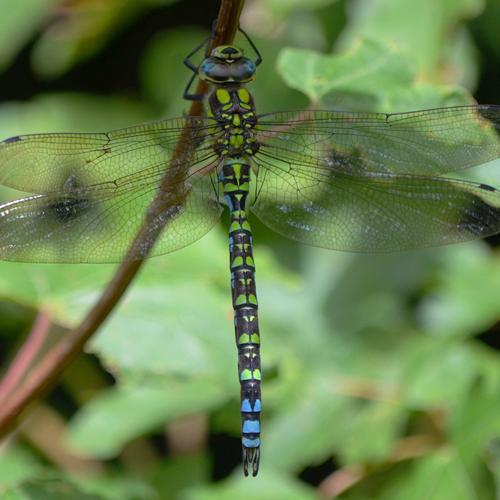} & 
\begin{tabular}{ccc}
    Insect & Dragonfly & Wings \\
    Striped & Branch & \\
\end{tabular} \\
\hline
\hline
00066\_dreidel & \includegraphics[width=.2\textwidth]{ 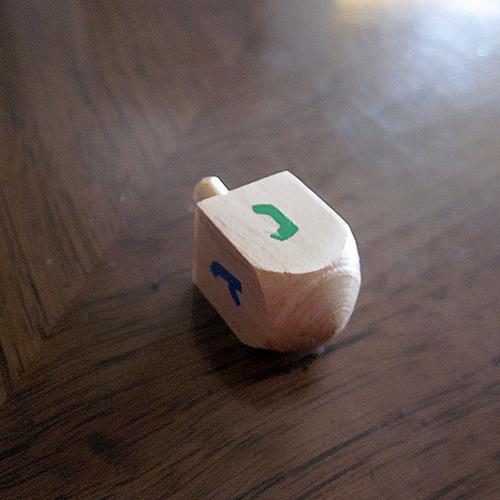} & 
\begin{tabular}{ccc}
    Toy & Dreidel & Wooden \\
    Spinning & Letters & \\
\end{tabular} \\
\hline
\hline
00067\_drum & \includegraphics[width=.2\textwidth]{ 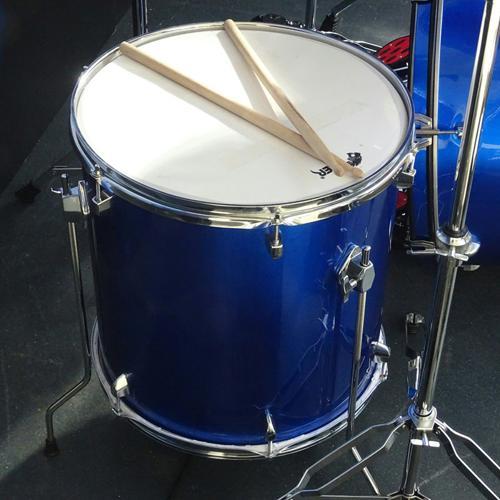} & 
\begin{tabular}{ccc}
    Instrument & Drum & Sticks \\
    Blue & Percussion & \\
\end{tabular} \\
\hline
\hline
00068\_duffel\_bag & \includegraphics[width=.2\textwidth]{ 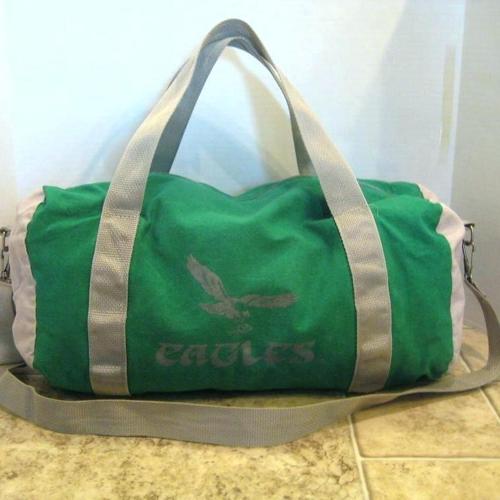} & 
\begin{tabular}{ccc}
    Bag & Container & Green \\
    Straps & Eagles & \\
\end{tabular} \\
\hline
\hline
00069\_eagle & \includegraphics[width=.2\textwidth]{ 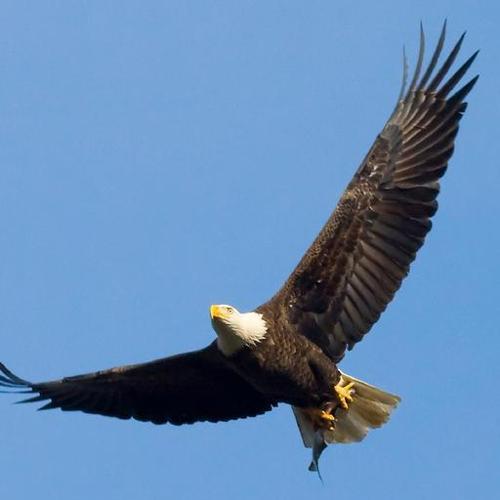} & 
\begin{tabular}{ccc}
    Bird & Eagle & Flight \\
    Wings & Sky & \\
\end{tabular} \\
\hline
\hline
00070\_eel & \includegraphics[width=.2\textwidth]{ 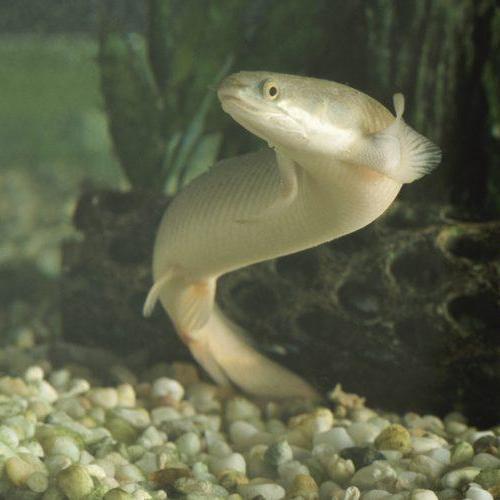} & 
\begin{tabular}{ccc}
   Fish  & Eel & Aquatic \\
   Tank  &  Gravel & \\
\end{tabular} \\
\hline
\hline
00071\_egg & \includegraphics[width=.2\textwidth]{ 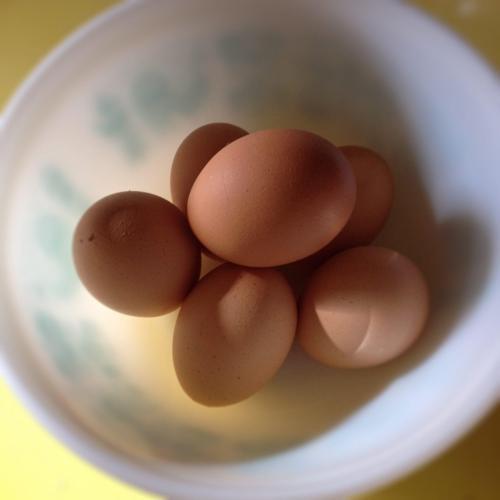} & 
\begin{tabular}{ccc}
    Eggs & Bowl & Brown \\
    Food & Shell & \\
\end{tabular} \\
\hline
\hline
00072\_elephant & \includegraphics[width=.2\textwidth]{ 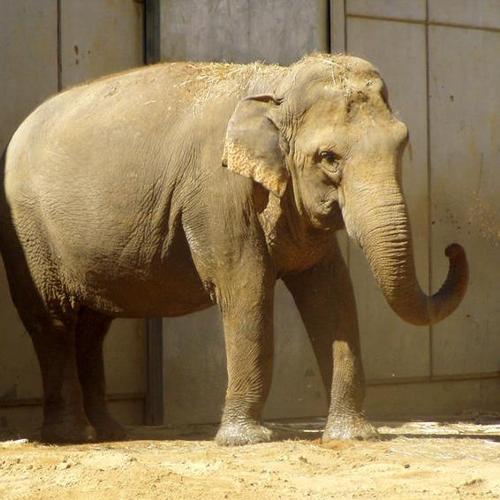} & 
\begin{tabular}{ccc}
   Animal & Elephant & Trunk \\
   Zoo  & Mammal & \\
\end{tabular} \\
\hline
\hline
00073\_espresso & \includegraphics[width=.2\textwidth]{ 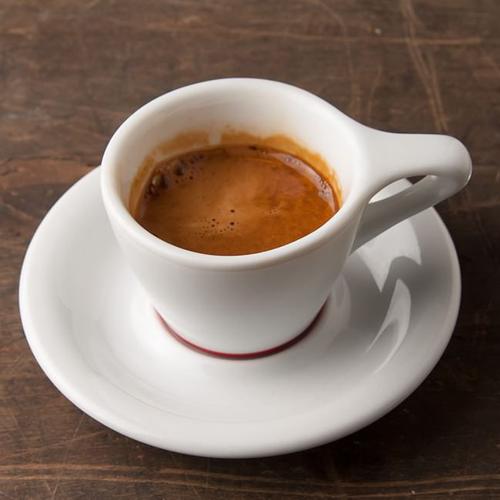} & 
\begin{tabular}{ccc}
   Drink  & Espresso & Cup \\
   Coffee  &  Saucer & \\
\end{tabular} \\
\hline
\hline
00074\_face\_mask & \includegraphics[width=.2\textwidth]{ 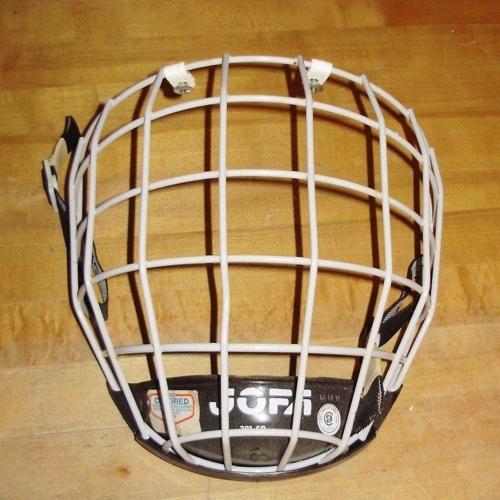} & 
\begin{tabular}{ccc}
   Gear & Mask & Helmet \\
   Cage  & Protection & \\
\end{tabular} \\
\hline
\hline
00075\_ferry & \includegraphics[width=.2\textwidth]{ 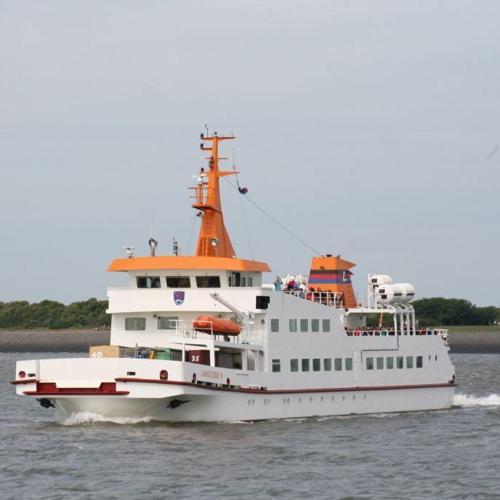} & 
\begin{tabular}{ccc}
   Ferry  & Boat & Transport \\
   Water  & Orange & \\
\end{tabular} \\
\hline
\hline
00076\_flamingo & \includegraphics[width=.2\textwidth]{ 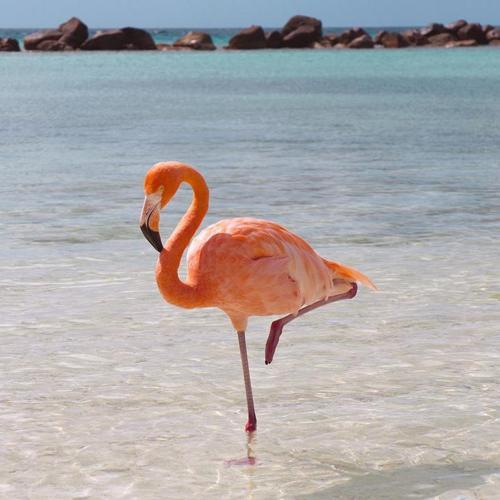} & 
\begin{tabular}{ccc}
    Bird & Flamingo & Pink \\
    Water & Beach & \\
\end{tabular} \\
\hline
\hline
00077\_folder & \includegraphics[width=.2\textwidth]{ 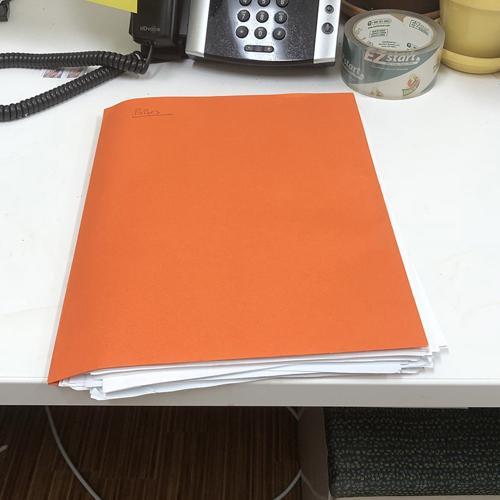} & 
\begin{tabular}{ccc}
   Folder & Office & Orange \\
   Papers  & Desk & \\
\end{tabular} \\
\hline
\hline
00078\_fork & \includegraphics[width=.2\textwidth]{ 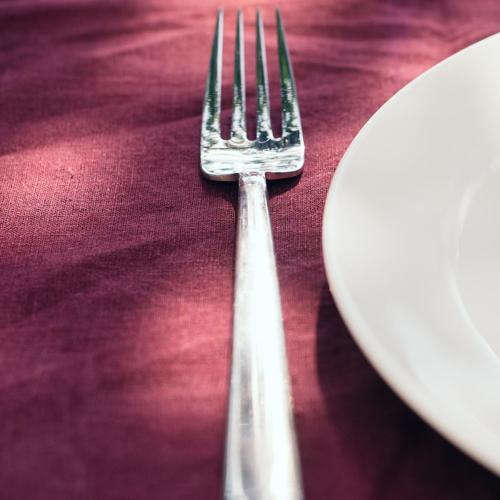} & 
\begin{tabular}{ccc}
    Utensil & Fork & Silver \\
    Plate & Tablecloth & \\
\end{tabular} \\
\hline
\hline
00079\_freezer & \includegraphics[width=.2\textwidth]{ 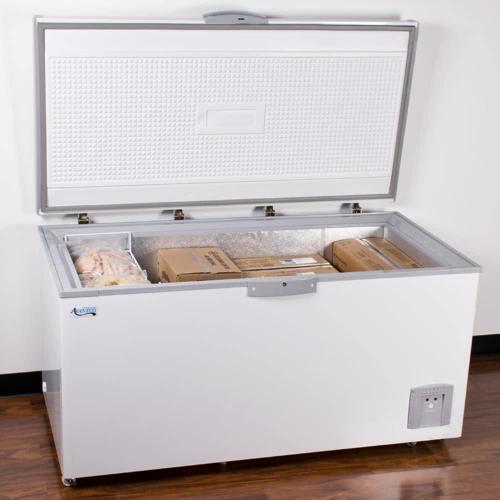} & 
\begin{tabular}{ccc}
   Appliance  & Freezer & Storage \\
    Cold & White & \\
\end{tabular} \\
\hline
\hline
00080\_french\_horn & \includegraphics[width=.2\textwidth]{ 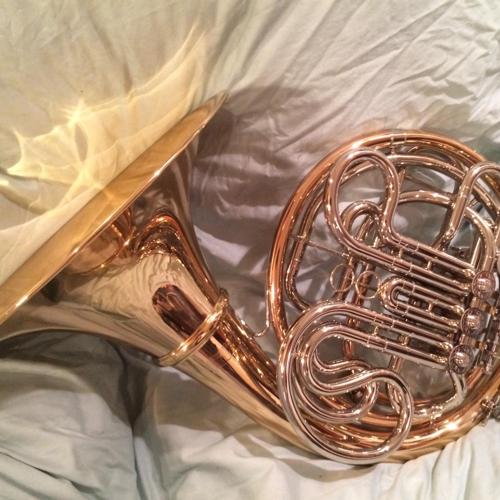} & 
\begin{tabular}{ccc}
    Instrument & Horn & Brass \\
    Coiled & Shiny & \\
\end{tabular} \\
\hline
\hline
00081\_fruit & \includegraphics[width=.2\textwidth]{ 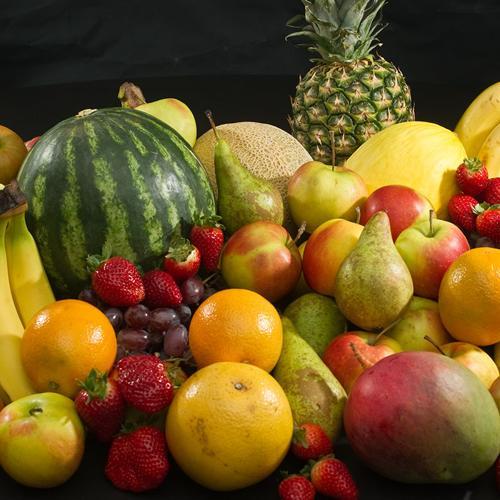} & 
\begin{tabular}{ccc}
   Fruits  & Assortment & Tropical \\
    Colorful & Fresh & \\
\end{tabular} \\
\hline
\hline
00082\_garlic & \includegraphics[width=.2\textwidth]{ 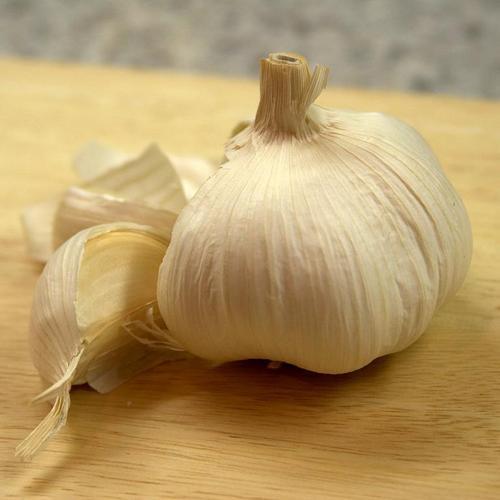} & 
\begin{tabular}{ccc}
    Garlic & Bulb & Cloves \\
    White & Peeled & \\
\end{tabular} \\
\hline
\hline
00083\_glove & \includegraphics[width=.2\textwidth]{ 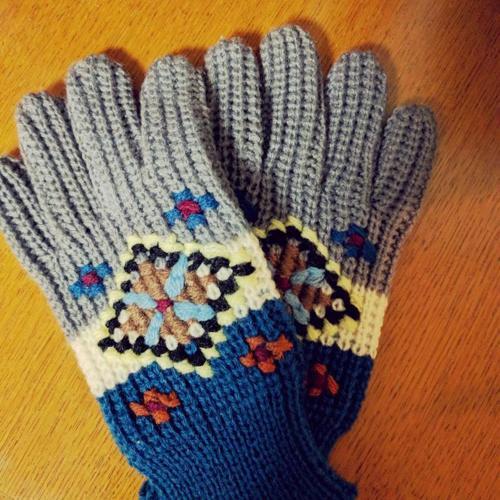} & 
\begin{tabular}{ccc}
   Gloves  & Knitted & Patterned \\
   Wool  & Gray & \\
\end{tabular} \\
\hline
\hline
00084\_golf\_cart & \includegraphics[width=.2\textwidth]{ 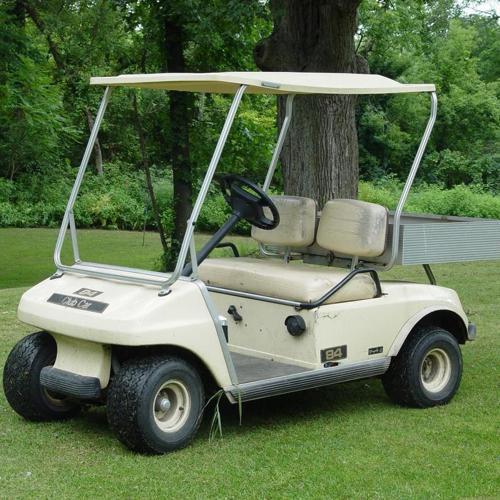} & 
\begin{tabular}{ccc}
    Vehicle & GolfCart & White \\
    Seats & Wheels & \\
\end{tabular} \\
\hline
\hline
00085\_gondola & \includegraphics[width=.2\textwidth]{ 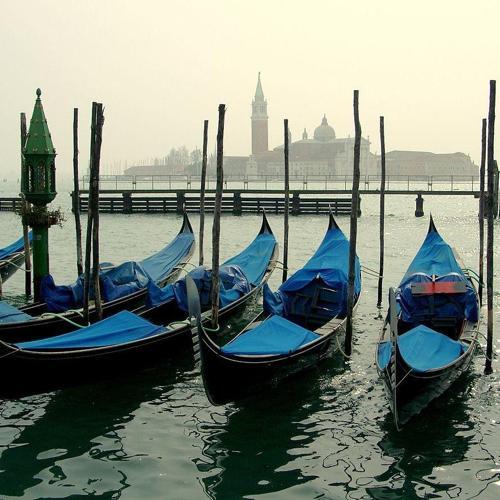} & 
\begin{tabular}{ccc}
   Boats  & Gondolas & Venice \\
   Water  &  Blue & \\
\end{tabular} \\
\hline
\hline
00086\_goose & \includegraphics[width=.2\textwidth]{ 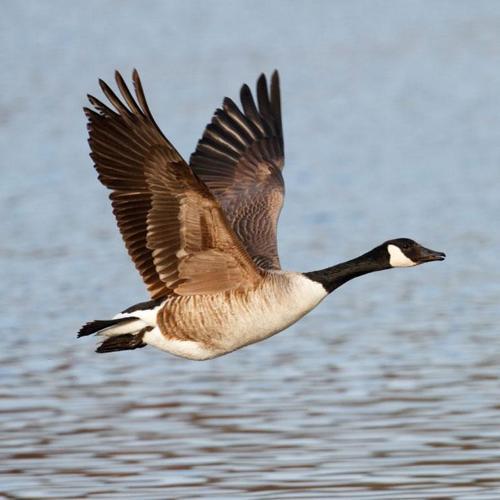} & 
\begin{tabular}{ccc}
    Bird & Goose & Flight \\
   Wings  & Lake & \\
\end{tabular} \\
\hline
\hline
00087\_gopher & \includegraphics[width=.2\textwidth]{ 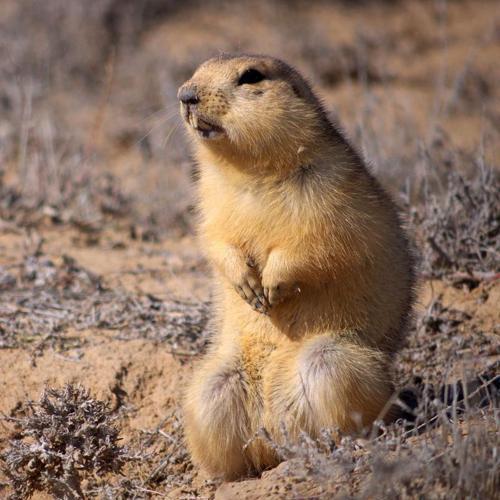} & 
\begin{tabular}{ccc}
   Animal  & Gopher & Furry \\
    Rodent & Field & \\
\end{tabular} \\
\hline
\hline
00088\_gorilla & \includegraphics[width=.2\textwidth]{ 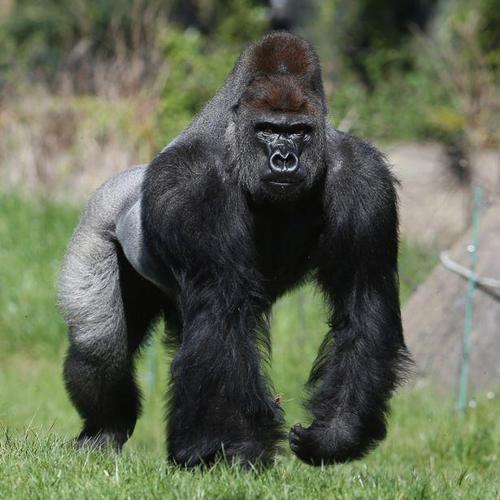} & 
\begin{tabular}{ccc}
   Animal  & Gorilla & Primates \\
    Silverback & Grass & \\
\end{tabular} \\
\hline
\hline
00089\_grasshopper & \includegraphics[width=.2\textwidth]{ 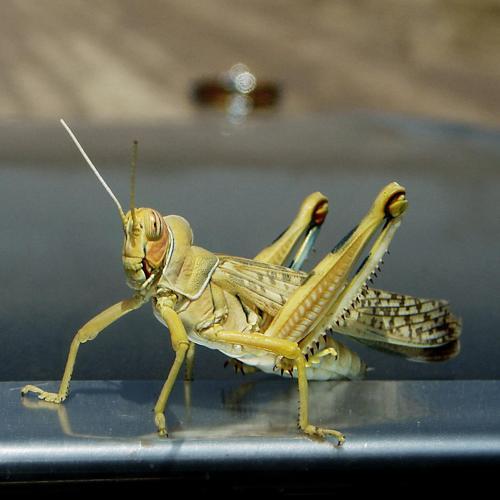} & 
\begin{tabular}{ccc}
   Insect  & Grasshopper & Antennae \\
    Legs & Green & \\
\end{tabular} \\
\hline
\hline
00090\_grenade & \includegraphics[width=.2\textwidth]{ 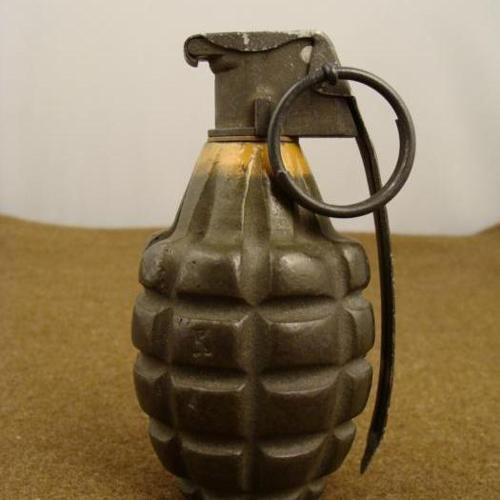} & 
\begin{tabular}{ccc}
    Weapon & Grenade & Metal \\
    Pin & Explosive & \\
\end{tabular} \\
\hline
\hline
00091\_hamburger & \includegraphics[width=.2\textwidth]{ 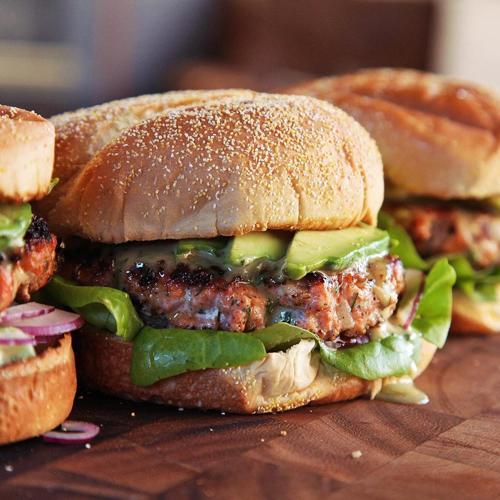} & 
\begin{tabular}{ccc}
    Food & Hamburger & Bun \\
    Lettuce & Grilled & \\
\end{tabular} \\
\hline
\hline
00092\_hammer & \includegraphics[width=.2\textwidth]{ 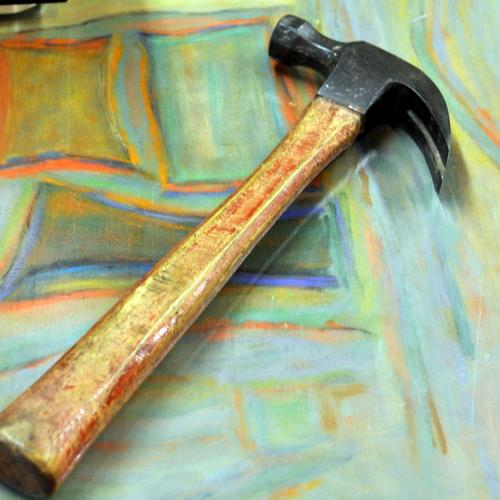} & 
\begin{tabular}{ccc}
    Tool & Hammer & Handle \\
    Metal & Claw & \\
\end{tabular} \\
\hline
\hline
00093\_handbrake & \includegraphics[width=.2\textwidth]{ 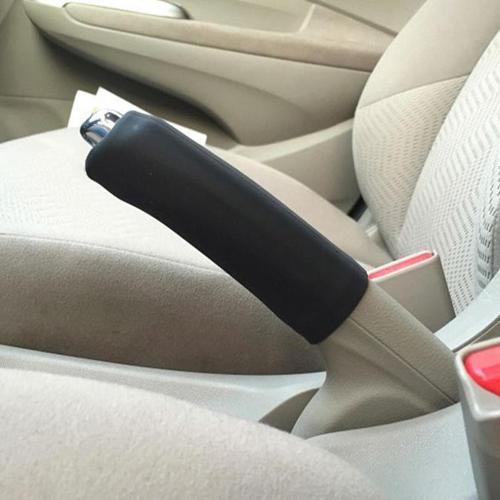} & 
\begin{tabular}{ccc}
   Automobile  & Interior & Handbrake \\
   Lever  & Grip & \\
\end{tabular} \\
\hline
\hline
00094\_headscarf & \includegraphics[width=.2\textwidth]{ 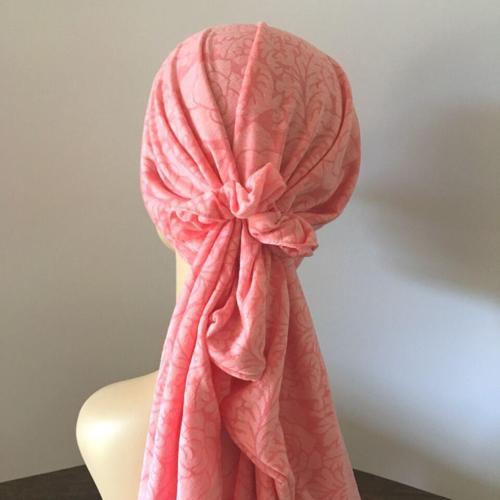} & 
\begin{tabular}{ccc}
    Headwear & Scarf & Fabric \\
    Pink & Wrap & \\
\end{tabular} \\
\hline
\hline
00095\_highchair & \includegraphics[width=.2\textwidth]{ 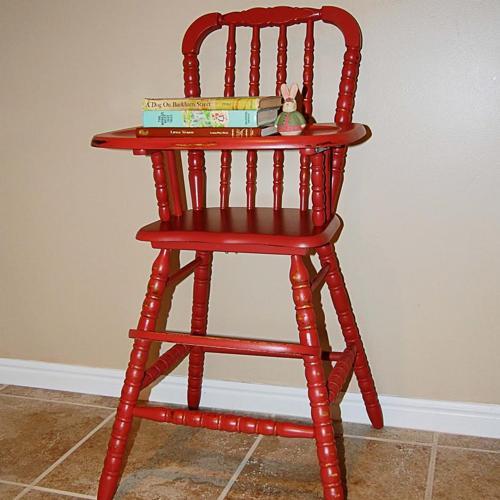} & 
\begin{tabular}{ccc}
    Red & Wooden & Chair \\
    Highchair & Furniture & \\
\end{tabular} \\
\hline
\hline
00096\_hoodie & \includegraphics[width=.2\textwidth]{ 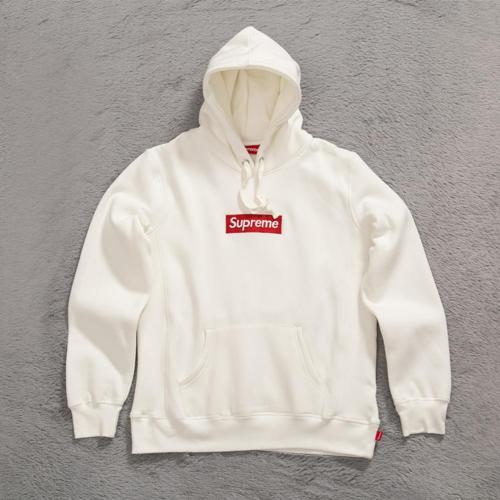} & 
\begin{tabular}{ccc}
    White & Hoodie & Ground \\
    Casual & Clothing & \\
\end{tabular} \\
\hline
\hline
00097\_hummingbird & \includegraphics[width=.2\textwidth]{ 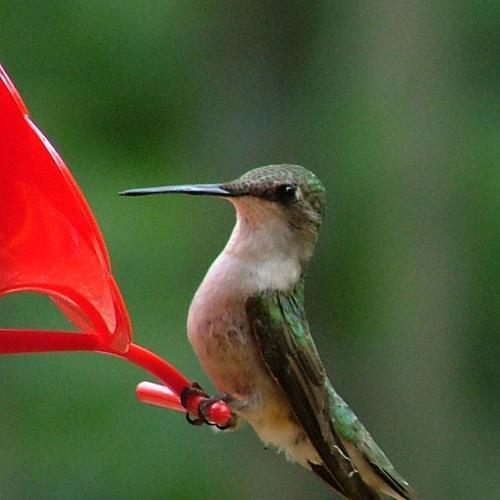} & 
\begin{tabular}{ccc}
    Hummingbird & Green & Feeder \\
    Small & Bird & \\
\end{tabular} \\
\hline
\hline
00098\_ice\_cube & \includegraphics[width=.2\textwidth]{ 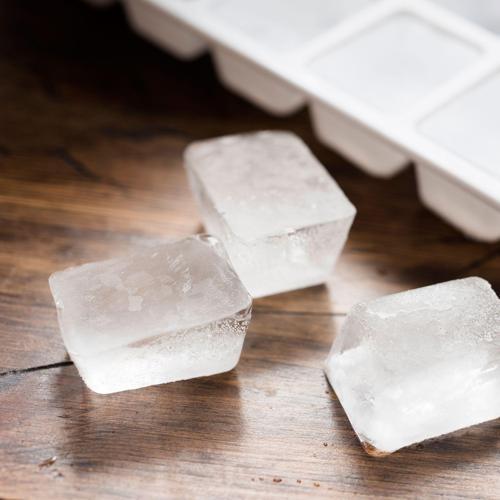} & 
\begin{tabular}{ccc}
   Ice  & Cold & Frozen \\
   Clear  & Cubes & \\
\end{tabular} \\
\hline
\hline
00099\_ice\_pack & \includegraphics[width=.2\textwidth]{ 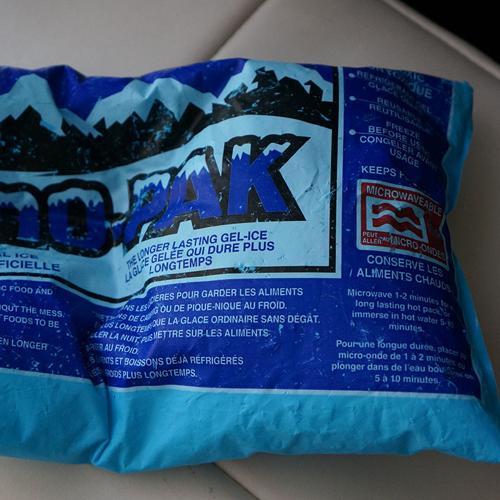} & 
\begin{tabular}{ccc}
   Gel  & Blue & Reusable \\
    Cold & Cooling & \\
\end{tabular} \\
\hline
\hline
00100\_jeep & \includegraphics[width=.2\textwidth]{ 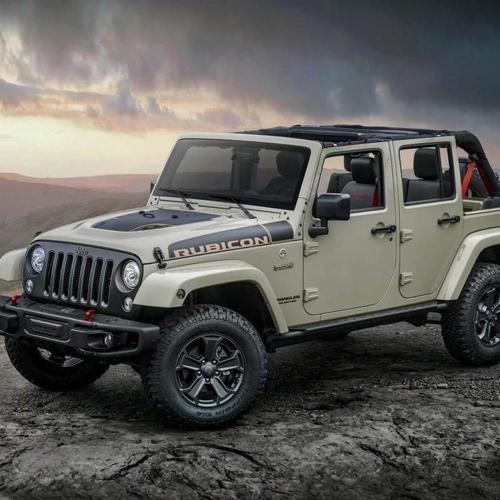} & 
\begin{tabular}{ccc}
    Off-road & Rugged & SUV \\
    Adventure & Durable & \\
\end{tabular} \\
\hline
\hline
00101\_jelly\_bean & \includegraphics[width=.2\textwidth]{ 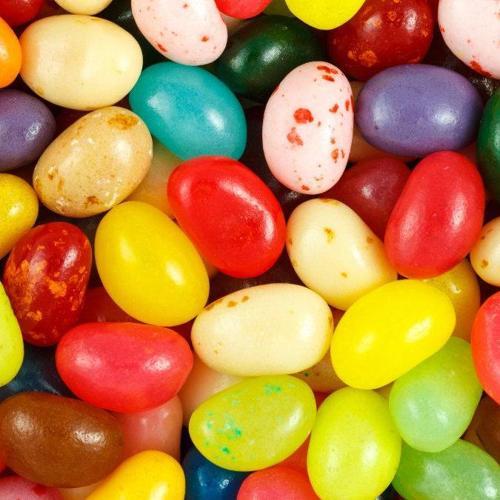} & 
\begin{tabular}{ccc}
    Colorful & Sweet & Candy \\
    Vibrant & Chewy & \\
\end{tabular} \\
\hline
\hline
00102\_jukebox & \includegraphics[width=.2\textwidth]{ 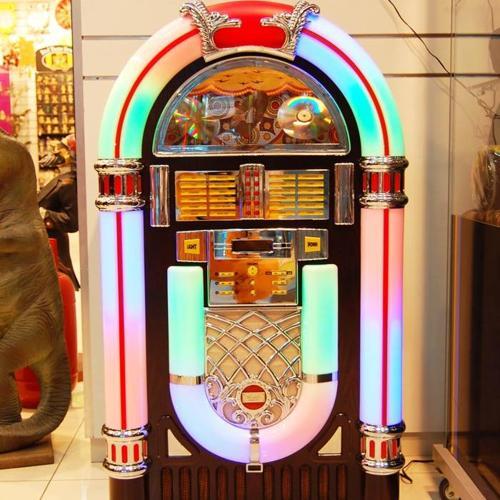} & 
\begin{tabular}{ccc}
    Retro & Vibrant & Music \\
    Neon & Classic & \\
\end{tabular} \\
\hline
\hline
00103\_kettle & \includegraphics[width=.2\textwidth]{ 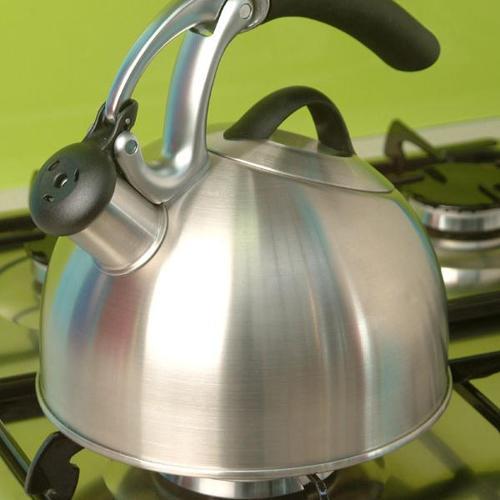} & 
\begin{tabular}{ccc}
    Shiny & Stovetop & Whistling \\
    Metallic & Classic & \\
\end{tabular} \\
\hline
\hline
00104\_kneepad & \includegraphics[width=.2\textwidth]{ 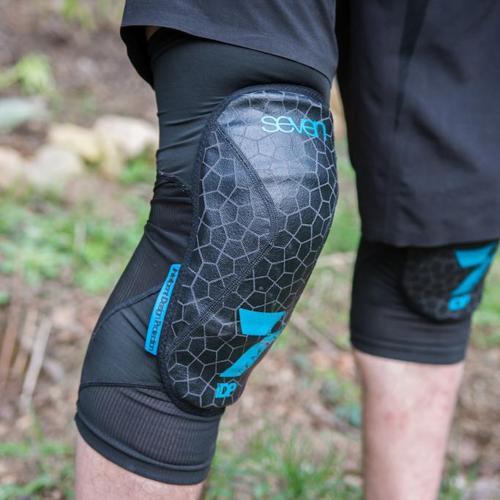} & 
\begin{tabular}{ccc}
    Protective & Sporty & Durable \\
    Cushioned & Ergonomic & \\
\end{tabular} \\
\hline
\hline
00105\_ladle & \includegraphics[width=.2\textwidth]{ 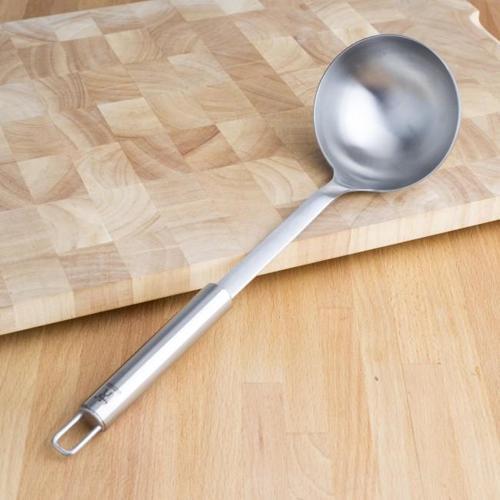} & 
\begin{tabular}{ccc}
    Stainless & Sleek & Functional \\
    Polished & Culinary & \\
\end{tabular} \\
\hline
\hline
00106\_lamb & \includegraphics[width=.2\textwidth]{ 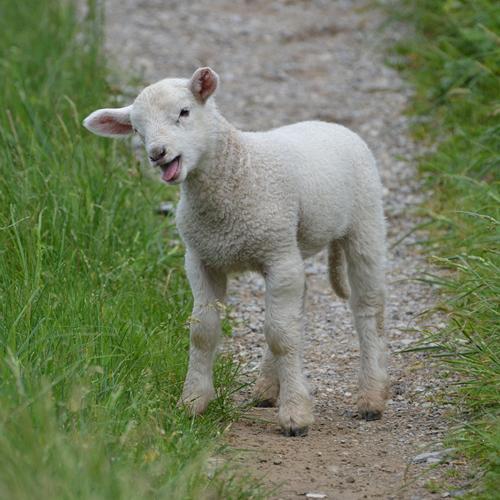} & 
\begin{tabular}{ccc}
    Adorable & Fluffy & Playful \\
    Animal & Lamb & \\
\end{tabular} \\
\hline
\hline
00107\_lampshade & \includegraphics[width=.2\textwidth]{ 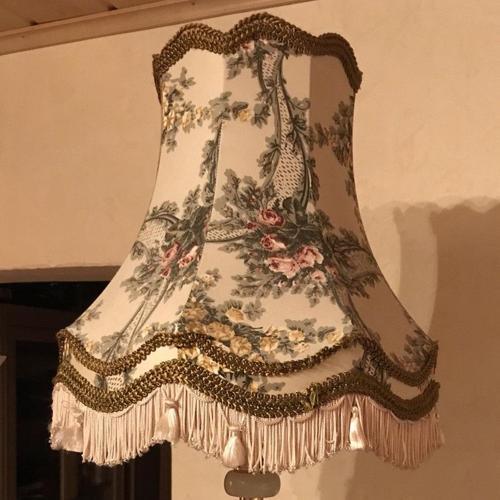} & 
\begin{tabular}{ccc}
    Vintage & Floral & Fabric \\
    Fringed & Ornate & \\
\end{tabular} \\
\hline
\hline
00108\_laundry\_basket & \includegraphics[width=.2\textwidth]{ 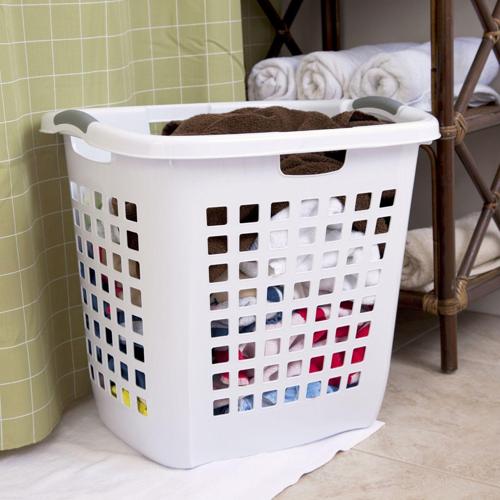} & 
\begin{tabular}{ccc}
    Laundry & Plastic & Basket \\
    Towels & Grid & \\
\end{tabular} \\
\hline
\hline
00109\_lettuce & \includegraphics[width=.2\textwidth]{ 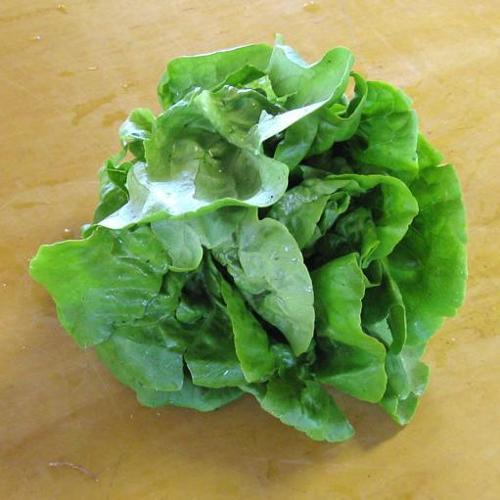} & 
\begin{tabular}{ccc}
    Vegetable & Lettuce & Leafy \\
    Fresh & Green & \\
\end{tabular} \\
\hline
\hline
00110\_lightning\_bug & \includegraphics[width=.2\textwidth]{ 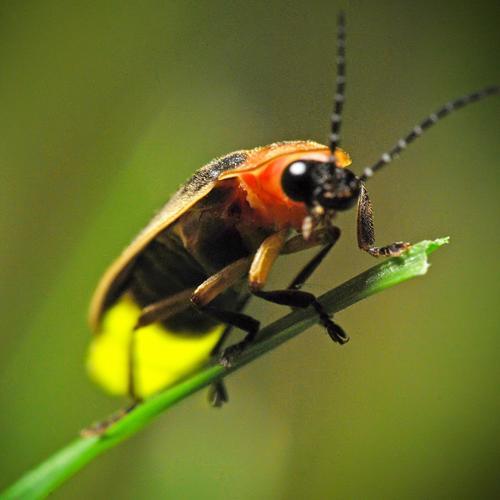} & 
\begin{tabular}{ccc}
   Insect  & Firefly & Antennae \\
   Glowing  & Segmented & \\
\end{tabular} \\
\hline
\hline
00111\_manatee & \includegraphics[width=.2\textwidth]{ 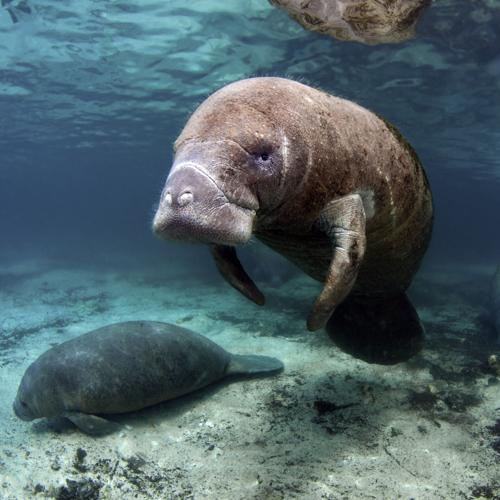} & 
\begin{tabular}{ccc}
    Aquatic & Manatee & Underwater \\
    Mammal & Floating & \\
\end{tabular} \\
\hline
\hline
00112\_marijuana & \includegraphics[width=.2\textwidth]{ 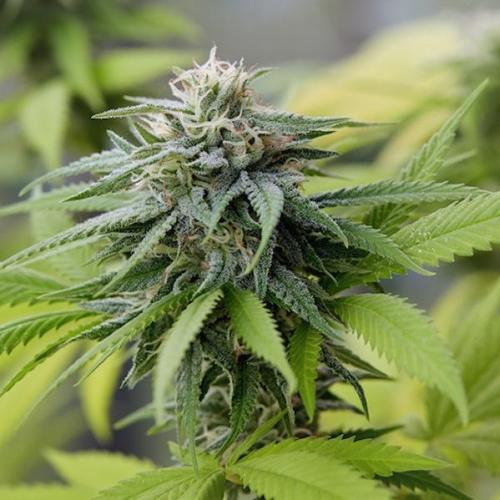} & 
\begin{tabular}{ccc}
   Cannabis  & Plant & Buds \\
   Leaves  & Green & \\
\end{tabular} \\
\hline
\hline
00113\_meatloaf & \includegraphics[width=.2\textwidth]{ 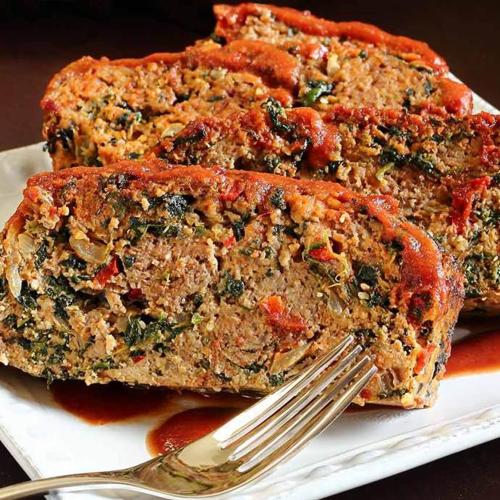} & 
\begin{tabular}{ccc}
   Food  & Meatloaf & Slice \\
   Sauce  & Hearty & \\
\end{tabular} \\
\hline
\hline
00114\_metal\_detector & \includegraphics[width=.2\textwidth]{ 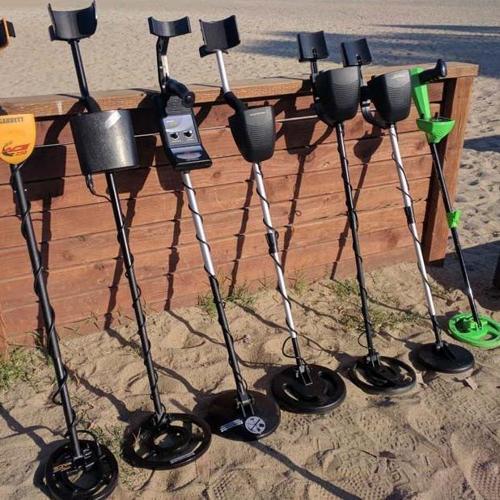} & 
\begin{tabular}{ccc}
   Equipment  & Detectors & Metal \\
   Beach  & Lineup & \\
\end{tabular} \\
\hline
\hline
00115\_minivan & \includegraphics[width=.2\textwidth]{ 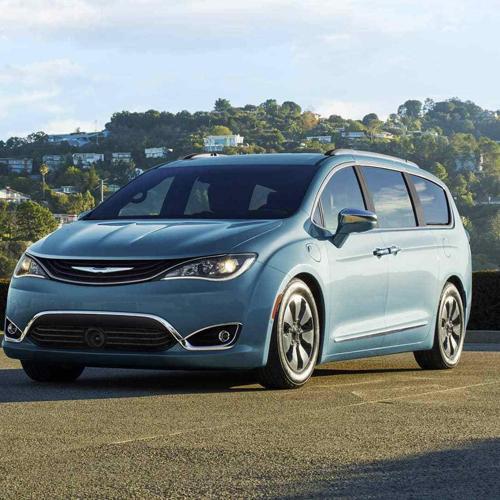} & 
\begin{tabular}{ccc}
   Vehicle  & Minivan & Car \\
    Blue & Electric & \\
\end{tabular} \\
\hline
\hline
00116\_modem & \includegraphics[width=.2\textwidth]{ 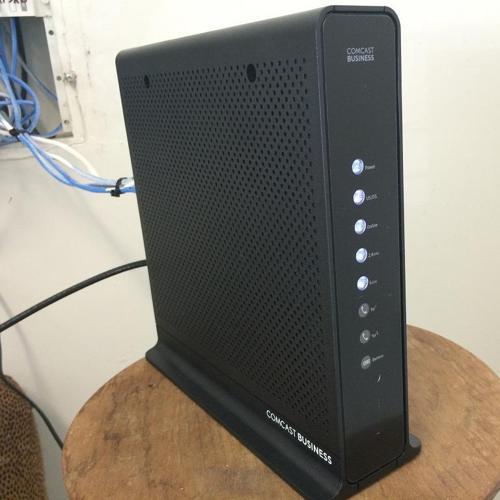} & 
\begin{tabular}{ccc}
   Device  & Modem & Router\\
    Black & Connectivity & \\
\end{tabular} \\
\hline
\hline
00117\_mosquito & \includegraphics[width=.2\textwidth]{ 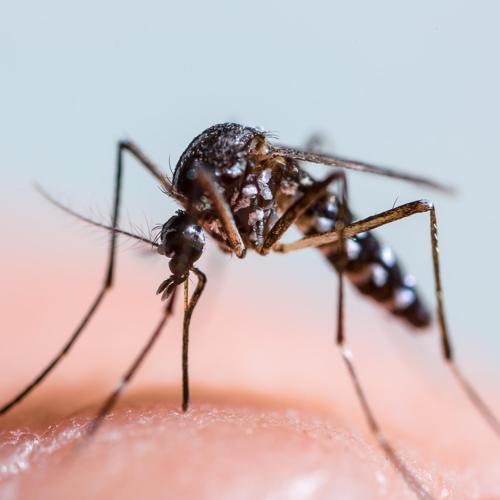} & 
\begin{tabular}{ccc}
   Insect  & Mosquito & Biting \\
    Legs &  Proboscis & \\
\end{tabular} \\
\hline
\hline
00118\_muff & \includegraphics[width=.2\textwidth]{ 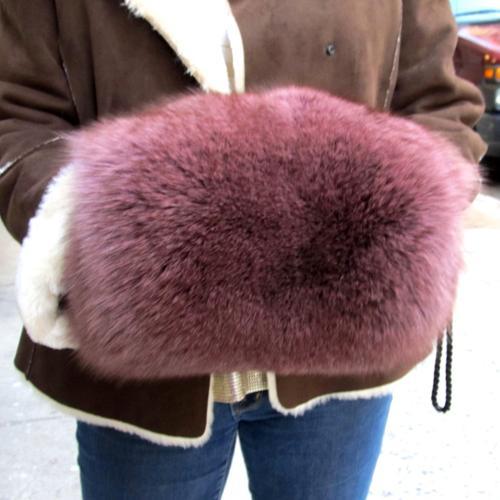} & 
\begin{tabular}{ccc}
  Accessory   & Muff &  Fur\\
    Warm & Pink & \\
\end{tabular} \\
\hline
\hline
00119\_music\_box & \includegraphics[width=.2\textwidth]{ 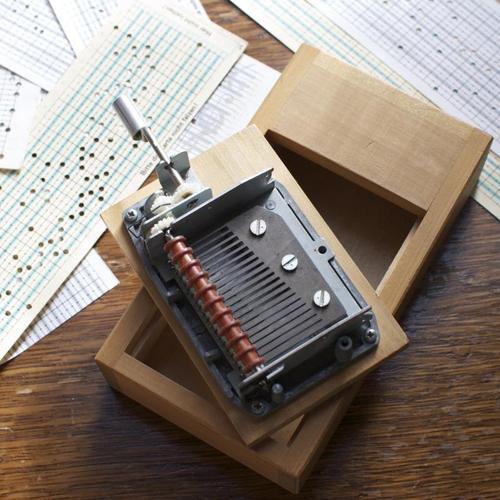} & 
\begin{tabular}{ccc}
   Device  & Music & Box \\
   Crank  & Punched & \\
\end{tabular} \\
\hline
\hline
00120\_mussel & \includegraphics[width=.2\textwidth]{ 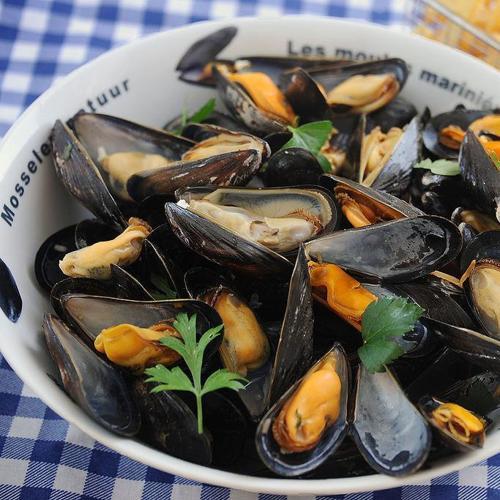} & 
\begin{tabular}{ccc}
    Seafood & Mussels & Shells \\
    Steamed & Parsley & \\
\end{tabular} \\
\hline
\hline
00121\_nightstand & \includegraphics[width=.2\textwidth]{ 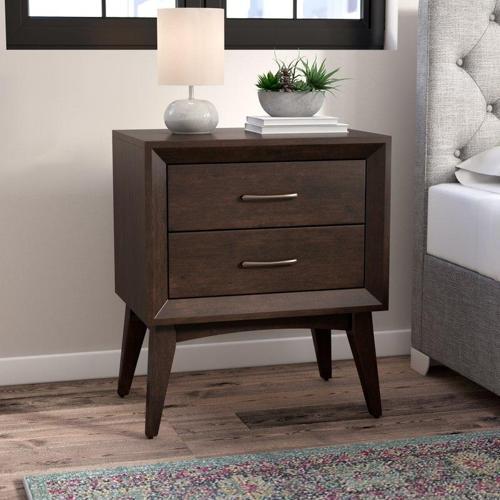} & 
\begin{tabular}{ccc}
    Furniture & Nightstand & Wooden \\
    Drawer &  Lamp & \\
\end{tabular} \\
\hline
\hline
00122\_okra & \includegraphics[width=.2\textwidth]{ 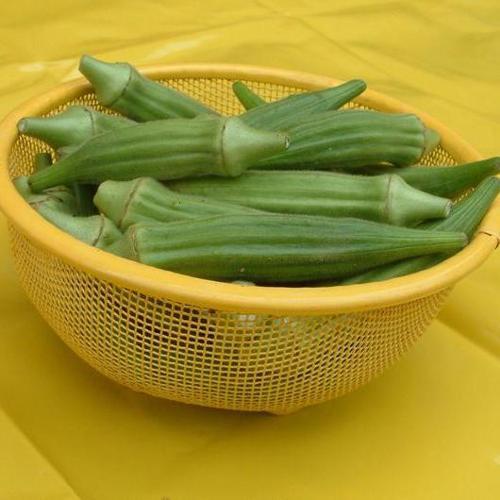} & 
\begin{tabular}{ccc}
    Vegetable & Okra & Green \\
    Basket & Fresh & \\
\end{tabular} \\
\hline
\hline
00123\_omelet & \includegraphics[width=.2\textwidth]{ 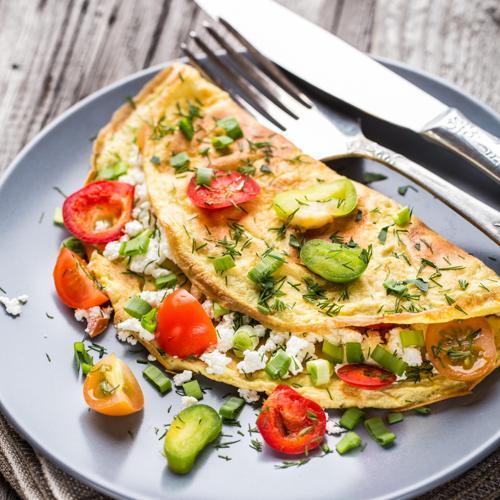} & 
\begin{tabular}{ccc}
   Breakfast  & Omelet & Vegetables \\
   Tomatoes  & Herbs & \\
\end{tabular} \\
\hline
\hline
00124\_onion & \includegraphics[width=.2\textwidth]{ 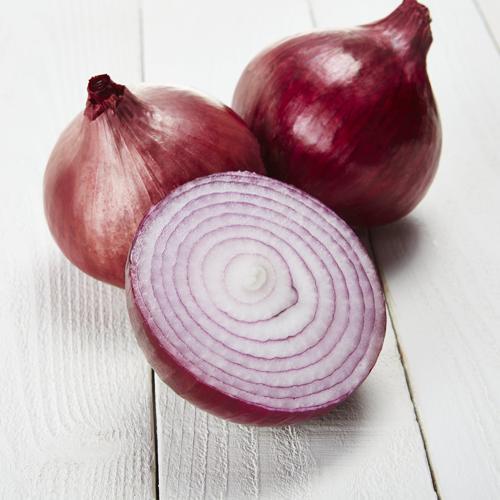} & 
\begin{tabular}{ccc}
   Vegetable  & Onion & Red \\
   Sliced  & Raw & \\
\end{tabular} \\
\hline
\hline
00125\_orange & \includegraphics[width=.2\textwidth]{ 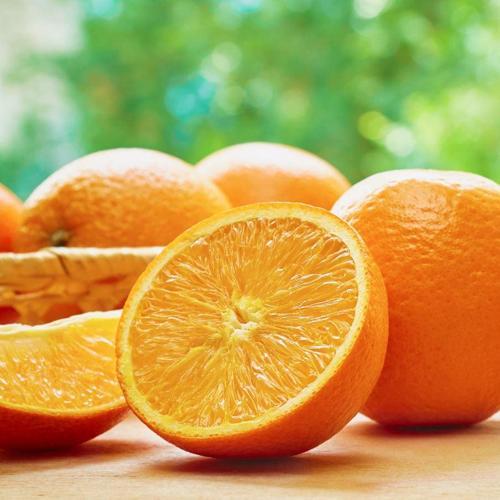} & 
\begin{tabular}{ccc}
   Fruit  & Orange & Citrus \\
   Sliced  & Juicy & \\
\end{tabular} \\
\hline
\hline
00126\_orchid & \includegraphics[width=.2\textwidth]{ 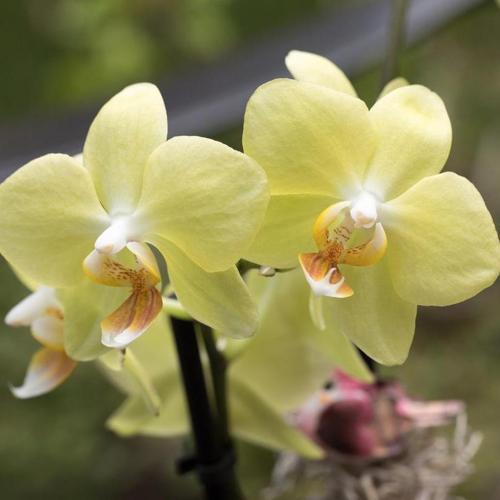} & 
\begin{tabular}{ccc}
   Flower  & Orchid & Yellow \\
   Bloom  & Petals & \\
\end{tabular} \\
\hline
\hline
00127\_ostrich & \includegraphics[width=.2\textwidth]{ 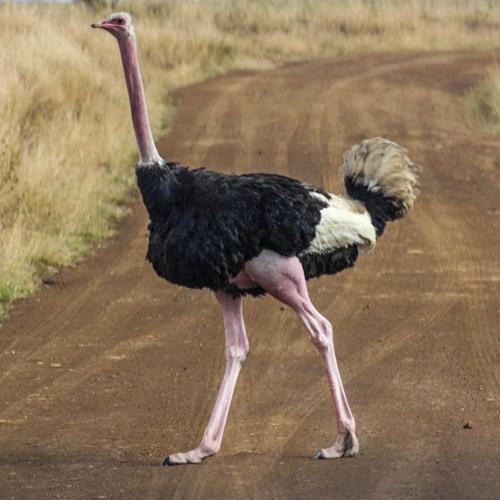} & 
\begin{tabular}{ccc}
   Bird  & Ostrich & Large \\
   Plumage  & Road & \\
\end{tabular} \\
\hline
\hline
00128\_pajamas & \includegraphics[width=.2\textwidth]{ 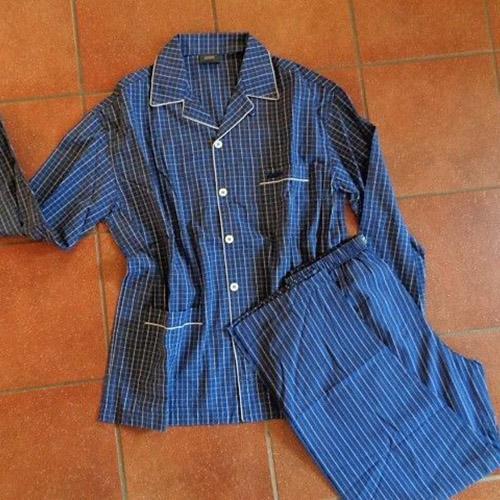} & 
\begin{tabular}{ccc}
   Clothing  & Pajamas & Striped \\
   Blue  & Fabric & \\
\end{tabular} \\
\hline
\hline
00129\_panther & \includegraphics[width=.2\textwidth]{ 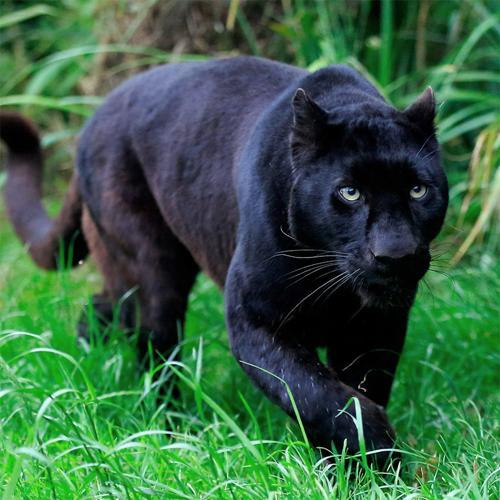} & 
\begin{tabular}{ccc}
   Animal  & Panther & Black \\
   Predator  & Stealthy & \\
\end{tabular} \\
\hline
\hline
00130\_paperweight & \includegraphics[width=.2\textwidth]{ 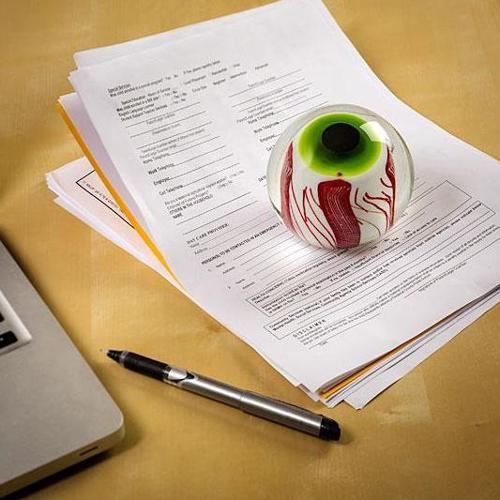} & 
\begin{tabular}{ccc}
   Office  & Paperwork & Paperweight \\
   Eyeball  & Documents & \\
\end{tabular} \\
\hline
\hline
00131\_pear & \includegraphics[width=.2\textwidth]{ 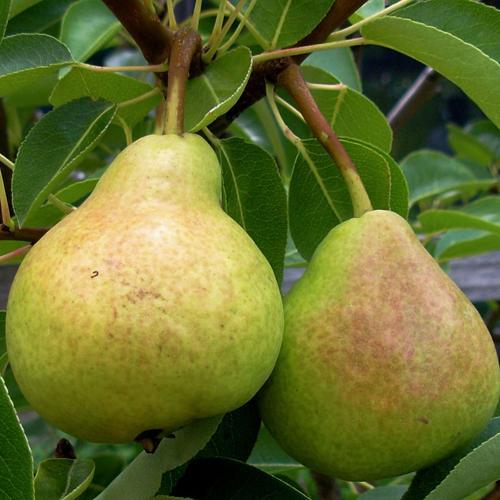} & 
\begin{tabular}{ccc}
    Fruit & Pear & Tree \\
    Green & Ripe & \\
\end{tabular} \\
\hline
\hline
00132\_pepper1 & \includegraphics[width=.2\textwidth]{ 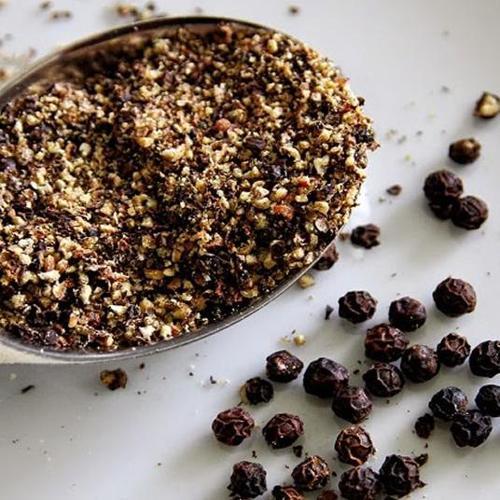} & 
\begin{tabular}{ccc}
    Spice & Pepper & Ground \\
    Black & Spoon & \\
\end{tabular} \\
\hline
\hline
00133\_pheasant & \includegraphics[width=.2\textwidth]{ 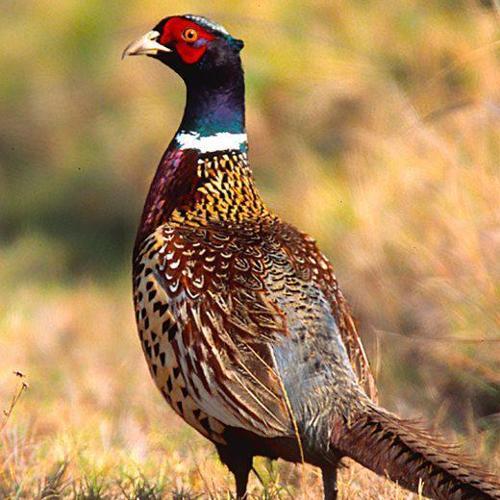} & 
\begin{tabular}{ccc}
    Bird & Pheasant & Feathers \\
   Colorful  & Wild & \\
\end{tabular} \\
\hline
\hline
00134\_pickax & \includegraphics[width=.2\textwidth]{ 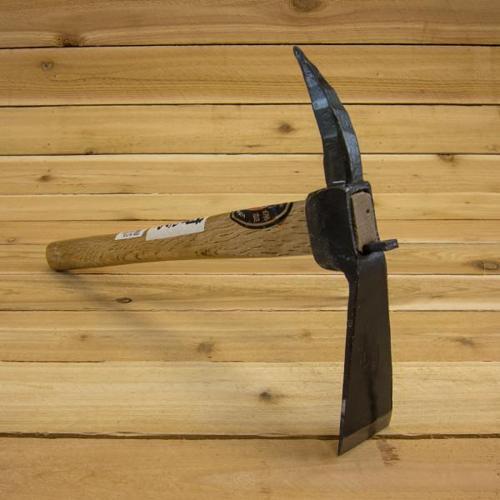} & 
\begin{tabular}{ccc}
   Tool  & Pickaxe &  Wooden\\
   Metal  & Digging & \\
\end{tabular} \\
\hline
\hline
00135\_pie & \includegraphics[width=.2\textwidth]{ 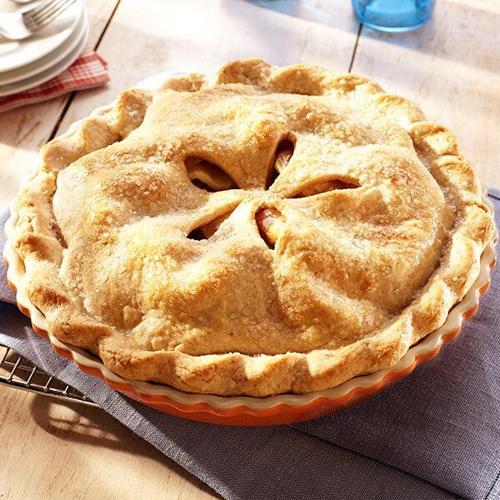} & 
\begin{tabular}{ccc}
   Dessert  & Pie & Baked \\
   Crust  & Golden & \\
\end{tabular} \\
\hline
\hline
00136\_pigeon & \includegraphics[width=.2\textwidth]{ 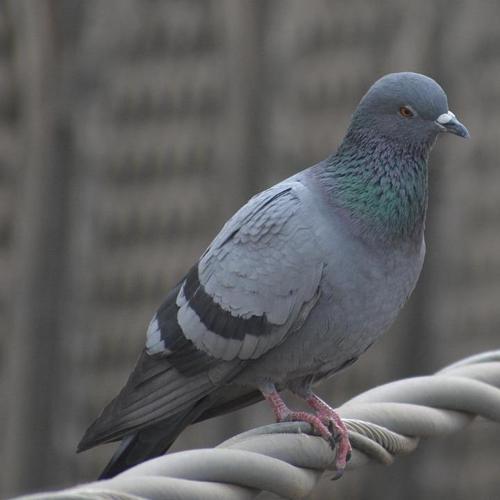} & 
\begin{tabular}{ccc}
    Bird & Pigeon & Grey \\
    Perched & Feathers & \\
\end{tabular} \\
\hline
\hline
00137\_piglet & \includegraphics[width=.2\textwidth]{ 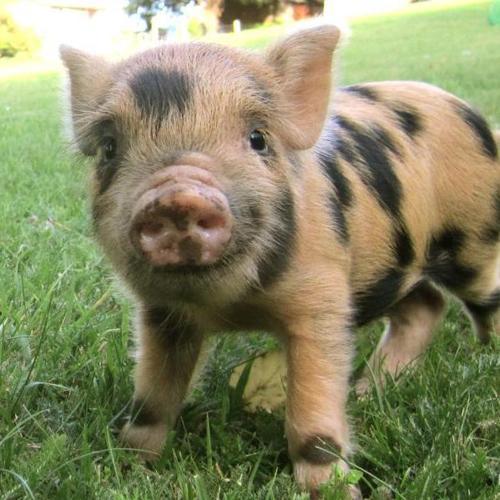} & 
\begin{tabular}{ccc}
    Animal & Piglet & Spotted \\
   Grass  & Cute & \\
\end{tabular} \\
\hline
\hline
00138\_pocket & \includegraphics[width=.2\textwidth]{ 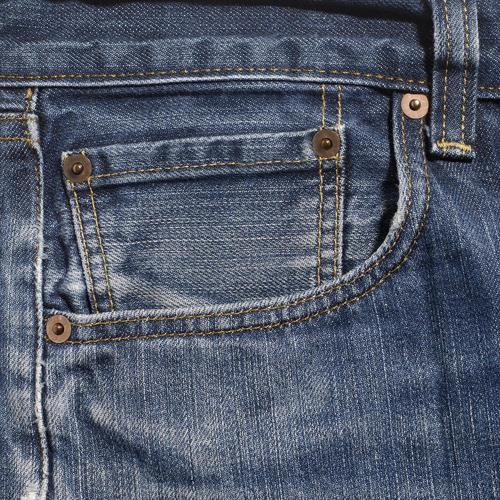} & 
\begin{tabular}{ccc}
   Clothing  & Jeans &  Pocket\\
   Denim  & Stitched & \\
\end{tabular} \\
\hline
\hline
00139\_pocketknife & \includegraphics[width=.2\textwidth]{ 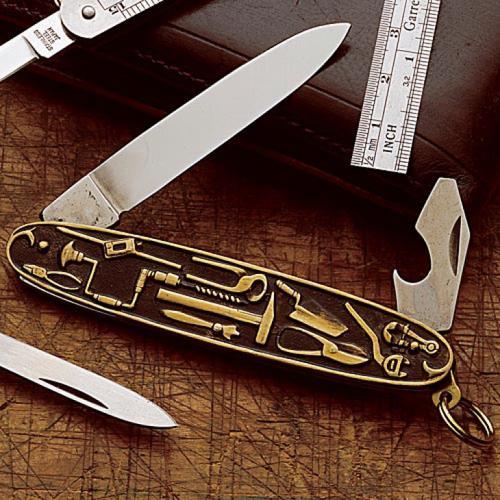} & 
\begin{tabular}{ccc}
   Tool  & Pocketknife & Blade \\
   Compact  & Multi-functional & \\
\end{tabular} \\
\hline
\hline
00140\_popcorn & \includegraphics[width=.2\textwidth]{ 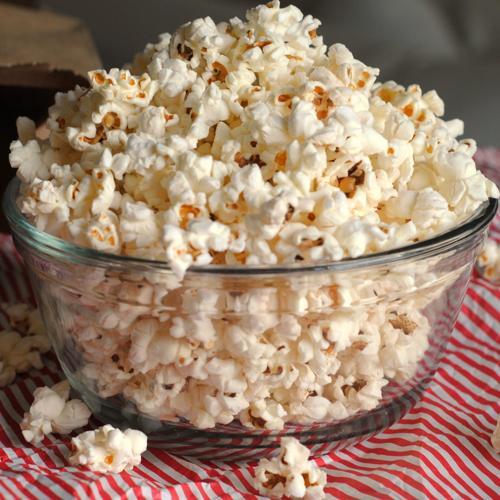} & 
\begin{tabular}{ccc}
   Snack  & Popcorn & Bowl \\
   Buttery  & Crispy & \\
\end{tabular} \\
\hline
\hline
00141\_popsicle & \includegraphics[width=.2\textwidth]{ 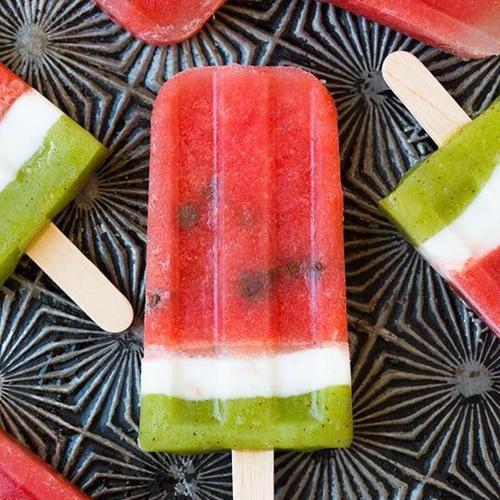} & 
\begin{tabular}{ccc}
   Dessert  & Popsicle & Colorful \\
   Frozen  & Fruit & \\
\end{tabular} \\
\hline
\hline
00142\_possum & \includegraphics[width=.2\textwidth]{ 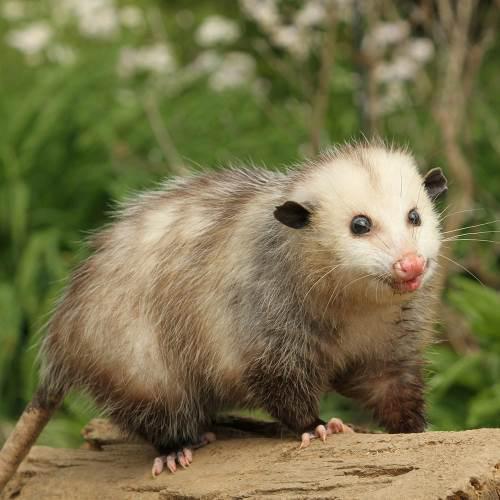} &
\begin{tabular}{ccc}
   Animal  & Possum &  Furry\\
   Marsupial  & Wild & \\
\end{tabular} \\
\hline
\hline
00143\_pretzel & \includegraphics[width=.2\textwidth]{ 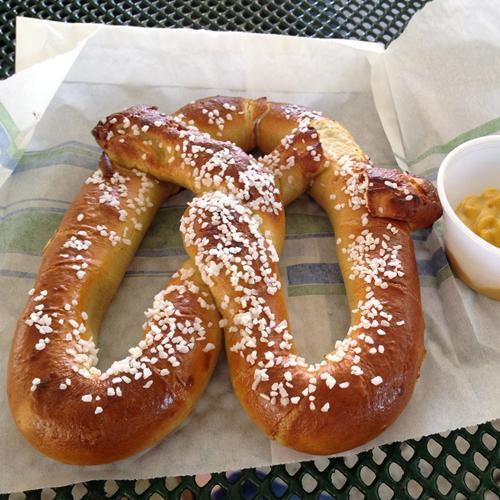} & 
\begin{tabular}{ccc}
   Snack  & Pretzel & Salted \\
   Baked  & Dough & \\
\end{tabular} \\
\hline
\hline
00144\_pug & \includegraphics[width=.2\textwidth]{ 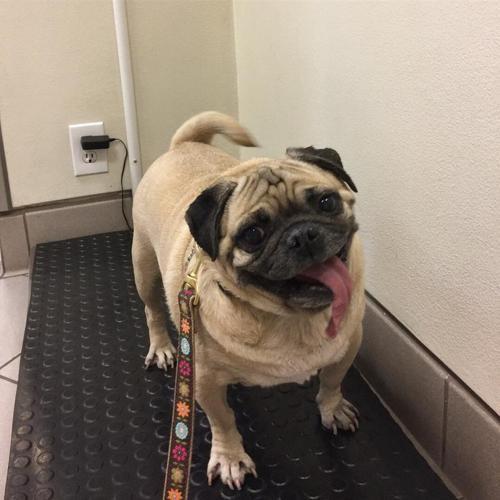} & 
\begin{tabular}{ccc}
  Animal   & Pug & Dog \\
   Leash  & Panting & \\
\end{tabular} \\
\hline
\hline
00145\_punch2 & \includegraphics[width=.2\textwidth]{ 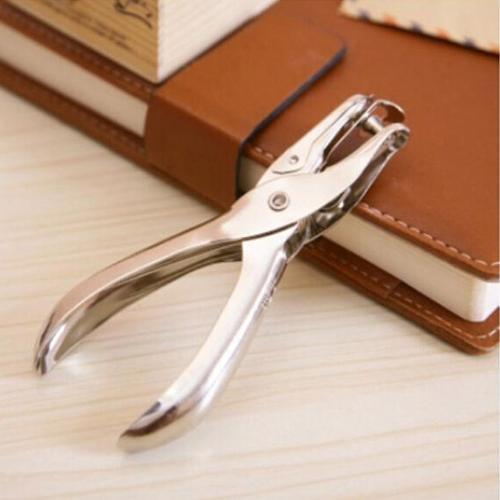} & 
\begin{tabular}{ccc}
    Tool & Punch & Metal \\
    Office & Desk & \\
\end{tabular} \\
\hline
\hline
00146\_purse & \includegraphics[width=.2\textwidth]{ 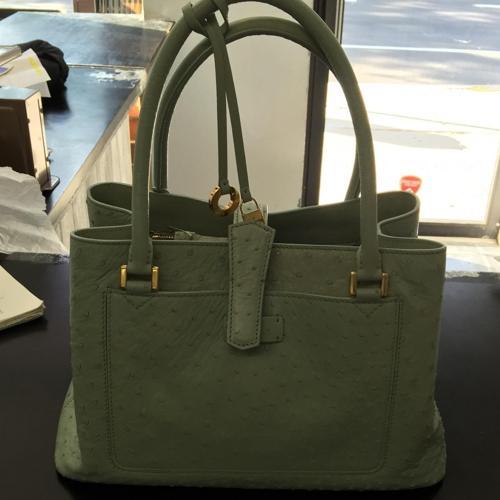} & 
\begin{tabular}{ccc}
   Accessory  & Purse & Leather \\
   Green  & Handles & \\
\end{tabular} \\
\hline
\hline
00147\_radish & \includegraphics[width=.2\textwidth]{ 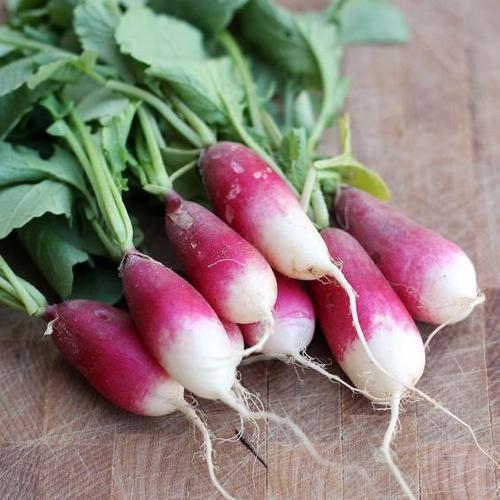} & 
\begin{tabular}{ccc}
    Vegetable & Radish & Root \\
    Fresh & Bunch & \\
\end{tabular} \\
\hline
\hline
00148\_raspberry & \includegraphics[width=.2\textwidth]{ 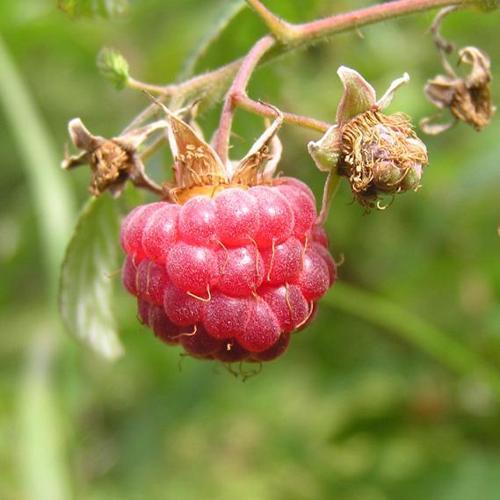} & 
\begin{tabular}{ccc}
   Fruit  & Raspberry & Red \\
   Berry  & Branch & \\
\end{tabular} \\
\hline
\hline
00149\_recorder & \includegraphics[width=.2\textwidth]{ 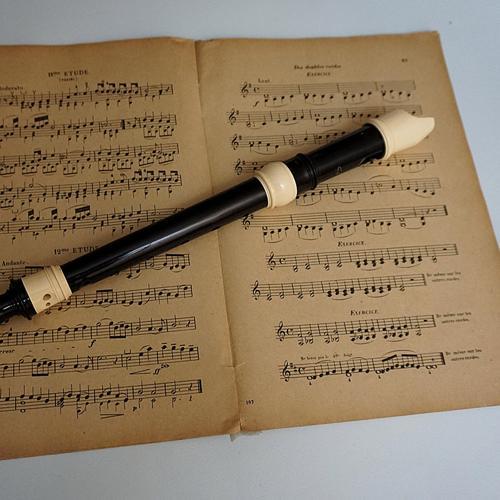} & 
\begin{tabular}{ccc}
   Instrument  & Recorder & Music \\
   Notes  & Sheet & \\
\end{tabular} \\
\hline
\hline
00150\_rhinoceros & \includegraphics[width=.2\textwidth]{ 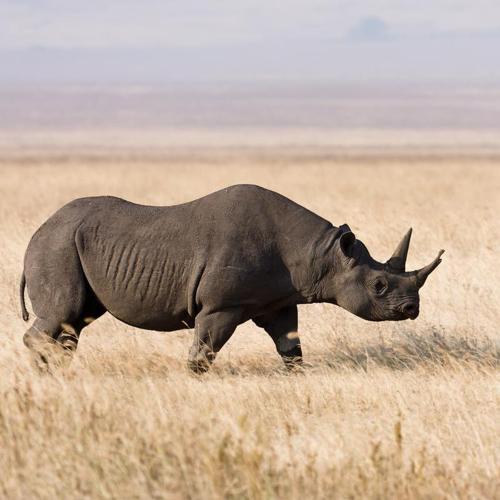} & 
\begin{tabular}{ccc}
    Animal & Rhinoceros &  Horned\\
    Savanna & Wild & \\
\end{tabular} \\
\hline
\hline
00151\_robot & \includegraphics[width=.2\textwidth]{ 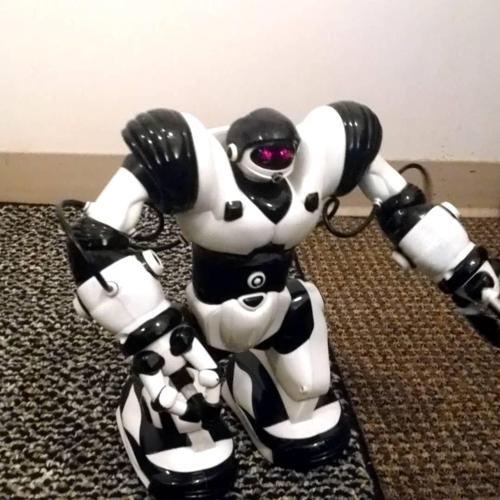} & 
\begin{tabular}{ccc}
    Robot & Toy & Humanoid \\
    Black & White & \\
\end{tabular} \\
\hline
\hline
00152\_rooster & \includegraphics[width=.2\textwidth]{ 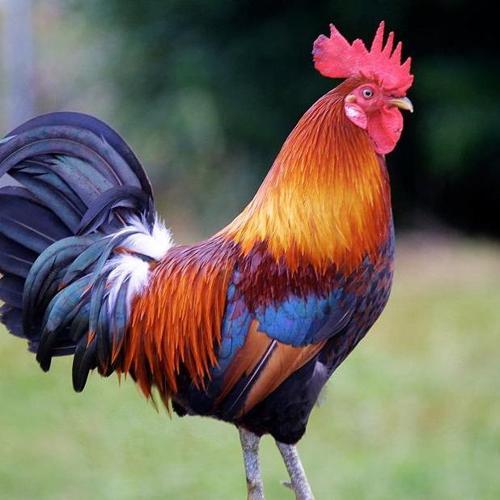} & 
\begin{tabular}{ccc}
    Bird & Rooster & Feathers \\
    Colorful & Comb & \\
\end{tabular} \\
\hline
\hline
00153\_rug & \includegraphics[width=.2\textwidth]{ 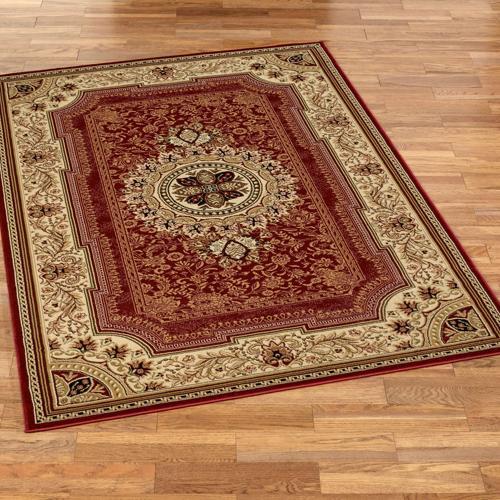} & 
\begin{tabular}{ccc}
    Furniture & Rug  & Patterned \\
    Red & Ornate & \\
\end{tabular} \\
\hline
\hline
00154\_sailboat & \includegraphics[width=.2\textwidth]{ 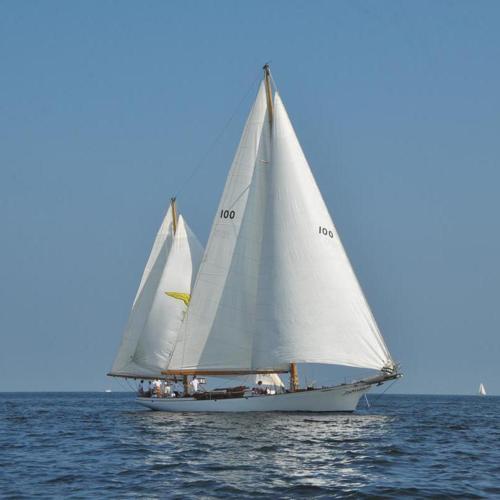} & 
\begin{tabular}{ccc}
    Boat & Sailboat & Ocean \\
    White & Wind & \\
\end{tabular} \\
\hline
\hline
00155\_sandal & \includegraphics[width=.2\textwidth]{ 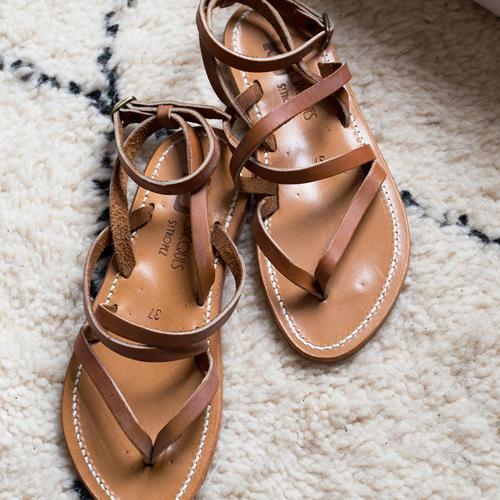} & 
\begin{tabular}{ccc}
    Footwear & Sandals & Leather \\
    Straps & Brown & \\
\end{tabular} \\
\hline
\hline
00156\_sandpaper & \includegraphics[width=.2\textwidth]{ 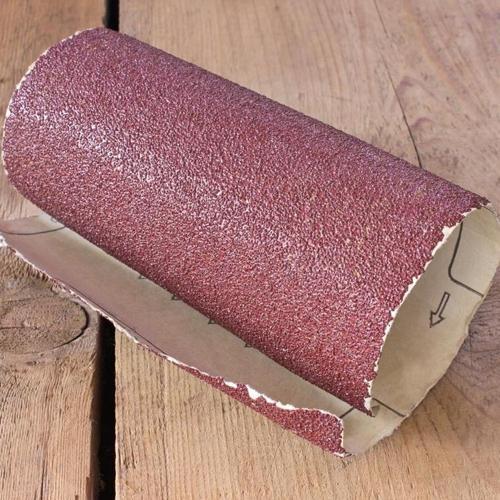} & 
\begin{tabular}{ccc}
    Tool & Sandpaper & Abrasive \\
    Roll & Rough  & \\
\end{tabular} \\
\hline
\hline
00157\_sausage & \includegraphics[width=.2\textwidth]{ 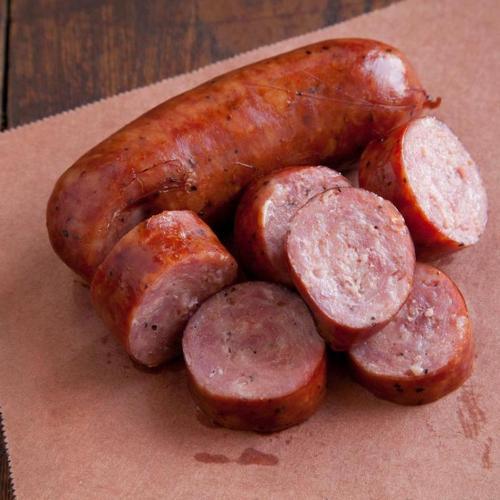} & 
\begin{tabular}{ccc}
     Food & Sausage & Sliced \\
     Smoked & Meat & \\
\end{tabular} \\
\hline
\hline
00158\_scallion & \includegraphics[width=.2\textwidth]{ 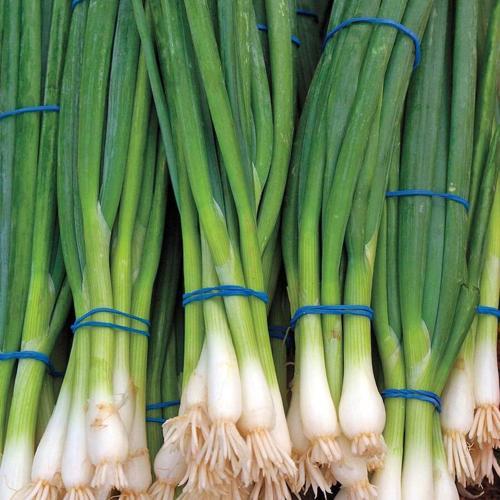} & 
\begin{tabular}{ccc}
     Vegetable & Scallion & Green \\
     Fresh & Bundle & \\
\end{tabular} \\
\hline
\hline
00159\_scallop & \includegraphics[width=.2\textwidth]{ 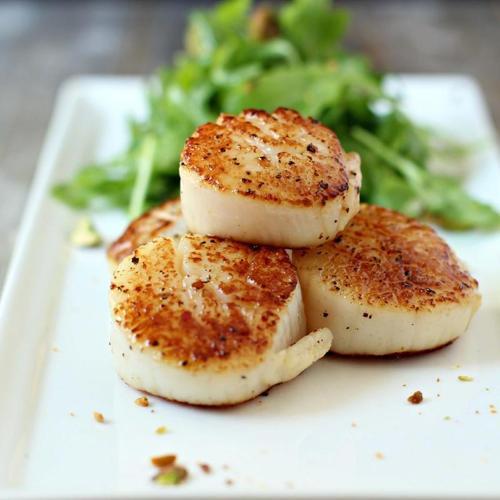} & 
\begin{tabular}{ccc}
    Seafood & Scallops & Seared \\
    Plate & Garnish & \\
\end{tabular} \\
\hline
\hline
00160\_scooter & \includegraphics[width=.2\textwidth]{ 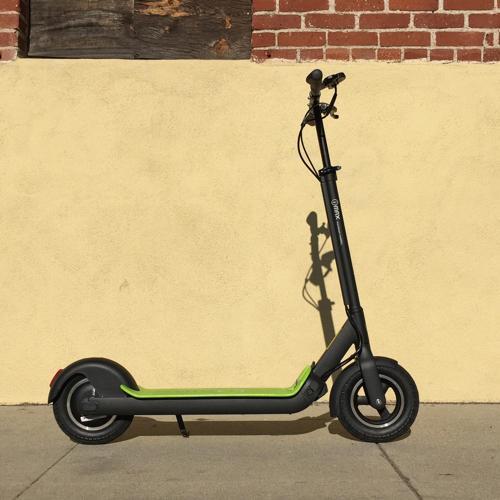} & 
\begin{tabular}{ccc}
     Vehicle & Scooter & Electric \\
     Green & Urban & \\
\end{tabular} \\
\hline
\hline
00161\_seagull & \includegraphics[width=.2\textwidth]{ 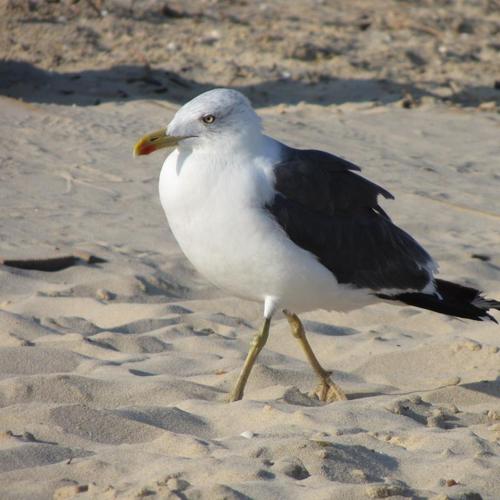} & 
\begin{tabular}{ccc}
     Bird & Seagull & Beach \\
     White & Walking & \\
\end{tabular} \\
\hline
\hline
00162\_seaweed & \includegraphics[width=.2\textwidth]{ 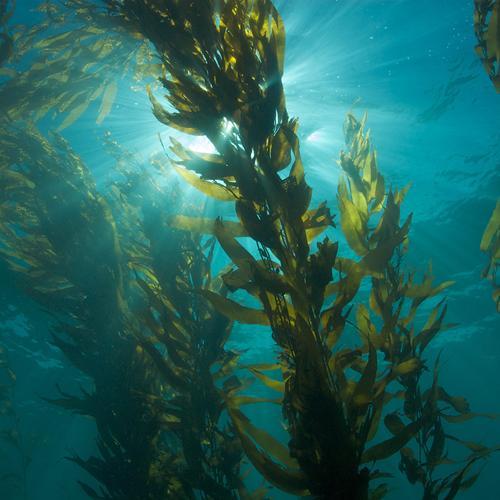} & 
\begin{tabular}{ccc}
     Marine & Seaweed &  Underwater \\
     Aquatic & Sunlight & \\
\end{tabular} \\
\hline
\hline
00163\_seed & \includegraphics[width=.2\textwidth]{ 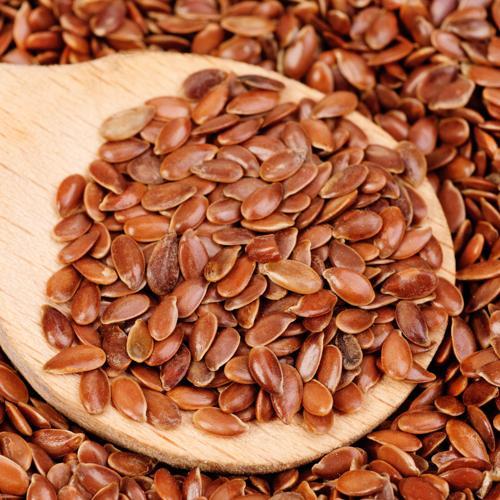} & 
\begin{tabular}{ccc}
     Food & Seeds & Flax \\
     Brown &  Spoon & \\
\end{tabular} \\
\hline
\hline
00164\_skateboard & \includegraphics[width=.2\textwidth]{ 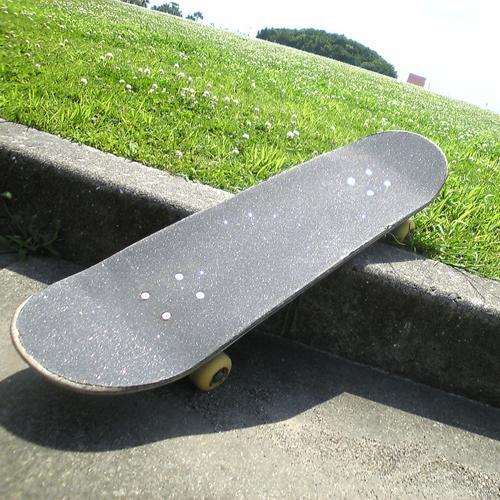} & 
\begin{tabular}{ccc}
    Sport & Skateboard & Wheels \\
    Outdoor & Deck & \\
\end{tabular} \\
\hline
\hline
00165\_sled & \includegraphics[width=.2\textwidth]{ 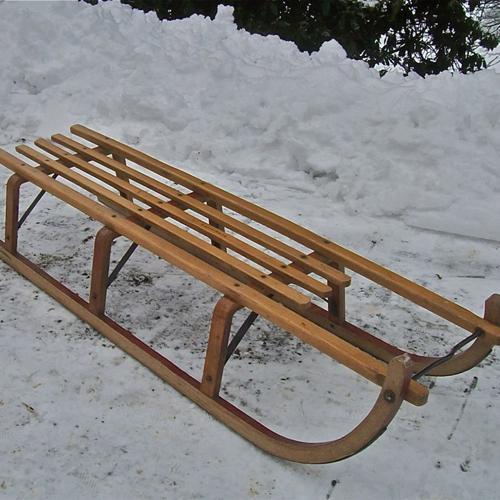} &
\begin{tabular}{ccc}
    Winter & Sled & Wooden \\
    Snow & Sleigh & \\
\end{tabular} \\
\hline
\hline
00166\_sleeping\_bag & \includegraphics[width=.2\textwidth]{ 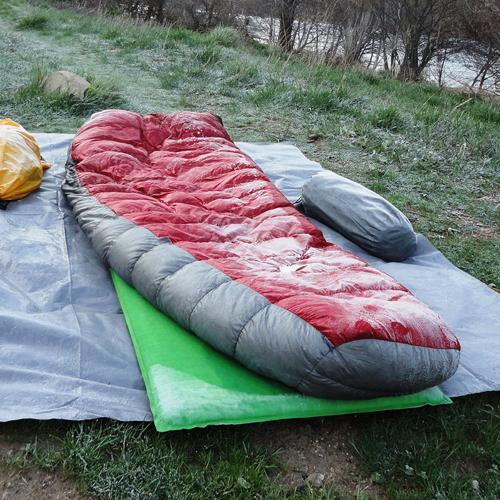} & 
\begin{tabular}{ccc}
    Camping & Sleeping & Bag \\
    Outdoor & Frost & \\
\end{tabular} \\
\hline
\hline
00167\_slide & \includegraphics[width=.2\textwidth]{ 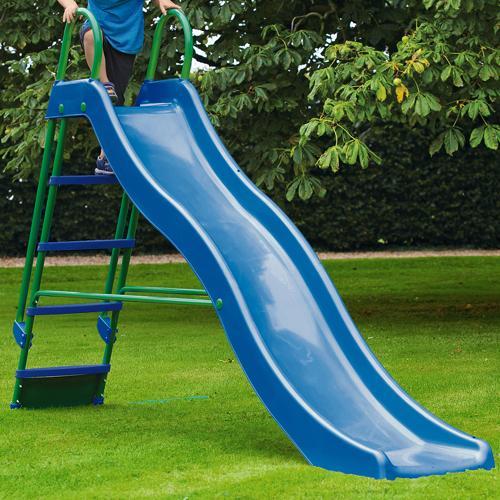} & 
\begin{tabular}{ccc}
    Playground & Slide & Blue \\
    Ladder & Outdoor & \\
\end{tabular} \\
\hline
\hline
00168\_slingshot & \includegraphics[width=.2\textwidth]{ 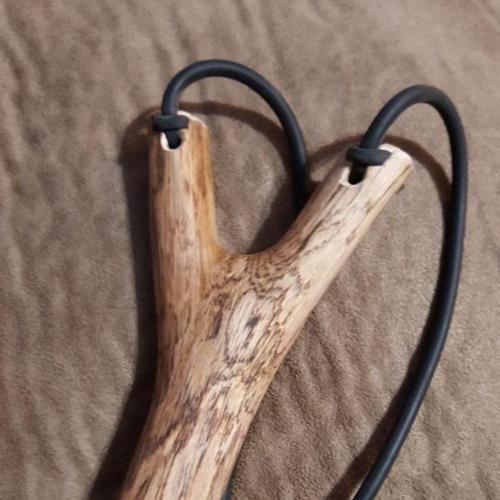} & 
\begin{tabular}{ccc}
    Tool & Slingshot & Wooden  \\
    Rubber & Y-shaped & \\
\end{tabular} \\
\hline
\hline
00169\_snowshoe & \includegraphics[width=.2\textwidth]{ 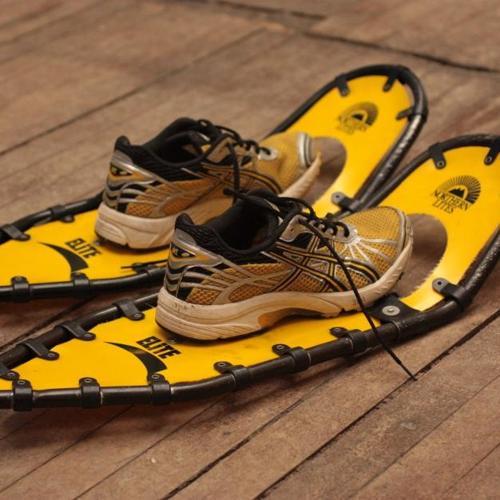} & 
\begin{tabular}{ccc}
    Footwear & Snowshoes & Yellow \\
    Running & Winter & \\
\end{tabular} \\
\hline
\hline
00170\_spatula & \includegraphics[width=.2\textwidth]{ 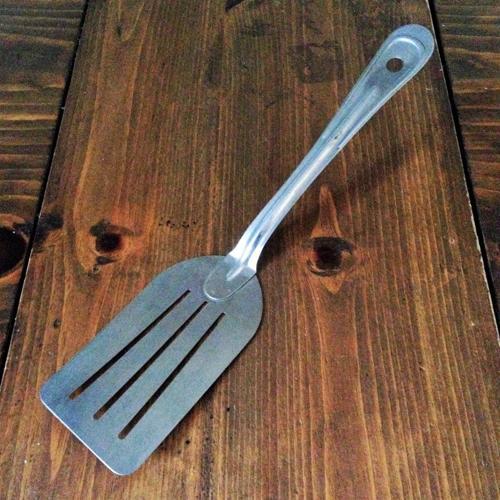} & 
\begin{tabular}{ccc}
   Utensil  & Spatula & Metal \\
   Slotted  & Handle & \\
\end{tabular} \\
\hline
\hline
00171\_spoon & \includegraphics[width=.2\textwidth]{ 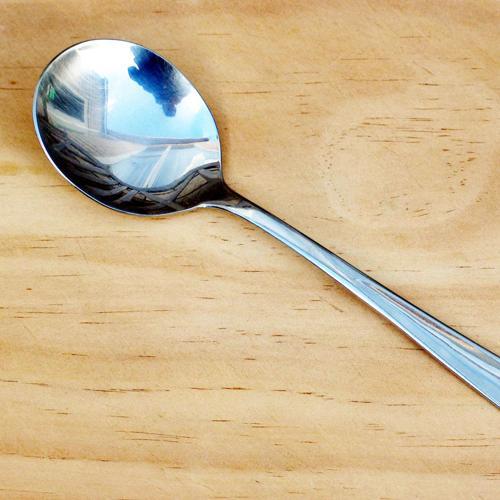} & 
\begin{tabular}{ccc}
   Utensil  & Spoon & Metal \\
    Reflection & Curved & \\
\end{tabular} \\
\hline
\hline
00172\_station\_wagon & \includegraphics[width=.2\textwidth]{ 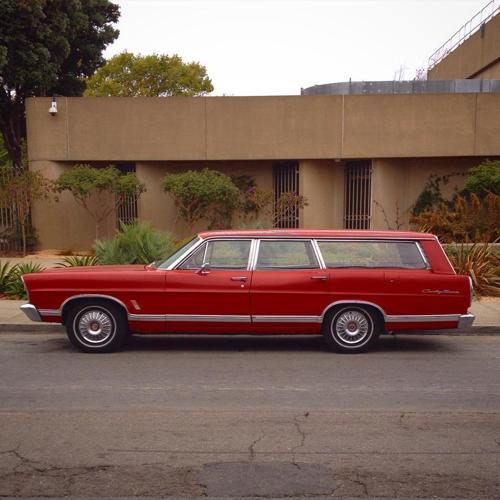} & 
\begin{tabular}{ccc}
   Vehicle  & Station & Wagon \\
    Red & Classic & \\
\end{tabular} \\
\hline
\hline
00173\_stethoscope & \includegraphics[width=.2\textwidth]{ 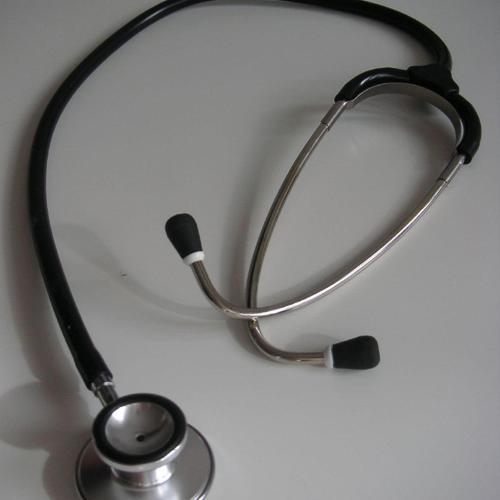} & 
\begin{tabular}{ccc}
    Medical & Stethoscope & Instrument \\
    Black & Diagnosis & \\
\end{tabular} \\
\hline
\hline
00174\_strawberry & \includegraphics[width=.2\textwidth]{ 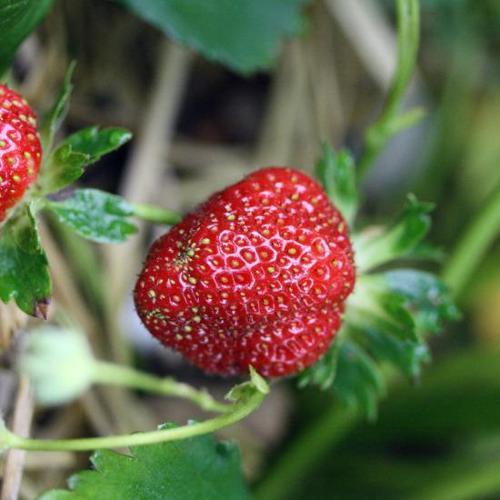} & 
\begin{tabular}{ccc}
    Fruit & Strawberry & Red \\
    Ripe & Plant & \\
\end{tabular} \\
\hline
\hline
00175\_submarine & \includegraphics[width=.2\textwidth]{ 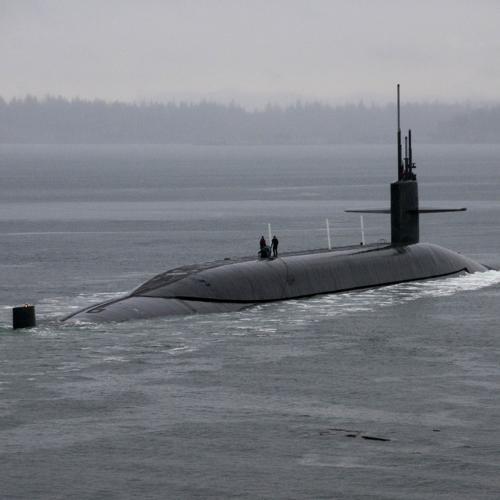} & 
\begin{tabular}{ccc}
    Vessel & Submarine & Navy \\
    Water & Stealth & \\
\end{tabular} \\
\hline
\hline
00176\_suit & \includegraphics[width=.2\textwidth]{ 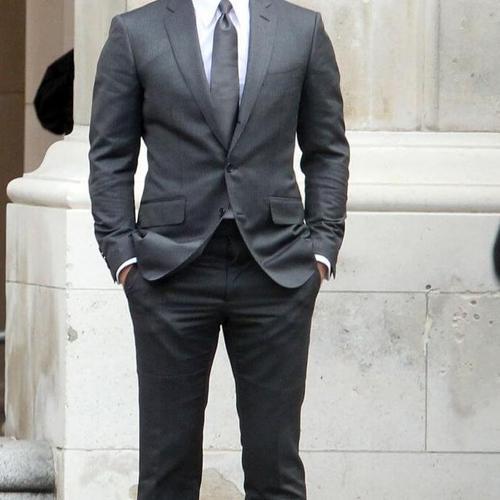} & 
\begin{tabular}{ccc}
    Clothing & Suit & Formal \\
    Business & Tailored & \\
\end{tabular} \\
\hline
\hline
00177\_t-shirt & \includegraphics[width=.2\textwidth]{ 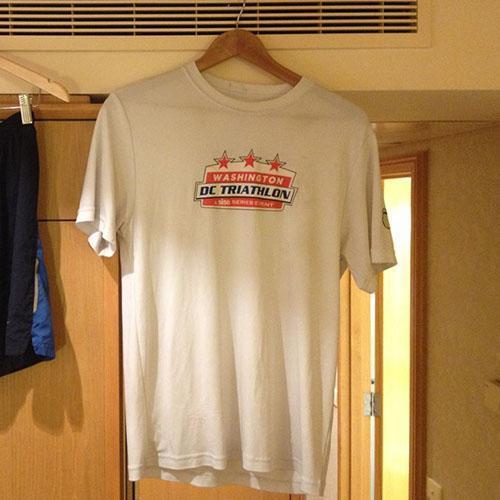} & 
\begin{tabular}{ccc}
    Clothing & T-shirt  & White \\
    Event & Hanger & \\
\end{tabular} \\
\hline
\hline
00178\_table & \includegraphics[width=.2\textwidth]{ 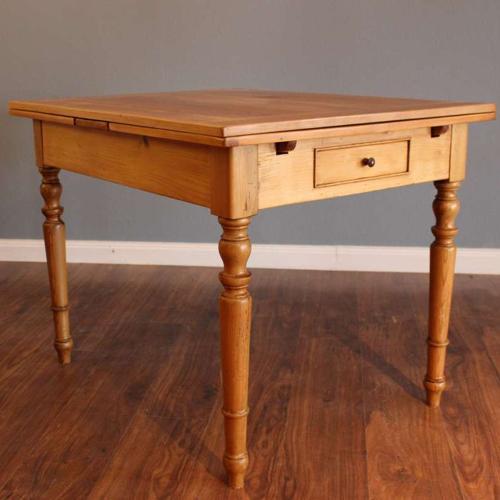} & 
\begin{tabular}{ccc}
   Furniture  & Table & Wooden \\
    Square & Drawer & \\
\end{tabular} \\
\hline
\hline
00179\_taillight & \includegraphics[width=.2\textwidth]{ 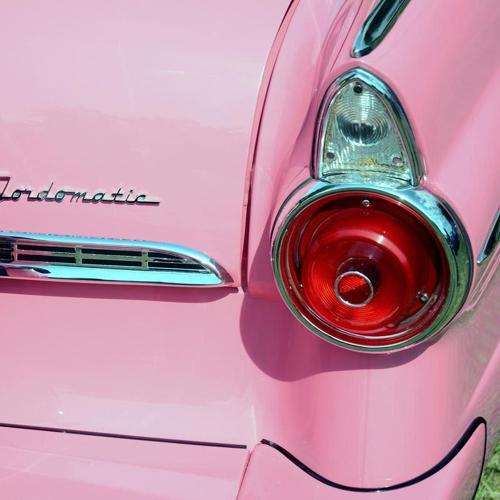} & 
\begin{tabular}{ccc}
   Vehicle  & Taillight & Pink \\
    Classic & Chrome & \\
\end{tabular} \\
\hline
\hline
00180\_tape\_recorder & \includegraphics[width=.2\textwidth]{ 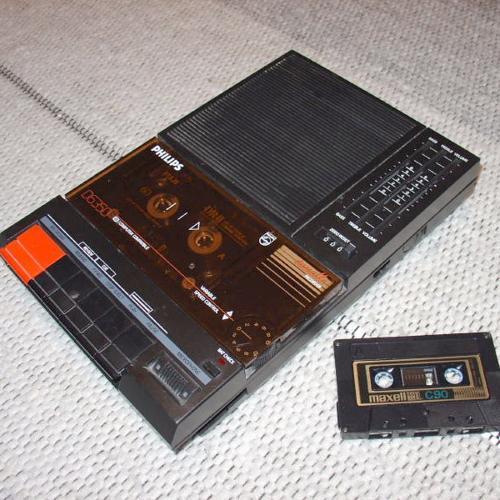} & 
\begin{tabular}{ccc}
   Device  & Recorder & Cassette \\
    Vintage & Audio & \\
\end{tabular} \\
\hline
\hline
00181\_television & \includegraphics[width=.2\textwidth]{ 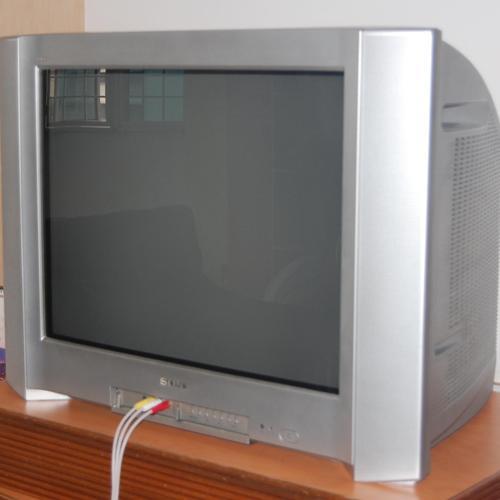} & 
\begin{tabular}{ccc}
    Electronics & Television & CRT \\
    Screen & Retro & \\
\end{tabular} \\
\hline
\hline
00182\_tiara & \includegraphics[width=.2\textwidth]{ 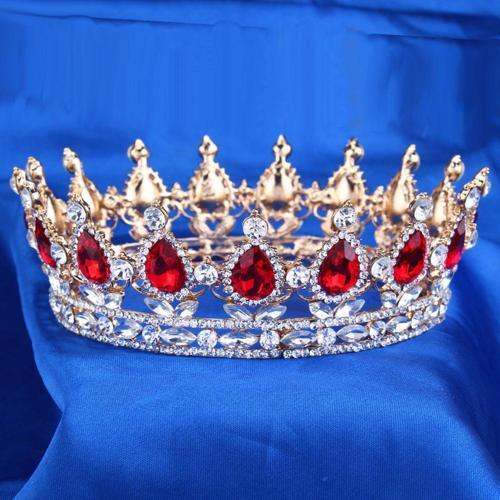} & 
\begin{tabular}{ccc}
    Crown & Tiara & Gold \\
     Jewels & Red & \\
\end{tabular} \\
\hline
\hline
00183\_tick & \includegraphics[width=.2\textwidth]{ 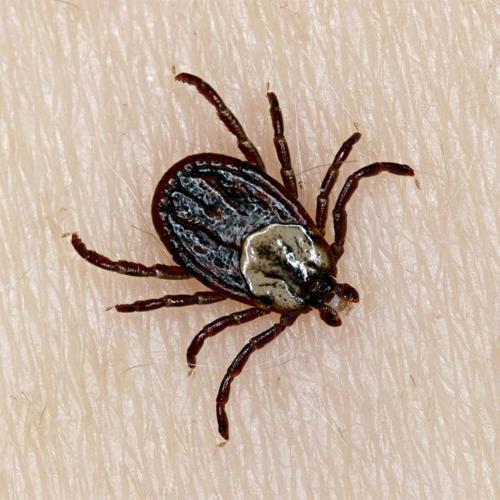} & 
\begin{tabular}{ccc}
   Insect  & Tick & Parasite \\
   Skin  & Tiny & \\
\end{tabular} \\
\hline
\hline
00184\_tomato\_sauce & \includegraphics[width=.2\textwidth]{ 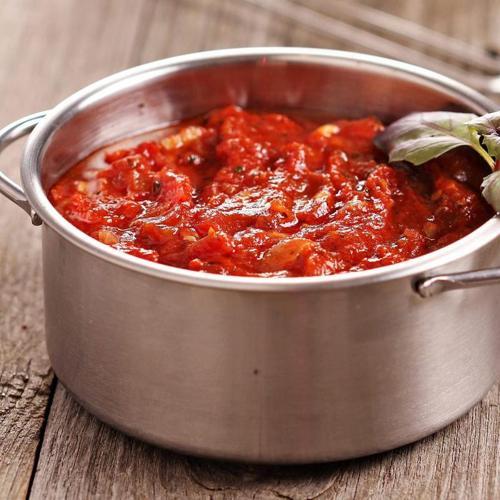} & 
\begin{tabular}{ccc}
   Food & Sauce & Tomato \\
   Pot  & Red & \\
\end{tabular} \\
\hline
\hline
00185\_tongs & \includegraphics[width=.2\textwidth]{ 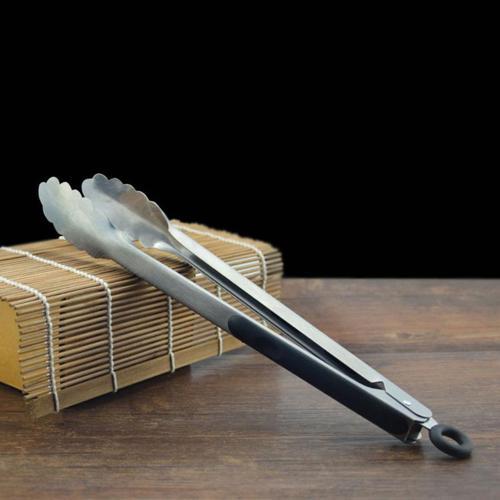} & 
\begin{tabular}{ccc}
    Utensil & Tongs & Metal \\
    Grip & Kitchen & \\
\end{tabular} \\
\hline
\hline
00186\_tool & \includegraphics[width=.2\textwidth]{ 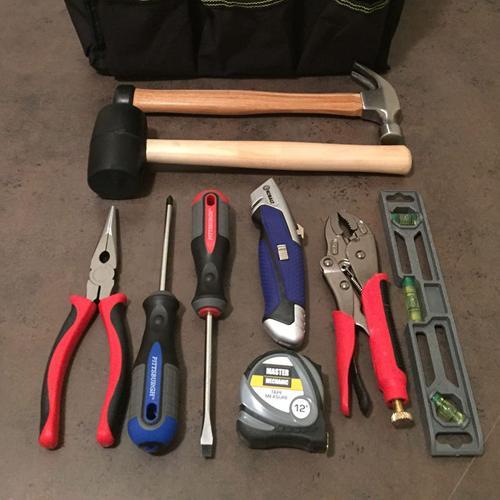} & 
\begin{tabular}{ccc}
    Tools & Hammer & Pliers \\
    Screwdriver & Utility & \\
\end{tabular} \\
\hline
\hline
00187\_top\_hat & \includegraphics[width=.2\textwidth]{ 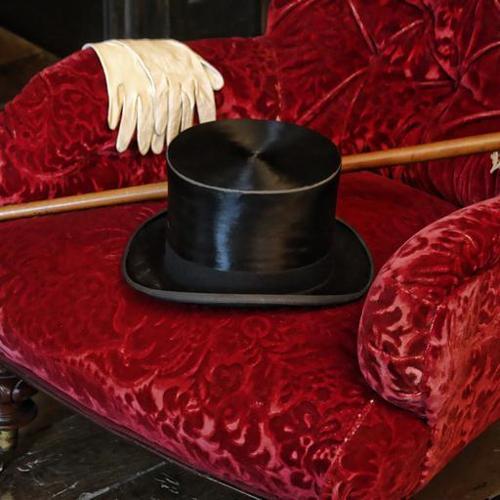} & 
\begin{tabular}{ccc}
    Accessory & Top-hat &  Cane \\
    Gloves & Velvet & \\
\end{tabular} \\
\hline
\hline
00188\_treadmill & \includegraphics[width=.2\textwidth]{ 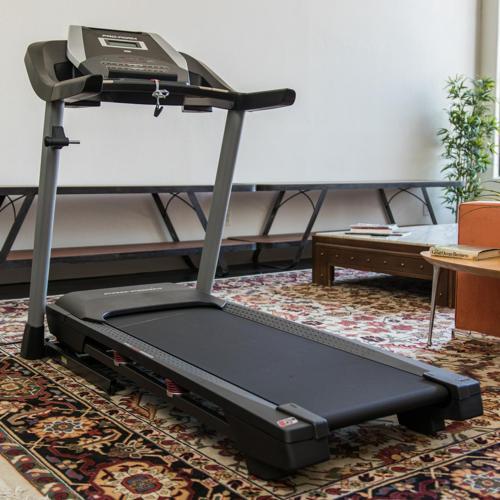} & 
\begin{tabular}{ccc}
   Exercise  & Treadmill & Machine \\
   Indoor  &  Fitness & \\
\end{tabular} \\
\hline
\hline
00189\_tube\_top & \includegraphics[width=.2\textwidth]{ 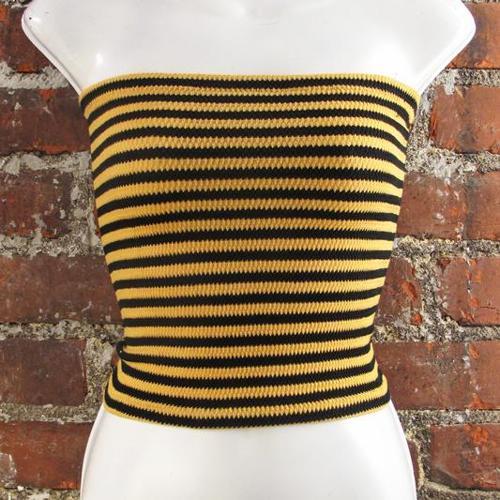} & 
\begin{tabular}{ccc}
   Clothing  & Top & Striped \\
   Yellow  & Knitted & \\
\end{tabular} \\
\hline
\hline
00190\_turkey & \includegraphics[width=.2\textwidth]{ 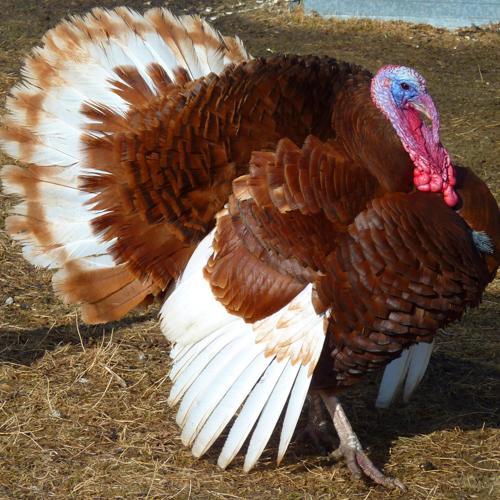} & 
\begin{tabular}{ccc}
   Bird  & Turkey & Feathers \\
   Fanned  & Brown & \\
\end{tabular} \\
\hline
\hline
00191\_unicycle & \includegraphics[width=.2\textwidth]{ 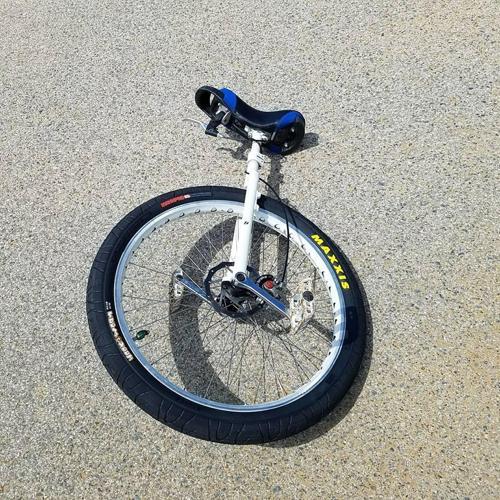} & 
\begin{tabular}{ccc}
   Vehicle  & Unicycle & Wheel \\
   Tire  & Seat & \\
\end{tabular} \\
\hline
\hline
00192\_vise & \includegraphics[width=.2\textwidth]{ 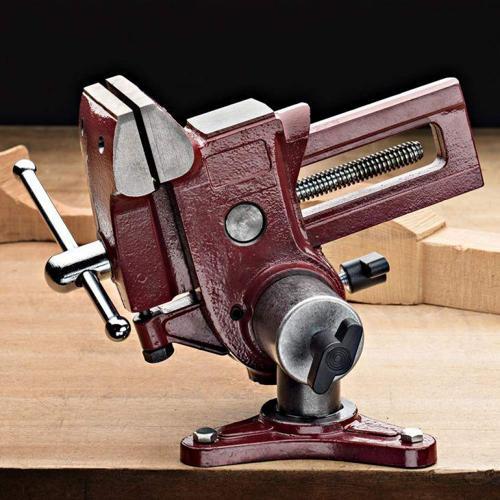} & 
\begin{tabular}{ccc}
    Tool & Vise  &  Metal\\
    Clamp & Adjustable & \\
\end{tabular} \\
\hline
\hline
00193\_volleyball & \includegraphics[width=.2\textwidth]{ 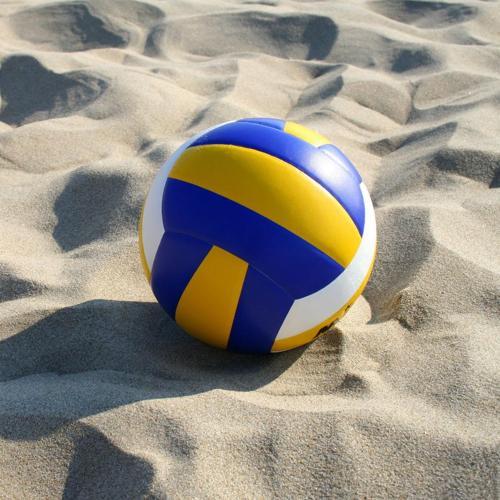} & 
\begin{tabular}{ccc}
   Sport  & Volleyball & Beach \\
   Ball  & Sand & \\
\end{tabular} \\
\hline
\hline
00194\_wallpaper & \includegraphics[width=.2\textwidth]{ 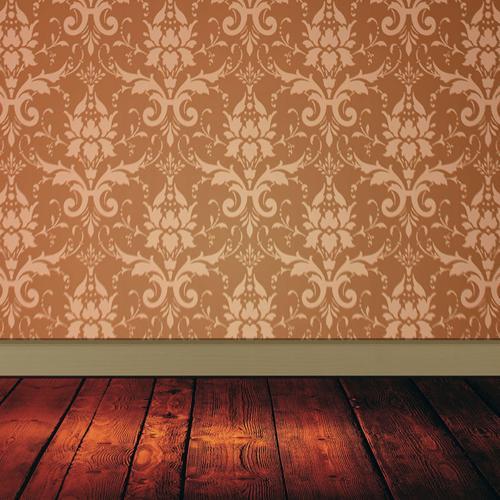} & 
\begin{tabular}{ccc}
    Interior & Wallpaper & Pattern \\
     Vintage & Wood & \\
\end{tabular} \\
\hline
\hline
00195\_walnut & \includegraphics[width=.2\textwidth]{ 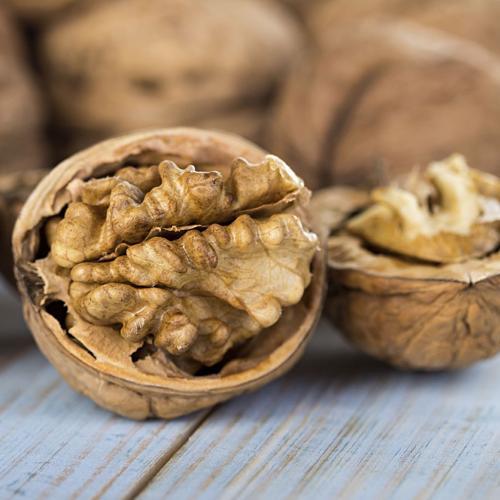} & 
\begin{tabular}{ccc}
   Food  & Walnut & Nut \\
    Shell & Brown & \\
\end{tabular} \\
\hline
\hline
00196\_wheat & \includegraphics[width=.2\textwidth]{ 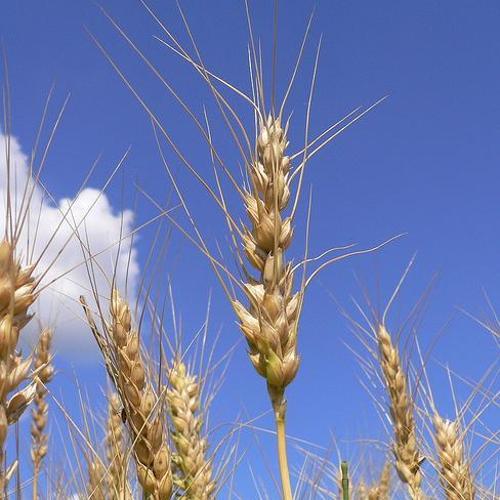} & 
\begin{tabular}{ccc}
   Crop  & Wheat & Grain \\
    Field & Stalk & \\
\end{tabular} \\
\hline
\hline
00197\_wheelchair & \includegraphics[width=.2\textwidth]{ 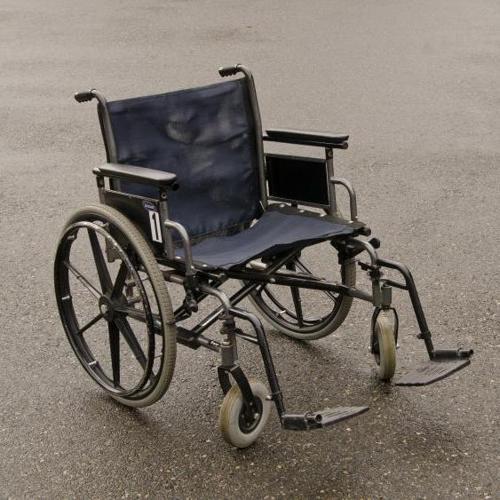} & 
\begin{tabular}{ccc}
   Mobility  & Wheelchair & Manual \\
    Wheels & Seat & \\
\end{tabular} \\
\hline
\hline
00198\_windshield & \includegraphics[width=.2\textwidth]{ 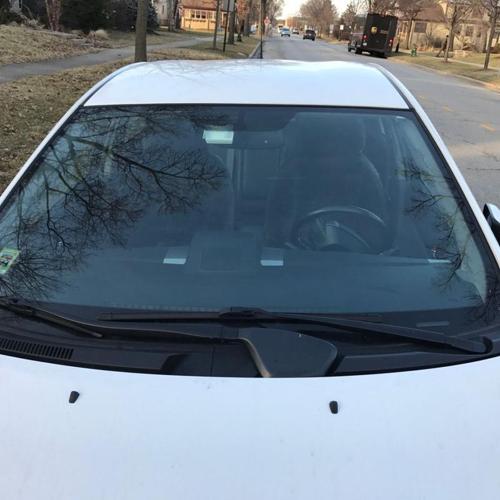} & 
\begin{tabular}{ccc}
   Vehicle  & Windshield & Glass \\
   Car  &  Street & \\
\end{tabular} \\
\hline
\hline
00199\_wine & \includegraphics[width=.2\textwidth]{ 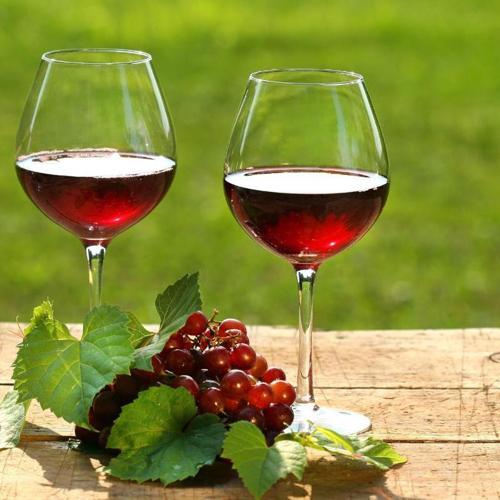} & 
\begin{tabular}{ccc}
   Beverage  & Wine & Glass \\
    Grapes & Red & \\
\end{tabular} \\
\hline
\hline
00200\_wok & \includegraphics[width=.2\textwidth]{ 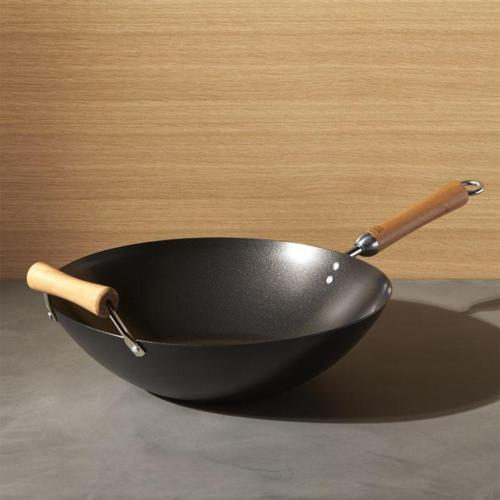} & 
\begin{tabular}{ccc}
   Cookware  & Wok & Pan \\
   Handles  & Black & \\
\end{tabular} \\
\hline

\end{longtable}

\section{The image generation results of NECOMIMI}
In this section, we will present all the images generated by various EEG encoders within the NECOMIMI framework using a fixed random seed. These images are generated using the testing set of the ThingsEEG dataset in a zero-shot setting, meaning that the model has not seen these categories during the EEG-Image contrastive learning training process. All the images illustrate the progression of visual representations generated using different embedding techniques in a diffusion model: (a) Top row: The original images shown to subjects (ground truth). (b) Second row: Images generated by the CLIP-ViT embeddings of the original images. It is only related to the seed and has nothing to do with the subject and EEG encoder. (c) Third row: Images generated by one-stage method using pure EEG embeddings with the EEG encoder. (d) Fourth row: Images generated by two-stage NECOMIMI method using pure EEG embeddings with EEG encoder.

\begin{figure}
    \centering
    \includegraphics[width=1\linewidth]{ 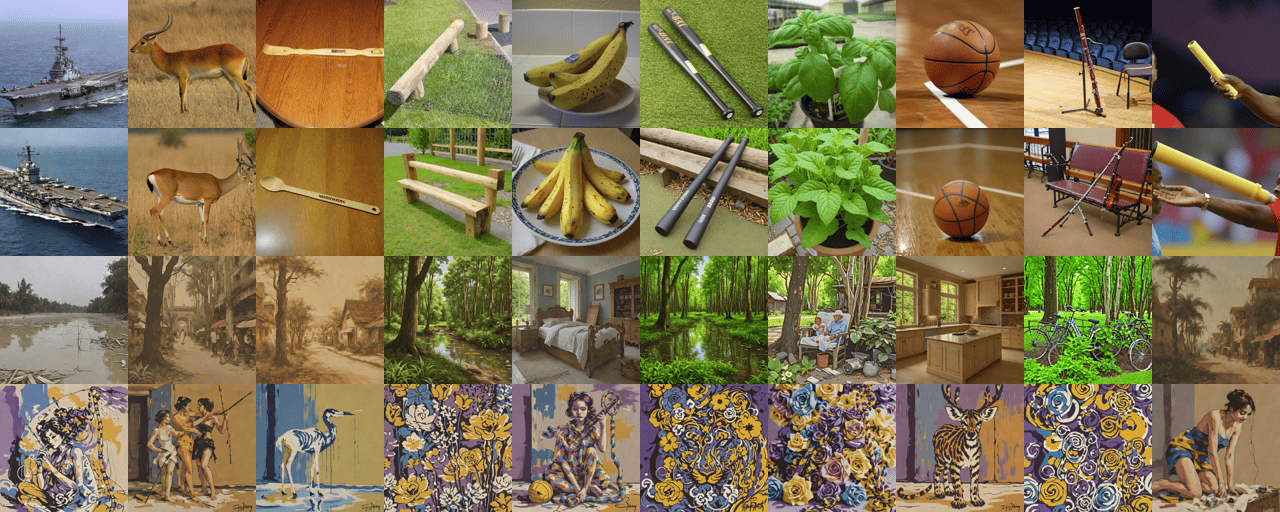}
    \caption{Random selected generated images in Subject 6 with NICE EEG encoder.}
    \label{fig:enter-label}
\end{figure}
\begin{figure}
    \centering
    \includegraphics[width=1\linewidth]{ 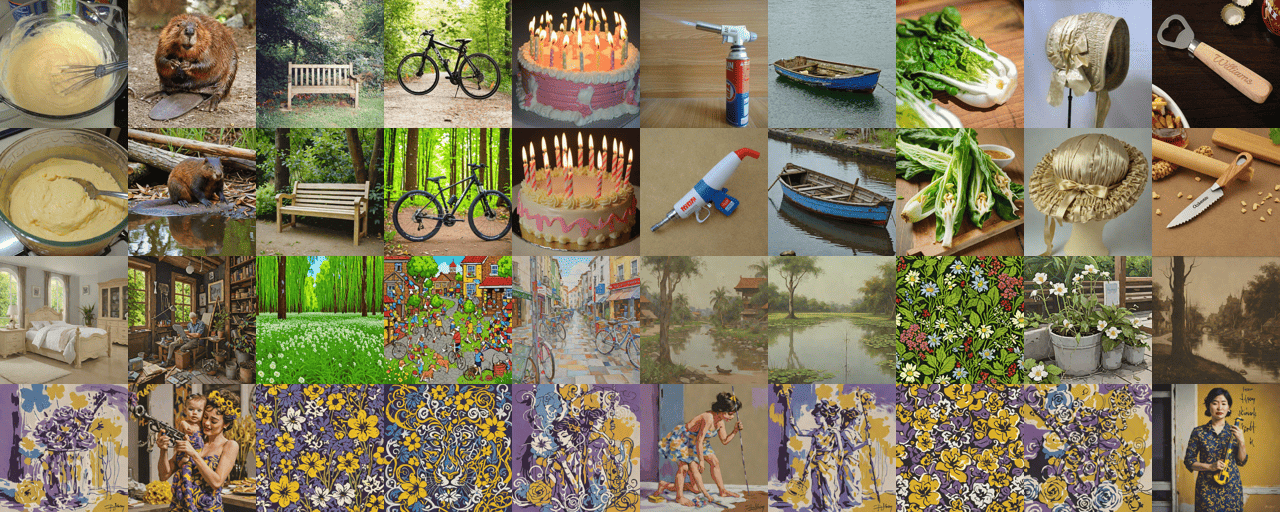}
    \caption{Random selected generated images in Subject 6 with NICE EEG encoder.}
    \label{fig:enter-label}
\end{figure}
\begin{figure}
    \centering
    \includegraphics[width=1\linewidth]{ 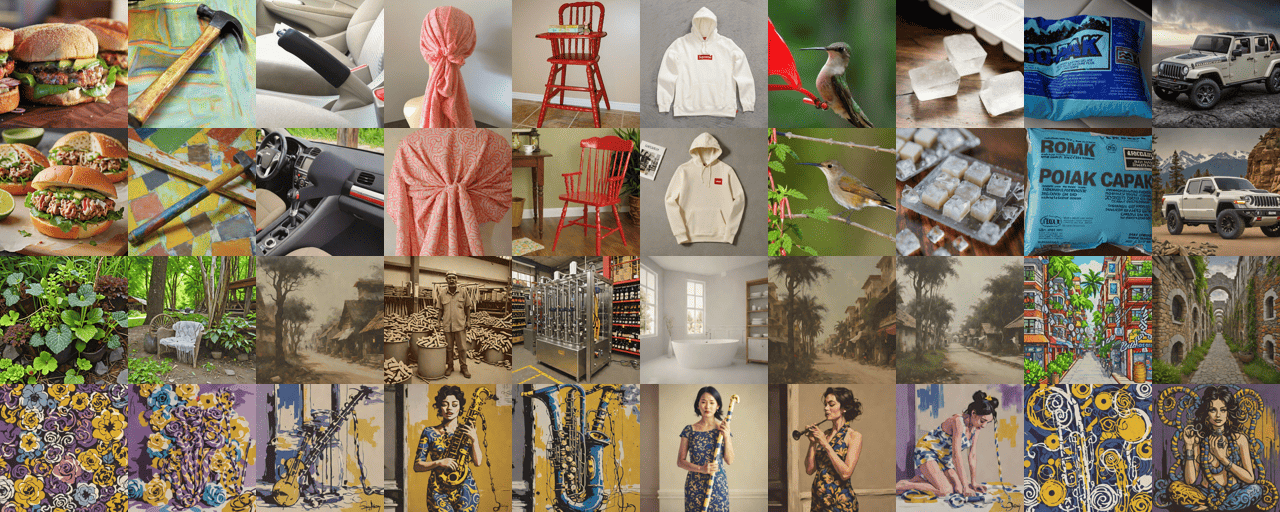}
    \caption{Random selected generated images in Subject 6 with NICE EEG encoder.}
    \label{fig:enter-label}
\end{figure}

\begin{figure}
    \centering
    \includegraphics[width=1\linewidth]{ 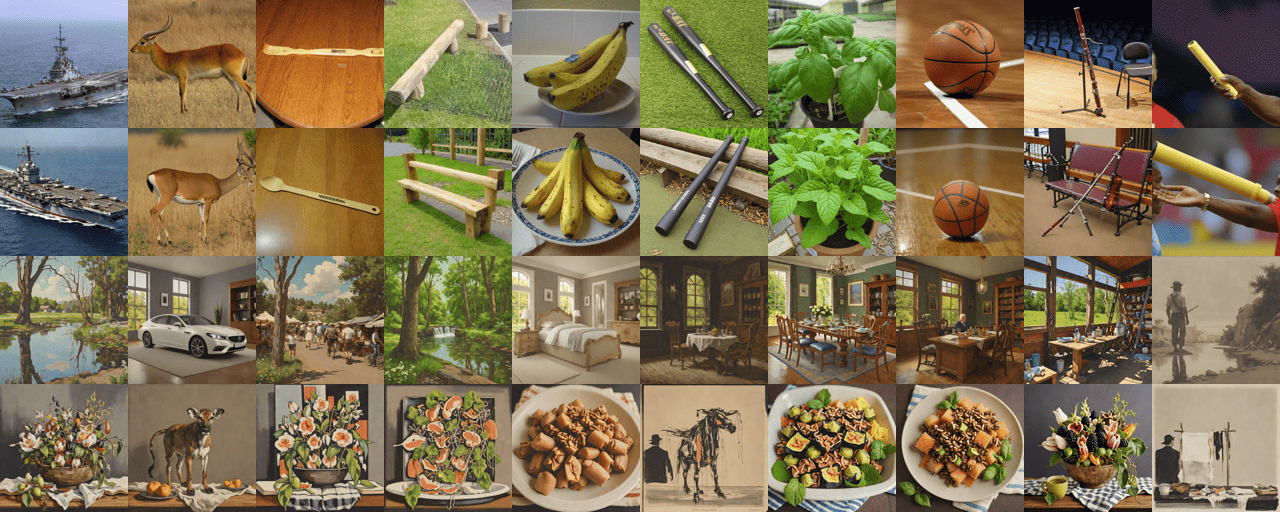}
    \caption{Random selected generated images in Subject 7 with NICE EEG encoder.}
    \label{fig:enter-label}
\end{figure}
\begin{figure}
    \centering
    \includegraphics[width=1\linewidth]{ 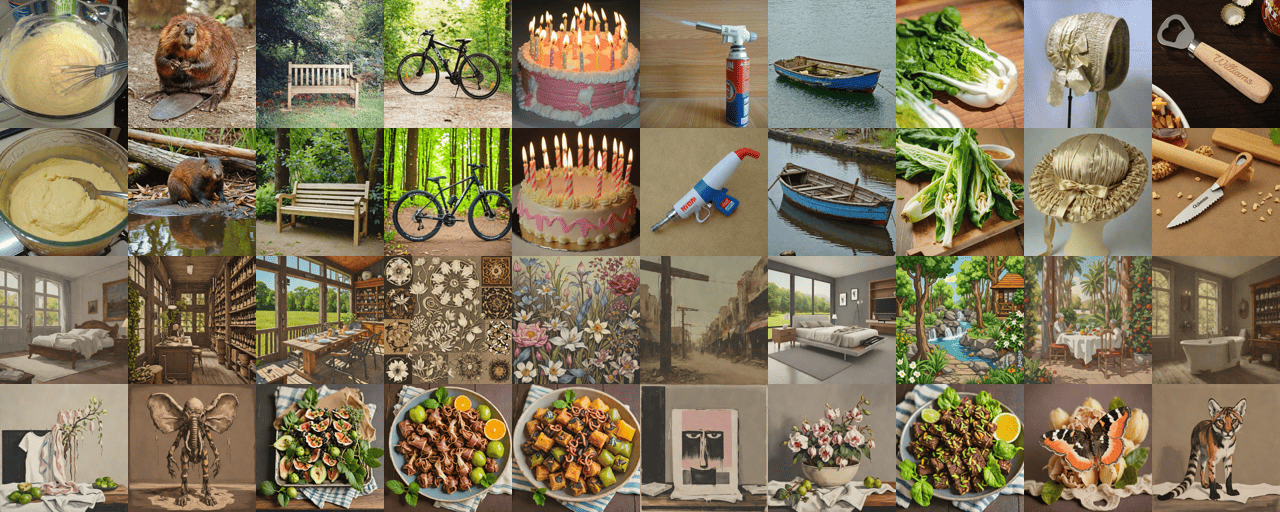}
    \caption{Random selected generated images in Subject 7 with NICE EEG encoder.}
    \label{fig:enter-label}
\end{figure}
\begin{figure}
    \centering
    \includegraphics[width=1\linewidth]{ 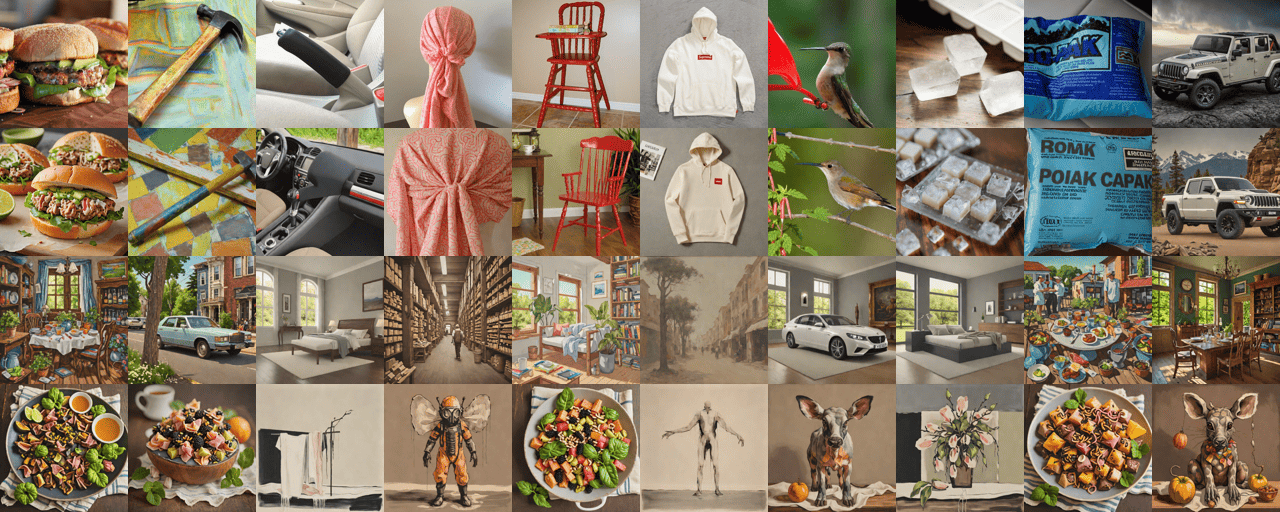}
    \caption{Random selected generated images in Subject 7 with NICE EEG encoder.}
    \label{fig:enter-label}
\end{figure}

\begin{figure}
    \centering
    \includegraphics[width=1\linewidth]{ 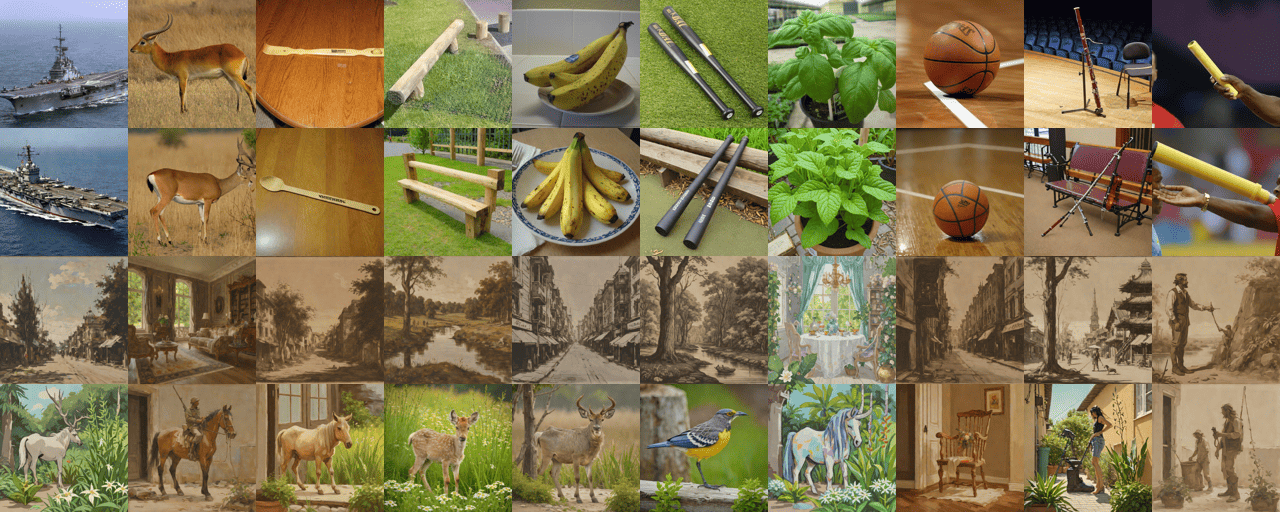}
    \caption{Random selected generated images in Subject 8 with NICE EEG encoder.}
    \label{fig:enter-label}
\end{figure}
\begin{figure}
    \centering
    \includegraphics[width=1\linewidth]{ 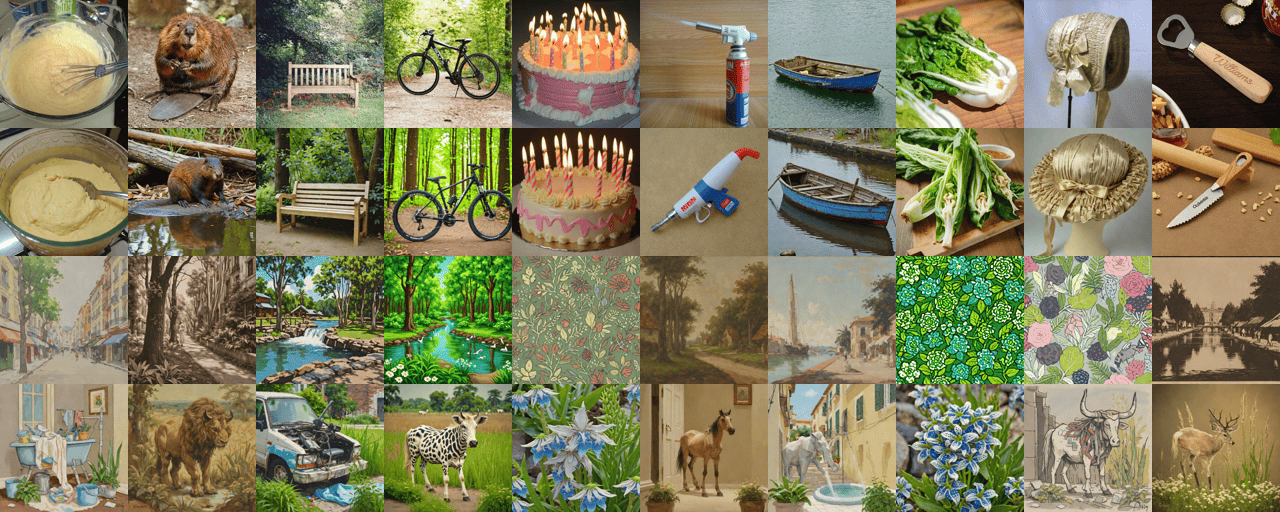}
    \caption{Random selected generated images in Subject 8 with NICE EEG encoder.}
    \label{fig:enter-label}
\end{figure}
\begin{figure}
    \centering
    \includegraphics[width=1\linewidth]{ 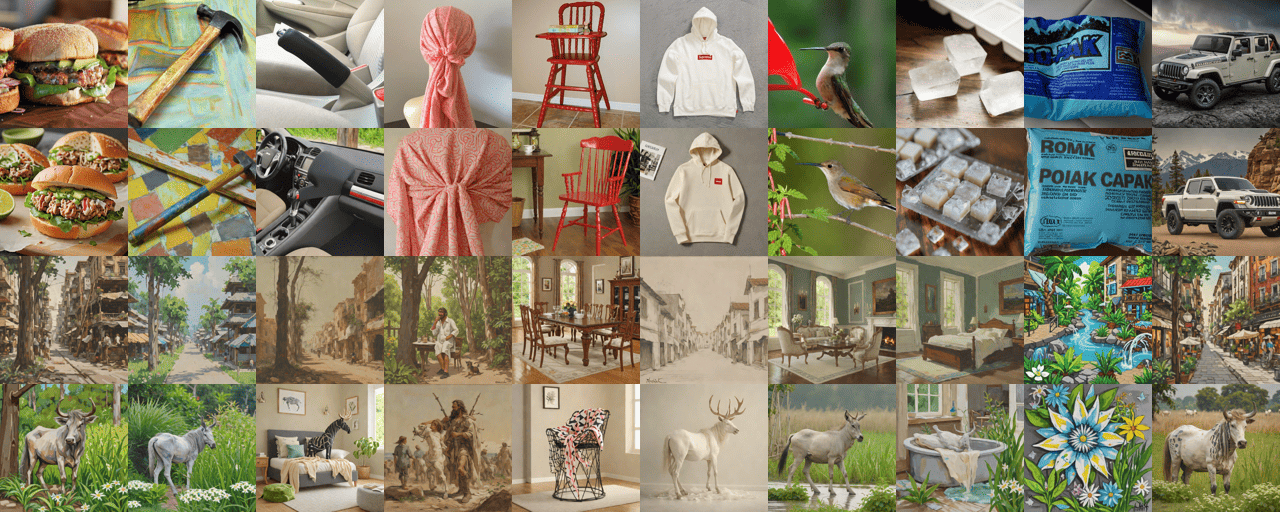}
    \caption{Random selected generated images in Subject 8 with NICE EEG encoder.}
    \label{fig:enter-label}
\end{figure}

\begin{figure}
    \centering
    \includegraphics[width=1\linewidth]{ 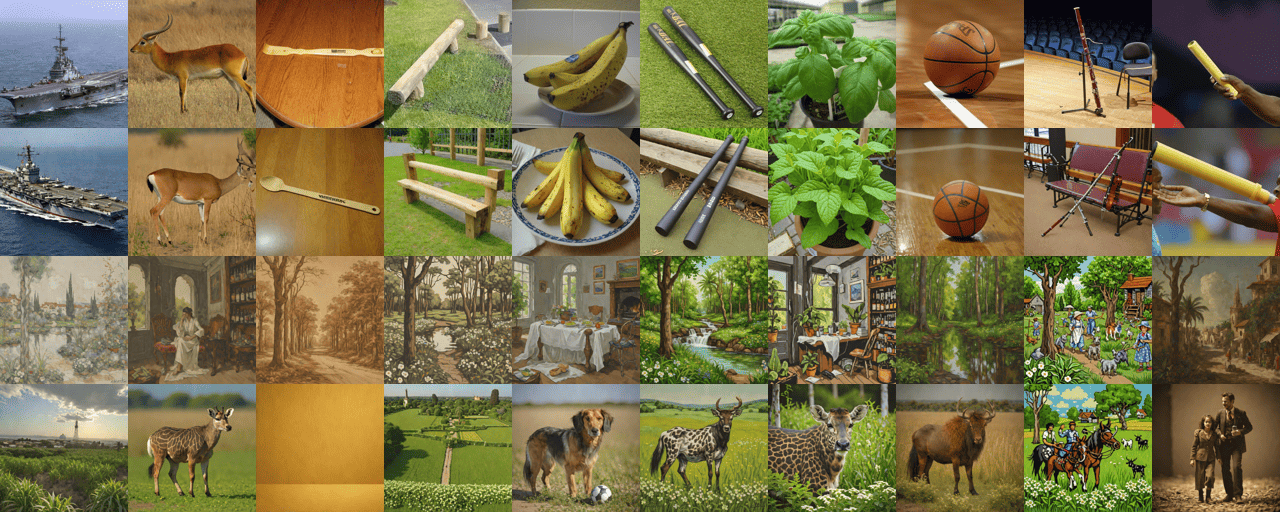}
    \caption{Random selected generated images in Subject 6 with Nervformer EEG encoder.}
    \label{fig:enter-label}
\end{figure}
\begin{figure}
    \centering
    \includegraphics[width=1\linewidth]{ 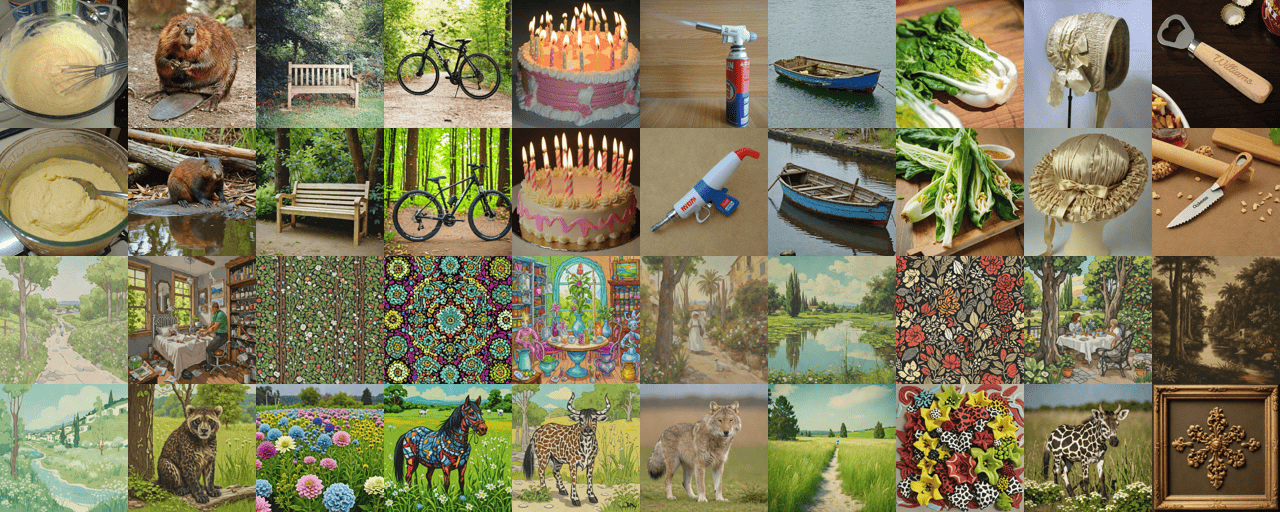}
    \caption{Random selected generated images in Subject 6 with Nervformer EEG encoder.}
    \label{fig:enter-label}
\end{figure}
\begin{figure}
    \centering
    \includegraphics[width=1\linewidth]{ 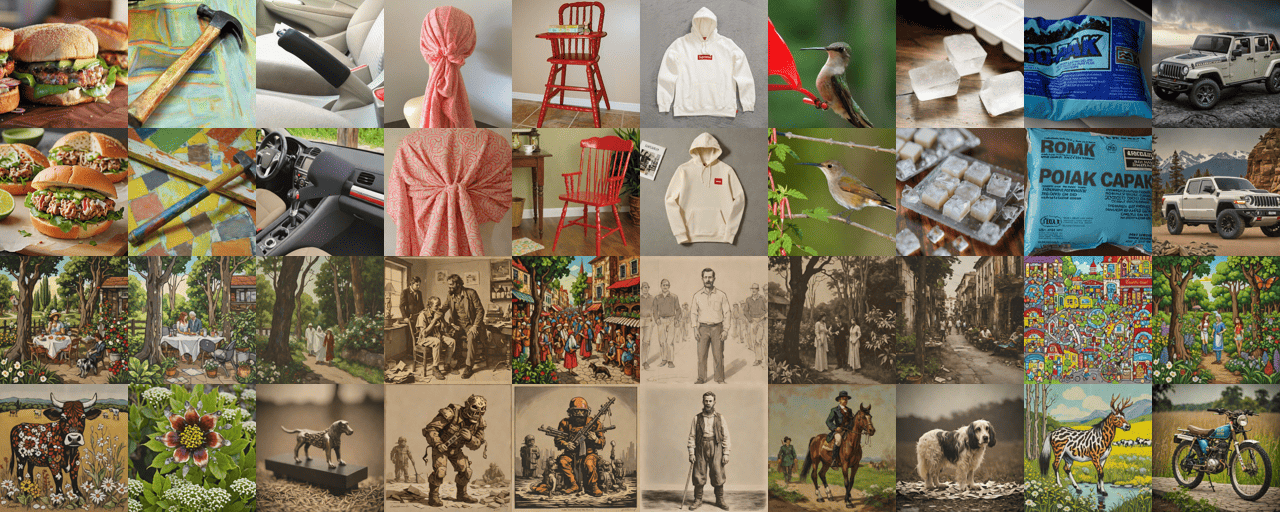}
    \caption{Random selected generated images in Subject 6 with Nervformer EEG encoder.}
    \label{fig:enter-label}
\end{figure}

\begin{figure}
    \centering
    \includegraphics[width=1\linewidth]{ 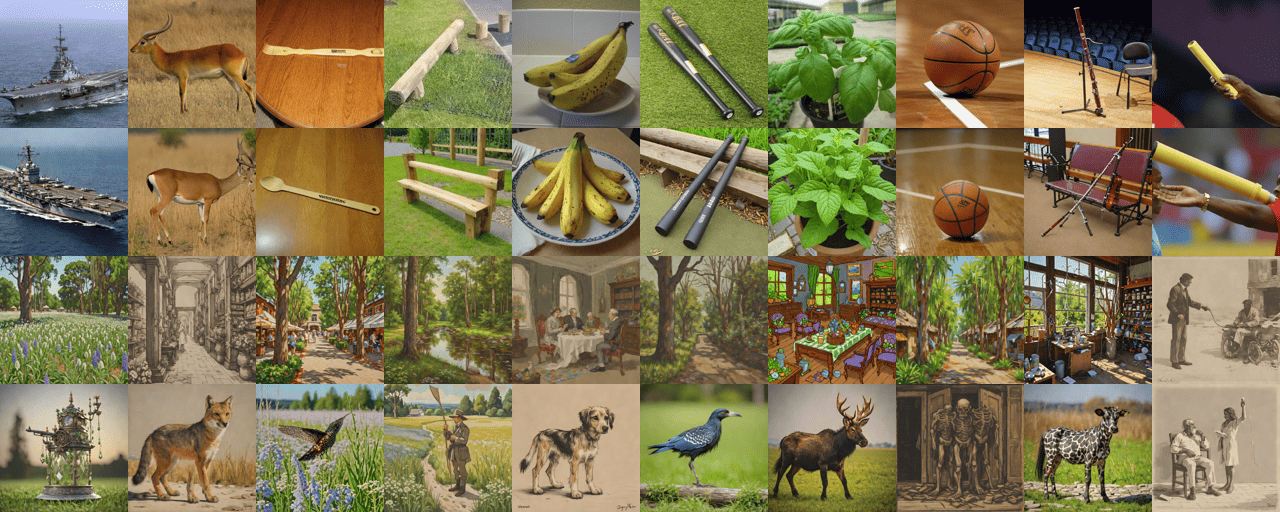}
    \caption{Random selected generated images in Subject 7 with Nervformer EEG encoder.}
    \label{fig:enter-label}
\end{figure}
\begin{figure}
    \centering
    \includegraphics[width=1\linewidth]{ 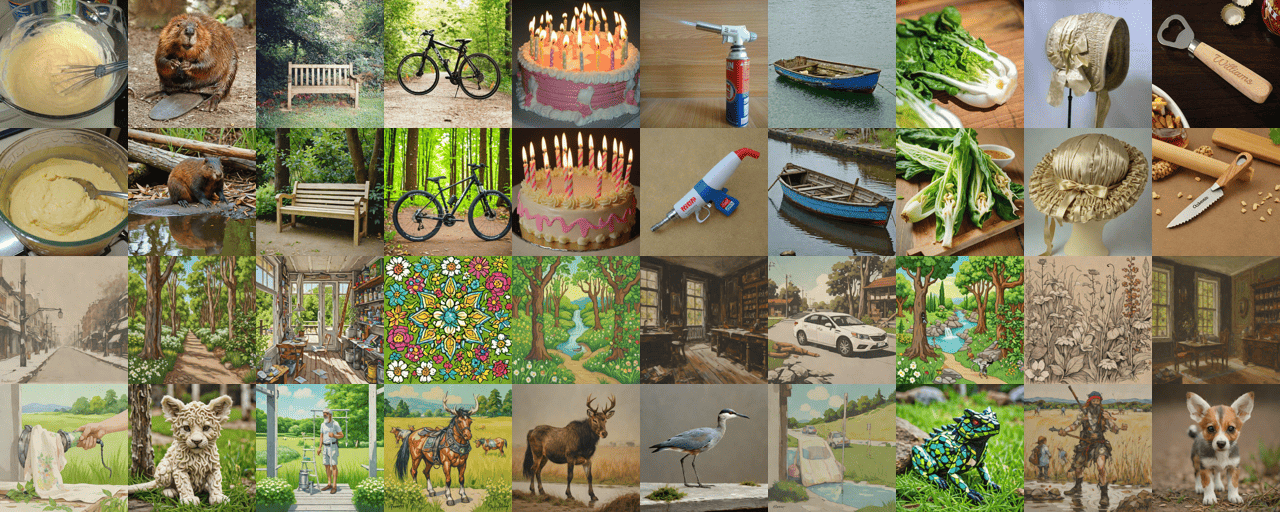}
    \caption{Random selected generated images in Subject 7 with Nervformer EEG encoder.}
    \label{fig:enter-label}
\end{figure}
\begin{figure}
    \centering
    \includegraphics[width=1\linewidth]{ 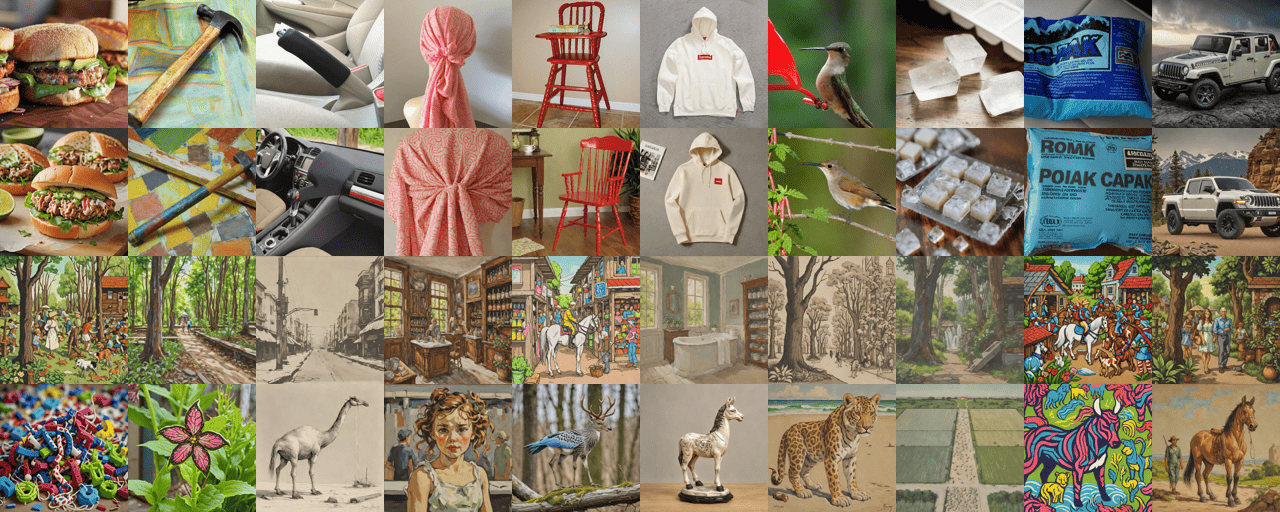}
    \caption{Random selected generated images in Subject 7 with Nervformer EEG encoder.}
    \label{fig:enter-label}
\end{figure}

\begin{figure}
    \centering
    \includegraphics[width=1\linewidth]{ 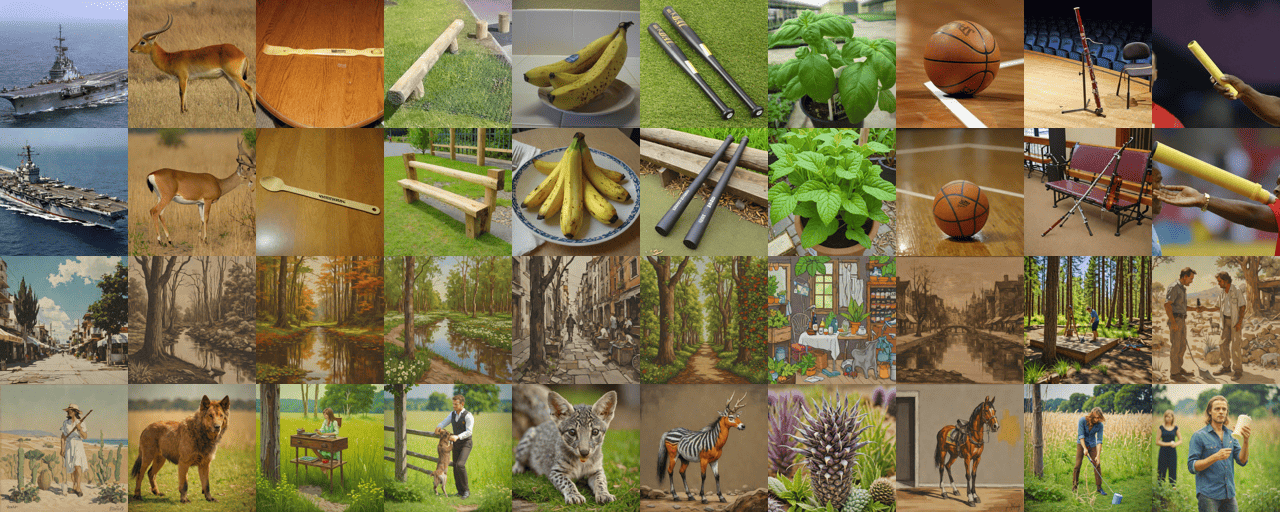}
    \caption{Random selected generated images in Subject 8 with Nervformer EEG encoder.}
    \label{fig:enter-label}
\end{figure}

\begin{figure}
    \centering
    \includegraphics[width=1\linewidth]{ 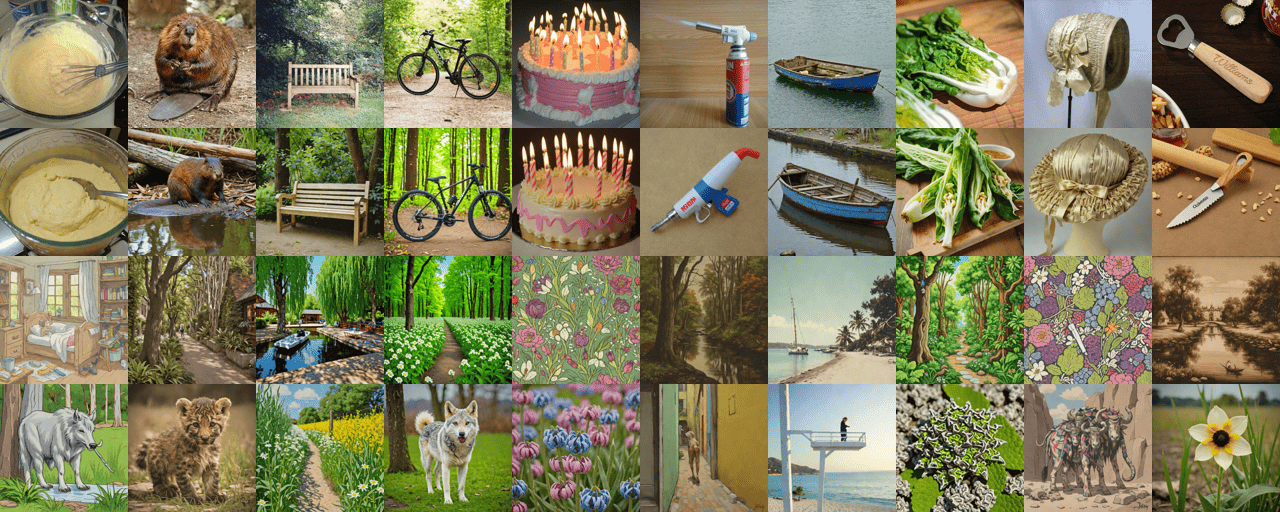}
    \caption{Random selected generated images in Subject 8 with Nervformer EEG encoder.}
    \label{fig:enter-label}
\end{figure}
\begin{figure}
    \centering
    \includegraphics[width=1\linewidth]{ 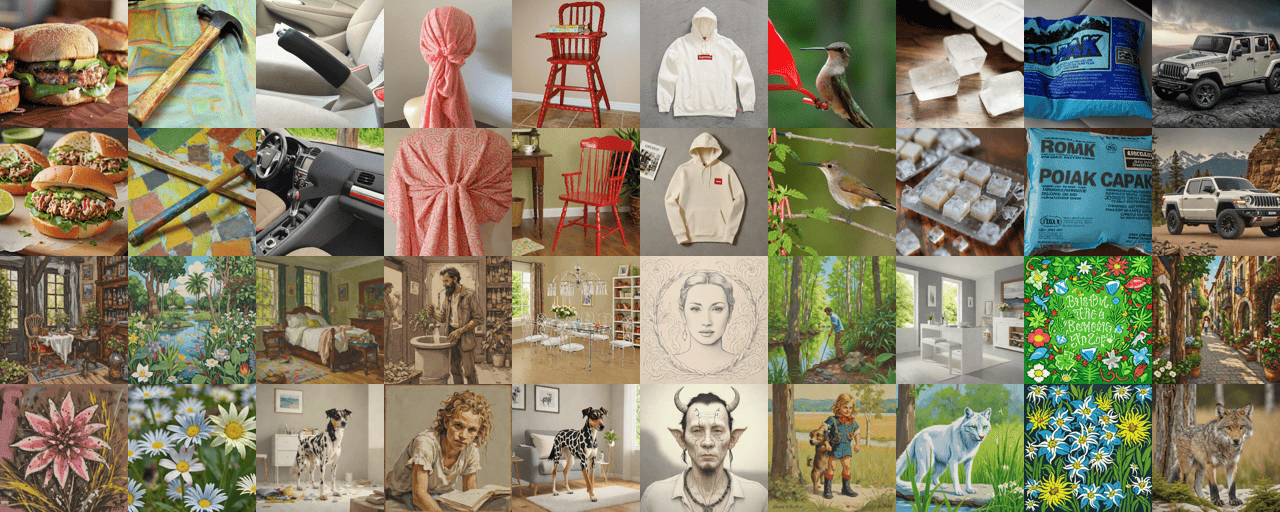}
    \caption{Random selected generated images in Subject 8 with Nervformer EEG encoder.}
    \label{fig:enter-label}
\end{figure}


\begin{figure}
    \centering
    \includegraphics[width=1\linewidth]{ 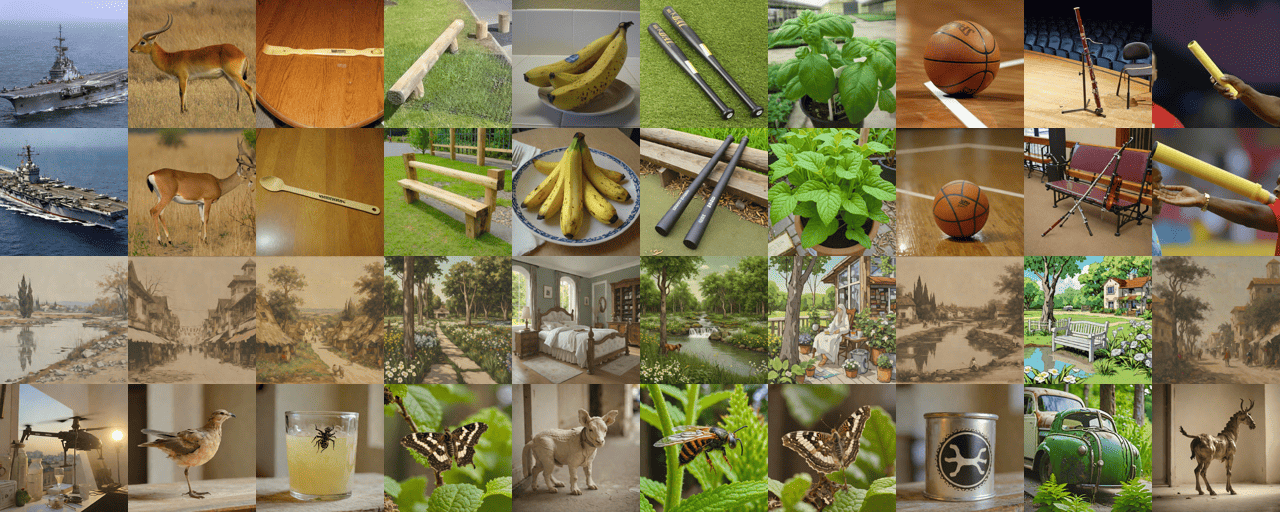}
    \caption{Random selected generated images in Subject 6 with MUSE EEG encoder.}
    \label{fig:enter-label}
\end{figure}
\begin{figure}
    \centering
    \includegraphics[width=1\linewidth]{ 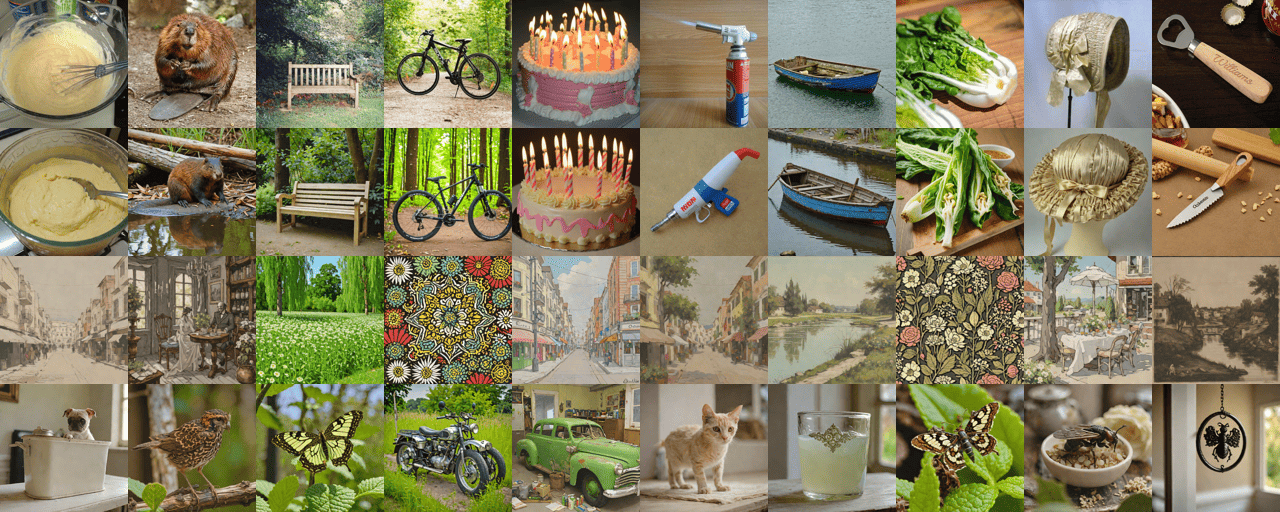}
    \caption{Random selected generated images in Subject 6 with MUSE EEG encoder.}
    \label{fig:enter-label}
\end{figure}
\begin{figure}
    \centering
    \includegraphics[width=1\linewidth]{ 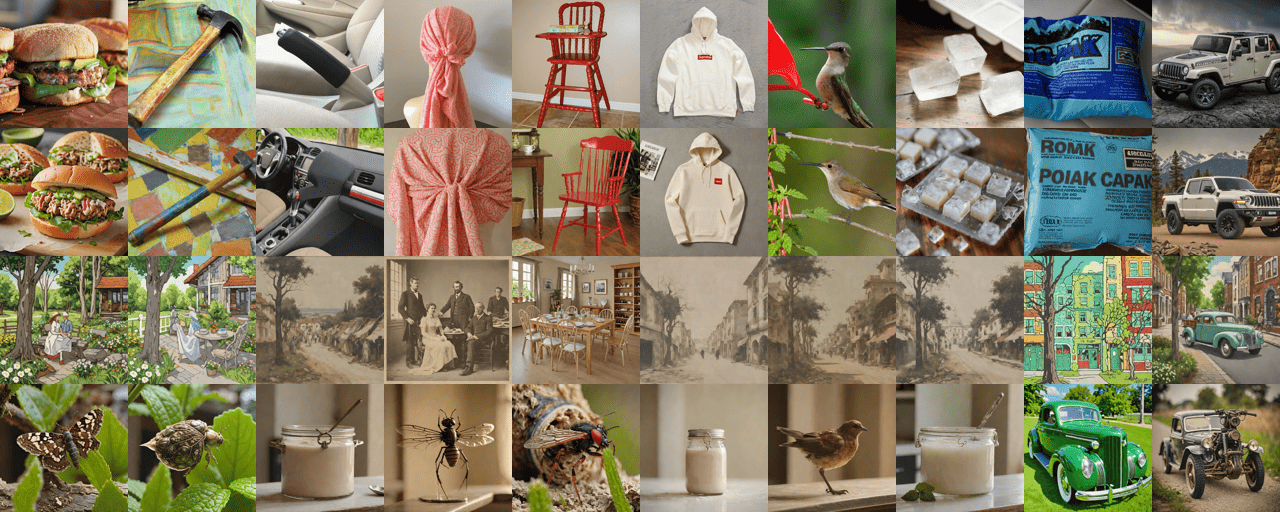}
    \caption{Random selected generated images in Subject 6 with MUSE EEG encoder.}
    \label{fig:enter-label}
\end{figure}

\begin{figure}
    \centering
    \includegraphics[width=1\linewidth]{ 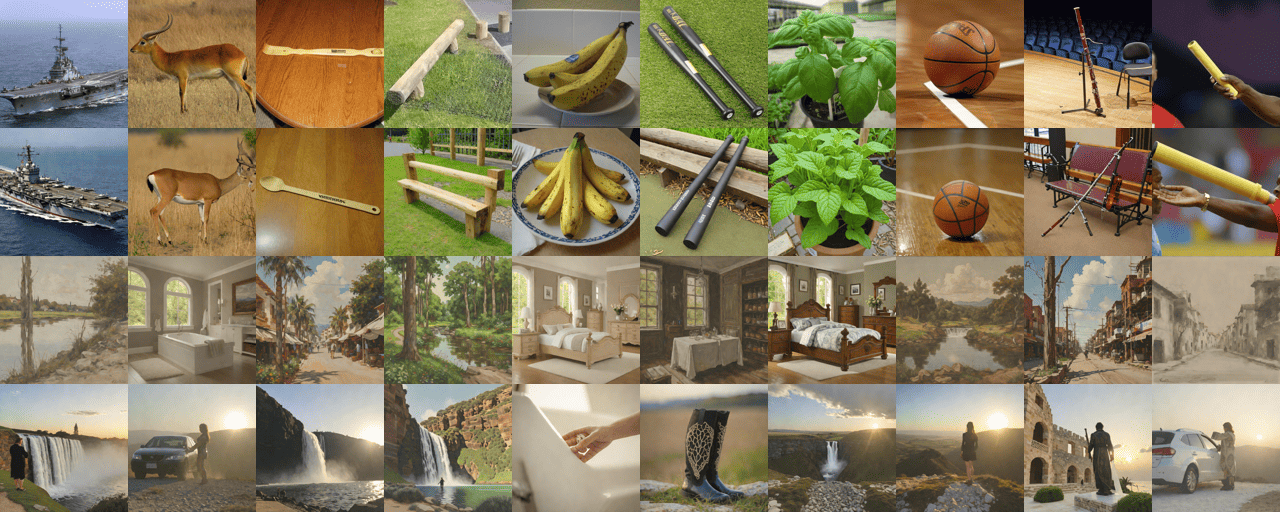}
    \caption{Random selected generated images in Subject 7 with MUSE EEG encoder.}
    \label{fig:enter-label}
\end{figure}
\begin{figure}
    \centering
    \includegraphics[width=1\linewidth]{ 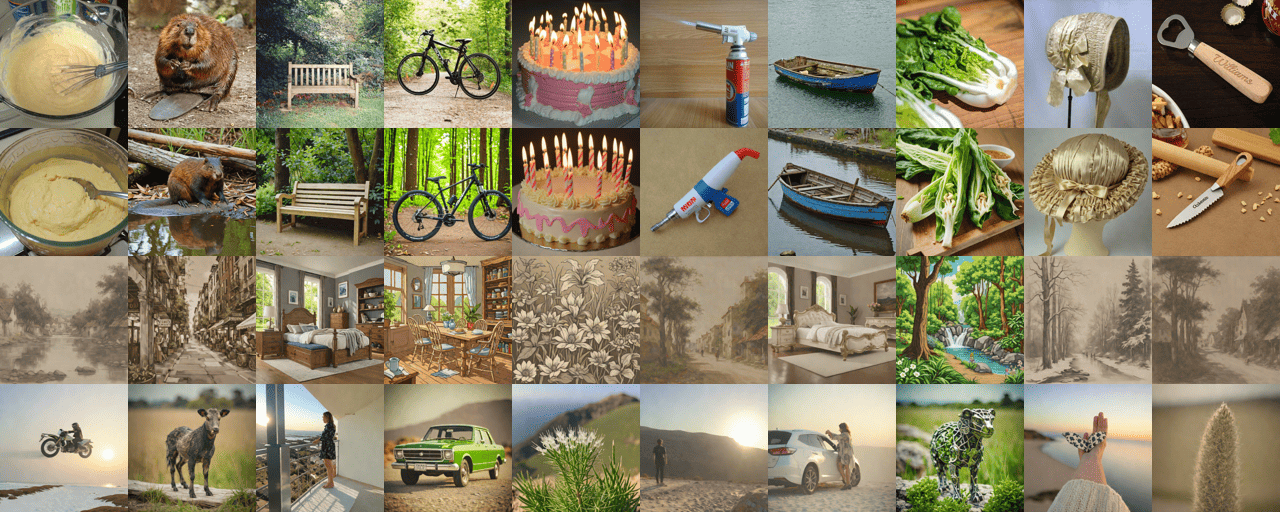}
    \caption{Random selected generated images in Subject 7 with MUSE EEG encoder.}
    \label{fig:enter-label}
\end{figure}
\begin{figure}
    \centering
    \includegraphics[width=1\linewidth]{ 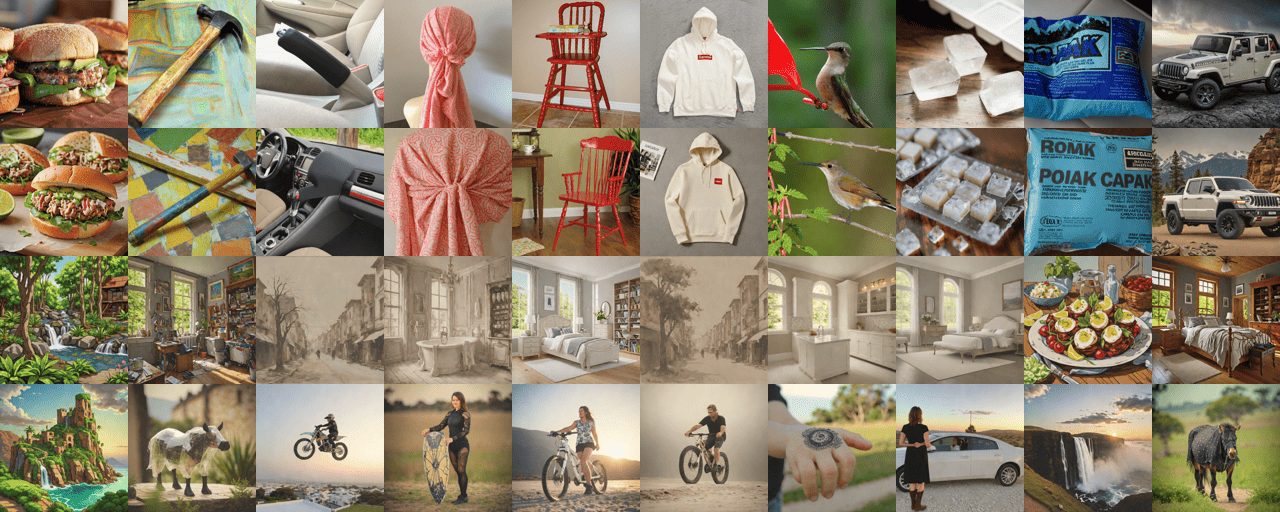}
    \caption{Random selected generated images in Subject 7 with MUSE EEG encoder.}
    \label{fig:enter-label}
\end{figure}

\begin{figure}
    \centering
    \includegraphics[width=1\linewidth]{ 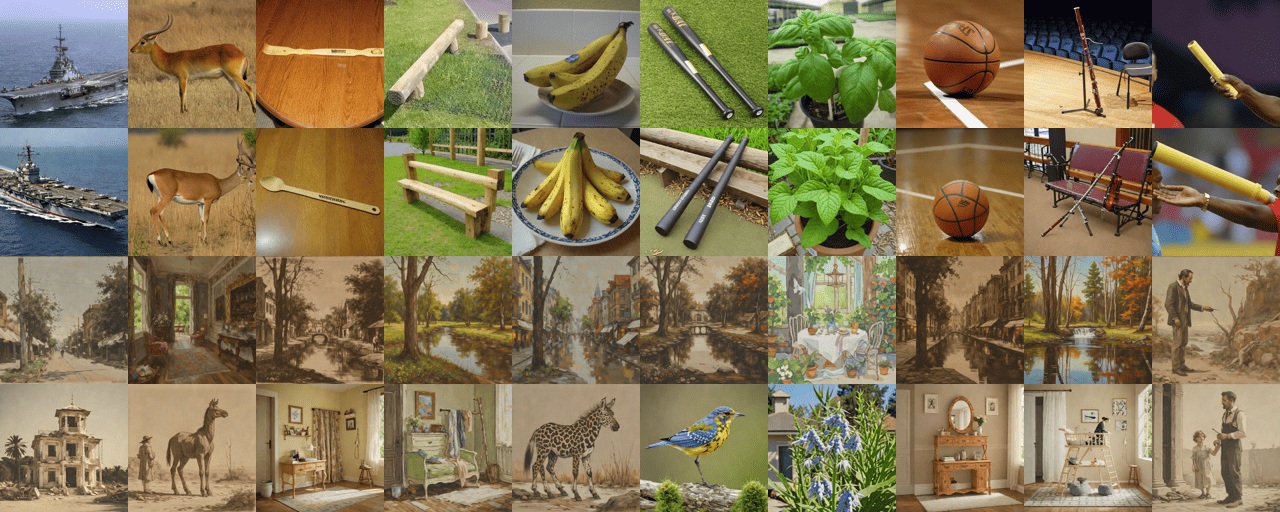}
    \caption{Random selected generated images in Subject 8 with MUSE EEG encoder.}
    \label{fig:enter-label}
\end{figure}
\begin{figure}
    \centering
    \includegraphics[width=1\linewidth]{ 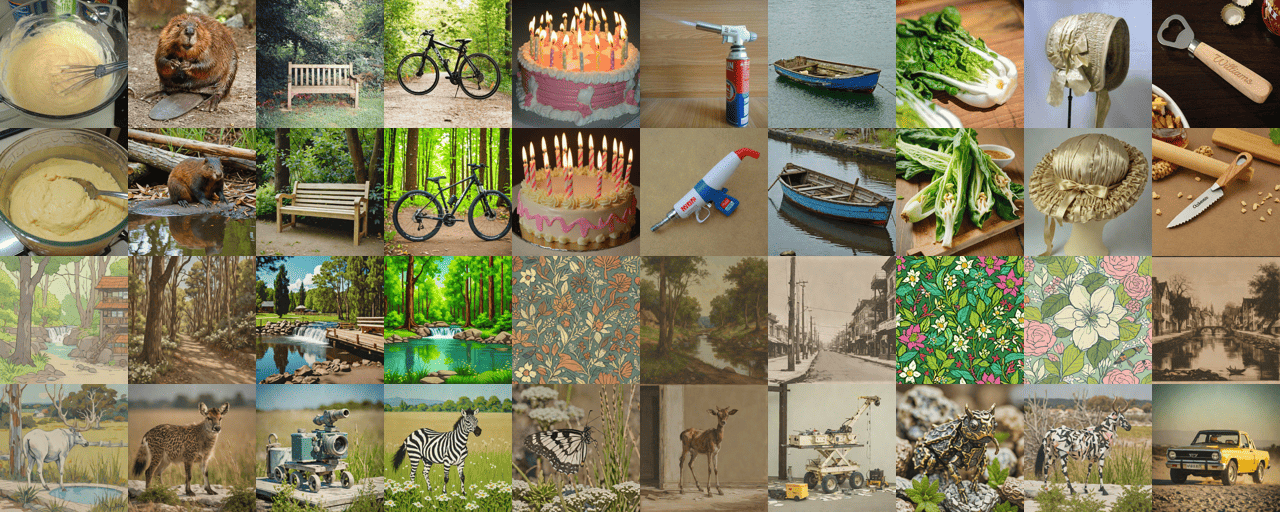}
    \caption{Random selected generated images in Subject 8 with MUSE EEG encoder.}
    \label{fig:enter-label}
\end{figure}
\begin{figure}
    \centering
    \includegraphics[width=1\linewidth]{ 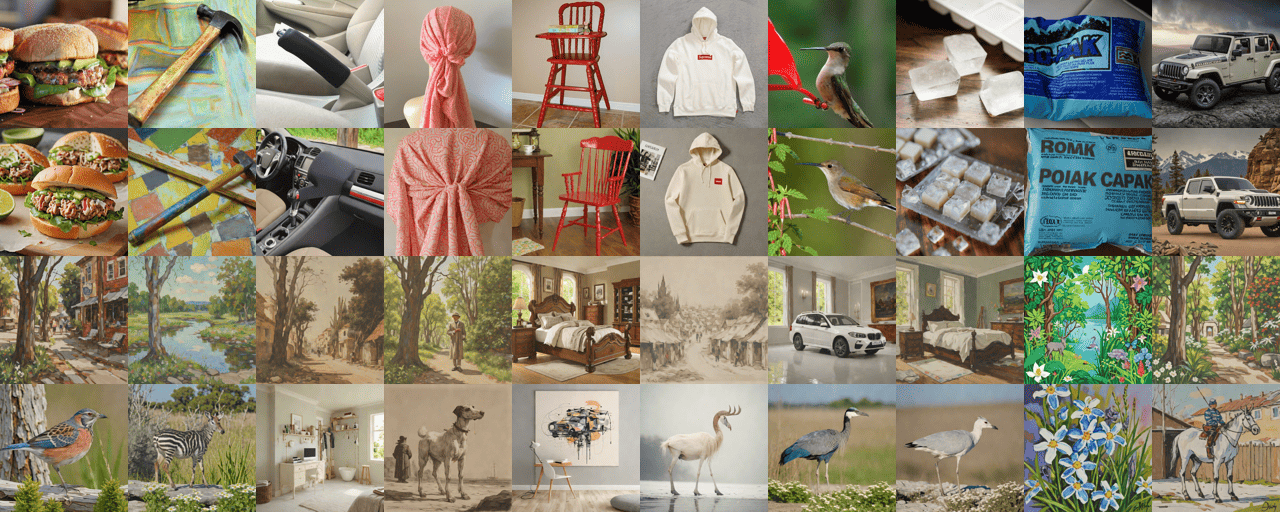}
    \caption{Random selected generated images in Subject 8 with MUSE EEG encoder.}
    \label{fig:enter-label}
\end{figure}

\begin{figure}
    \centering
    \includegraphics[width=1\linewidth]{ 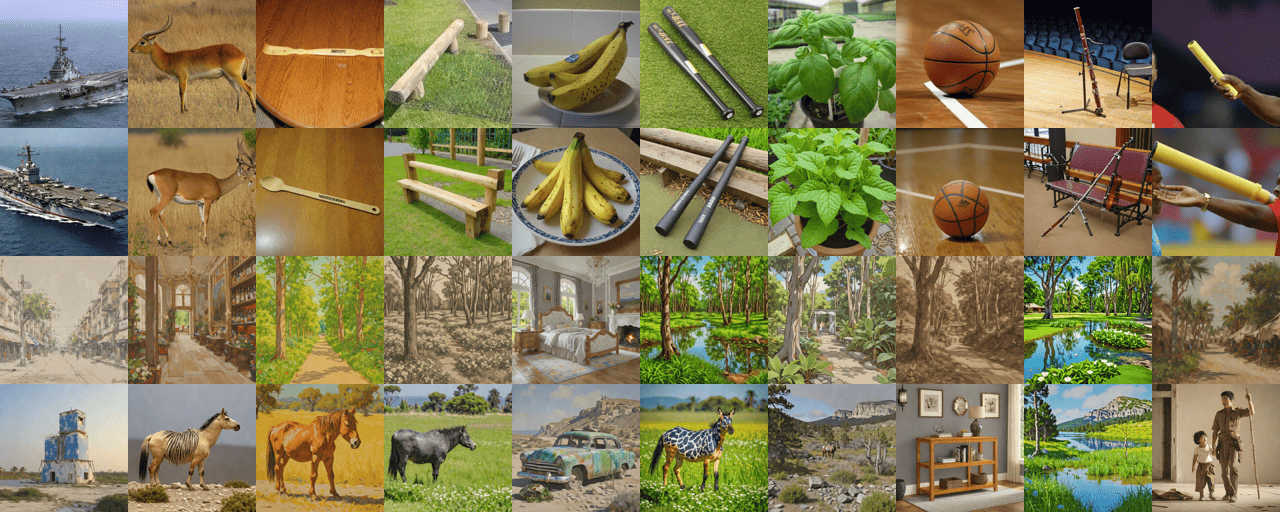}
    \caption{Random selected generated images in Subject 6 with ATM-S EEG encoder.}
    \label{fig:enter-label}
\end{figure}
\begin{figure}
    \centering
    \includegraphics[width=1\linewidth]{ 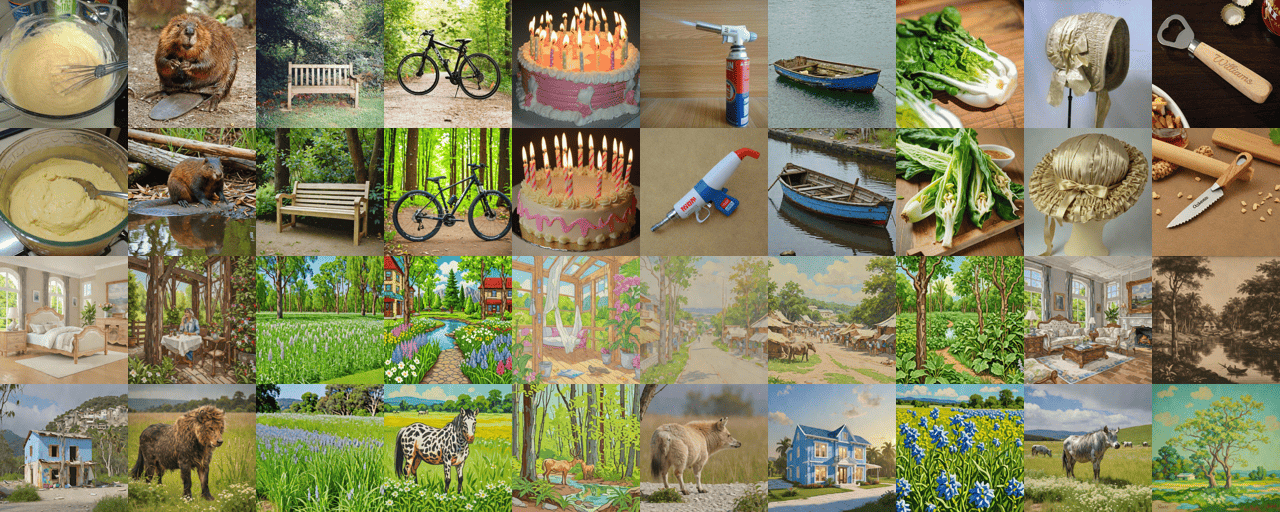}
    \caption{Random selected generated images in Subject 6 with ATM-S EEG encoder.}
    \label{fig:enter-label}
\end{figure}
\begin{figure}
    \centering
    \includegraphics[width=1\linewidth]{ 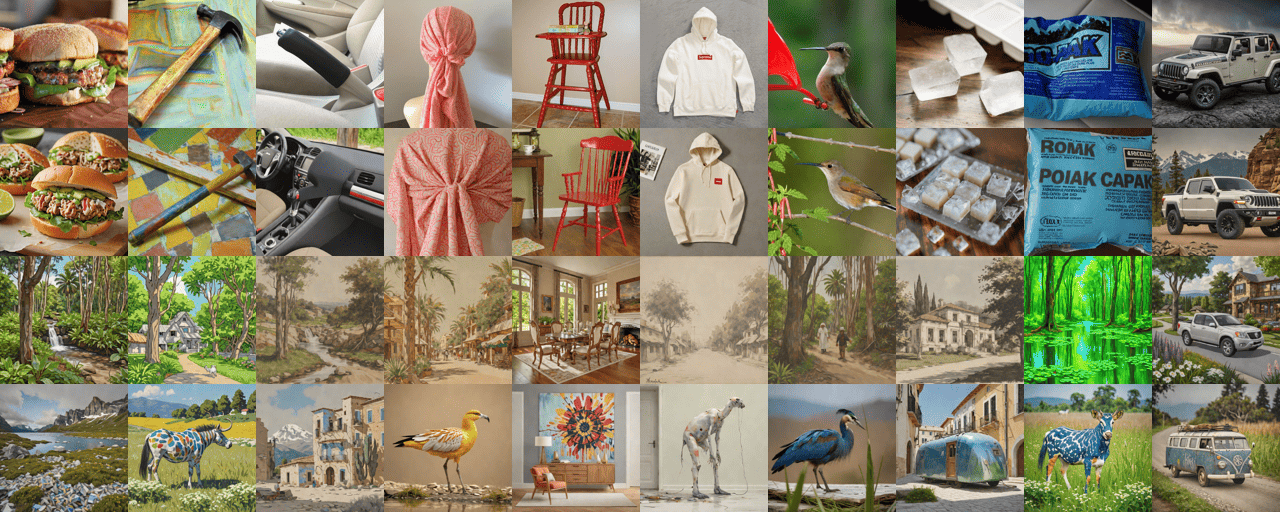}
    \caption{Random selected generated images in Subject 6 with ATM-S EEG encoder.}
    \label{fig:enter-label}
\end{figure}

\begin{figure}
    \centering
    \includegraphics[width=1\linewidth]{ 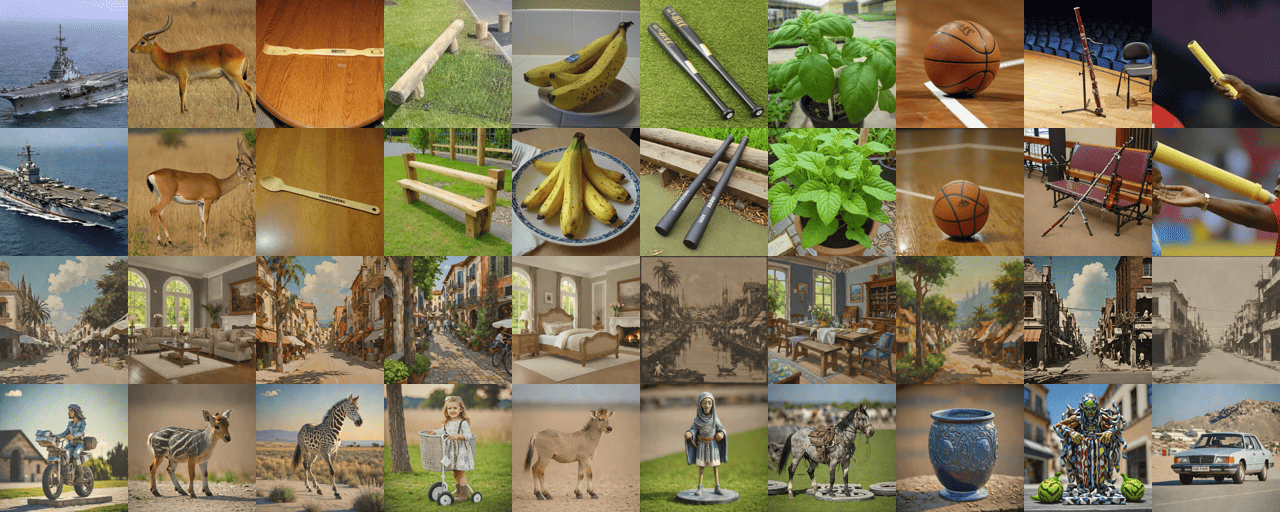}
    \caption{Random selected generated images in Subject 7 with ATM-S EEG encoder.}
    \label{fig:enter-label}
\end{figure}
\begin{figure}
    \centering
    \includegraphics[width=1\linewidth]{ 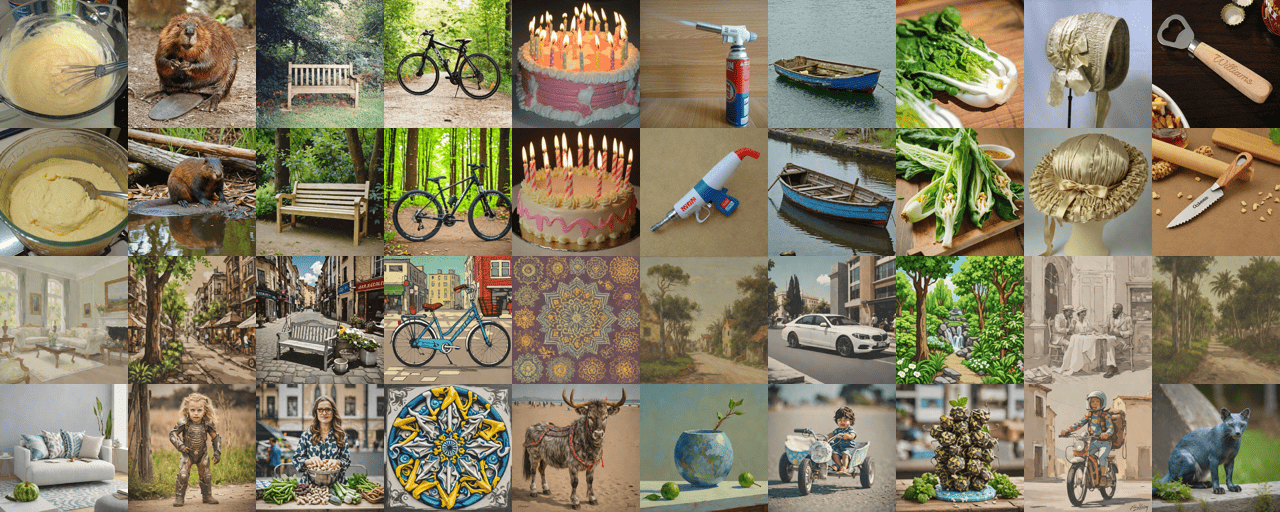}
    \caption{Random selected generated images in Subject 7 with ATM-S EEG encoder.}
    \label{fig:enter-label}
\end{figure}
\begin{figure}
    \centering
    \includegraphics[width=1\linewidth]{ 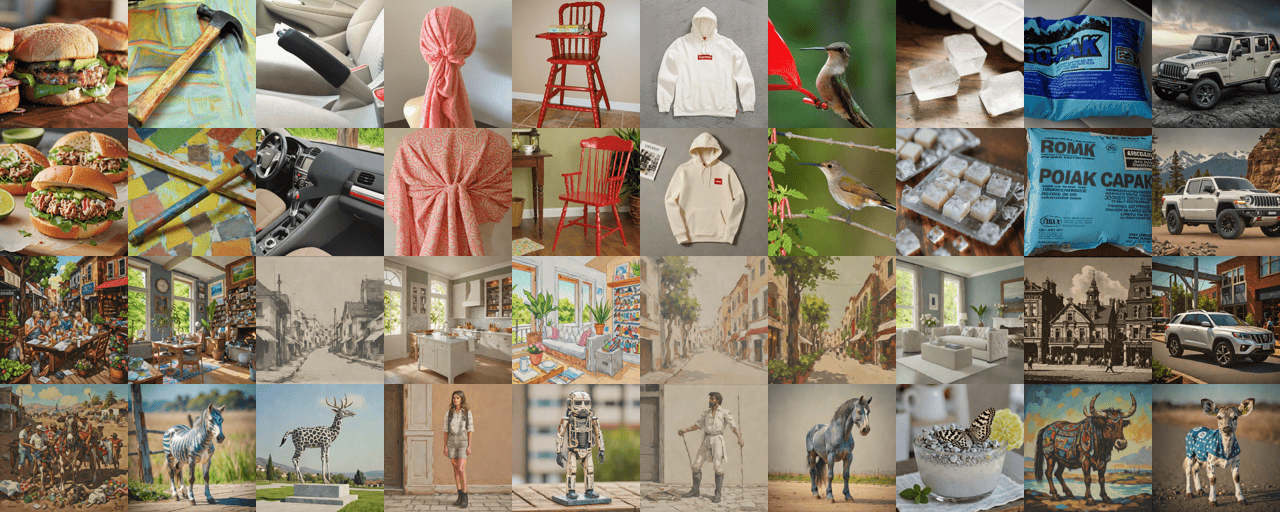}
    \caption{Random selected generated images in Subject 7 with ATM-S EEG encoder.}
    \label{fig:enter-label}
\end{figure}

\begin{figure}
    \centering
    \includegraphics[width=1\linewidth]{ 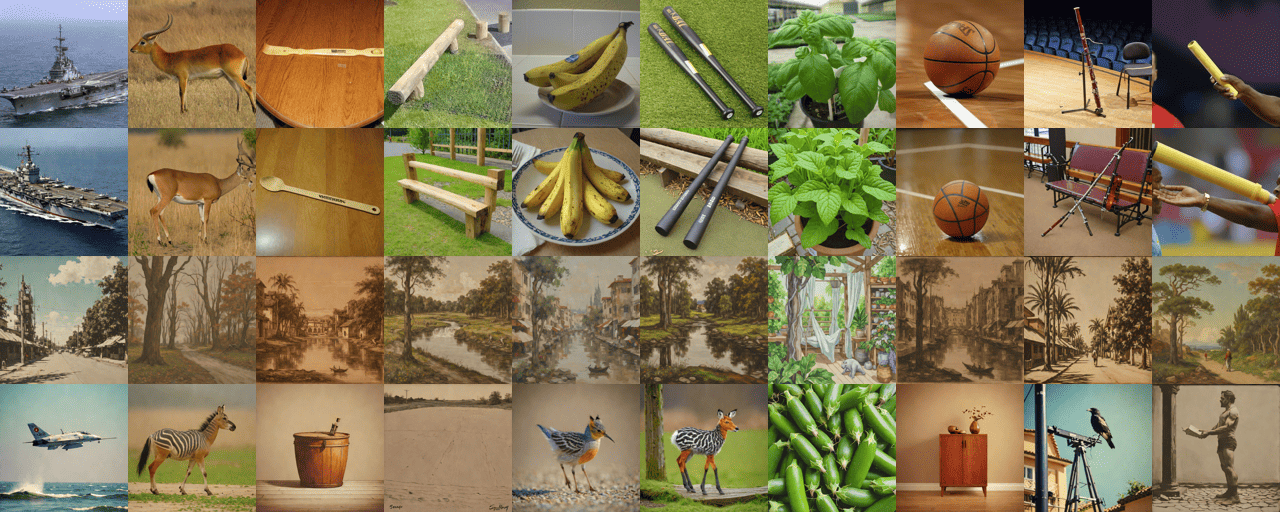}
    \caption{Random selected generated images in Subject 8 with ATM-S EEG encoder.}
    \label{fig:enter-label}
\end{figure}
\begin{figure}
    \centering
    \includegraphics[width=1\linewidth]{ 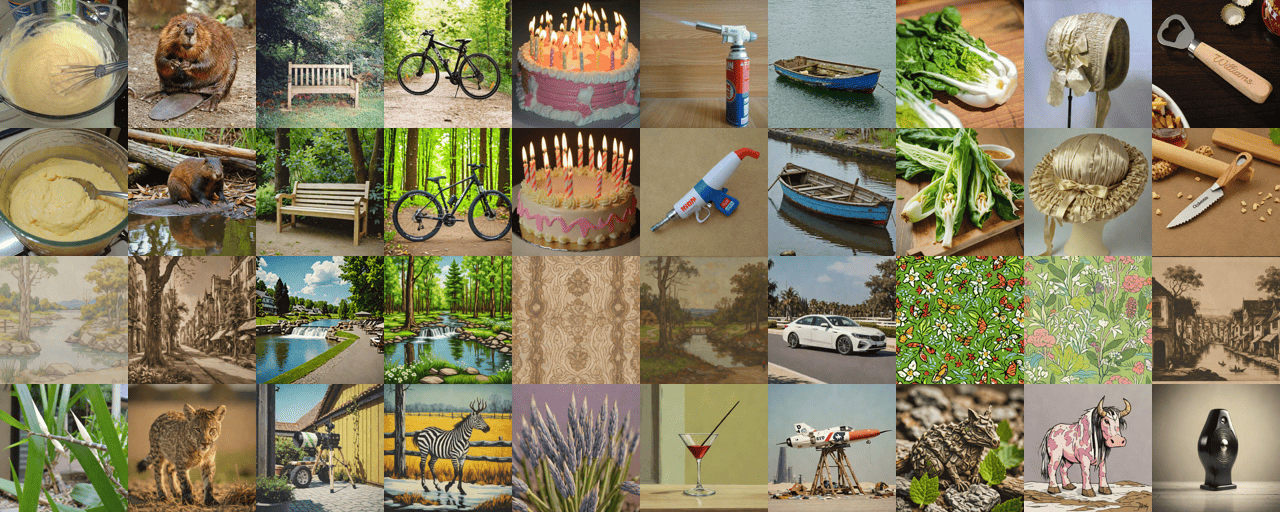}
    \caption{Random selected generated images in Subject 8 with ATM-S EEG encoder.}
    \label{fig:enter-label}
\end{figure}
\begin{figure}
    \centering
    \includegraphics[width=1\linewidth]{ 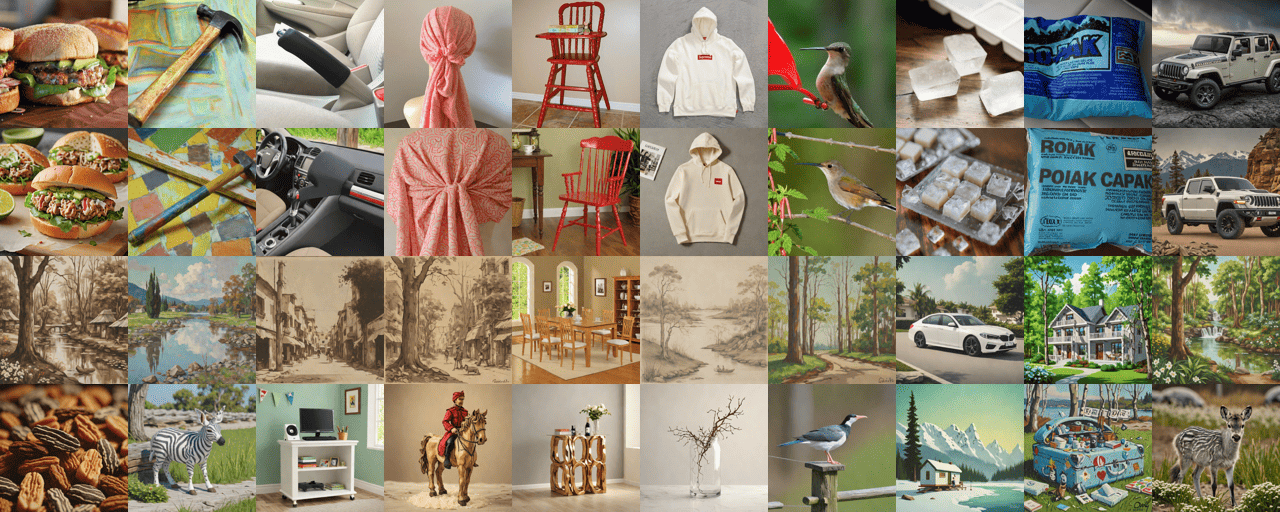}
    \caption{Random selected generated images in Subject 8 with ATM-S EEG encoder.}
    \label{fig:enter-label}
\end{figure}

\begin{figure}
    \centering
    \includegraphics[width=1\linewidth]{ 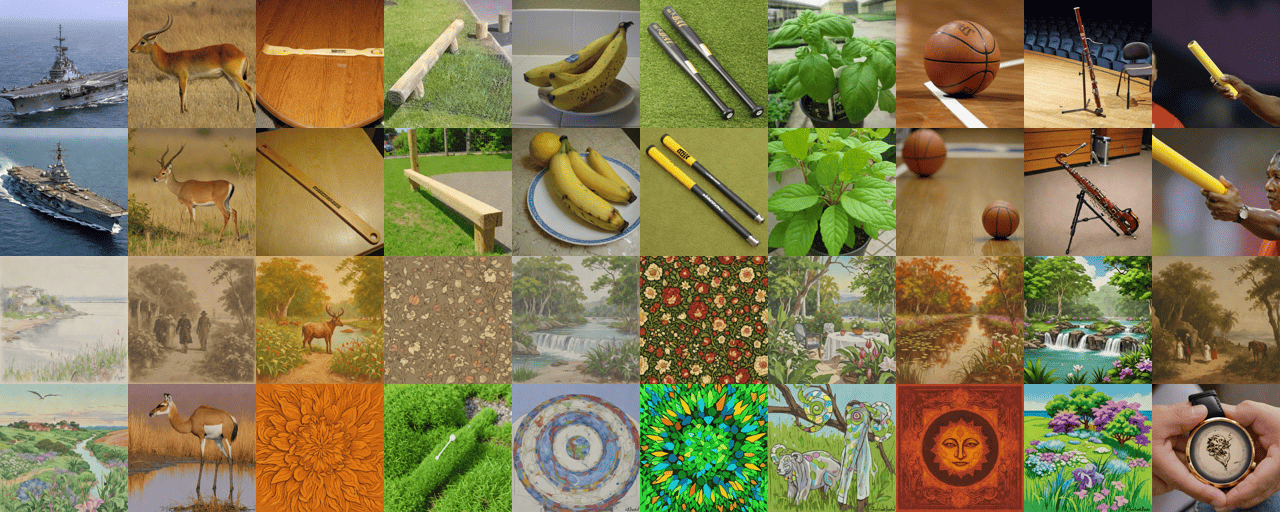}
    \caption{Random selected generated images in Subject 6 with NERV EEG encoder.}
    \label{fig:enter-label}
\end{figure}
\begin{figure}
    \centering
    \includegraphics[width=1\linewidth]{ 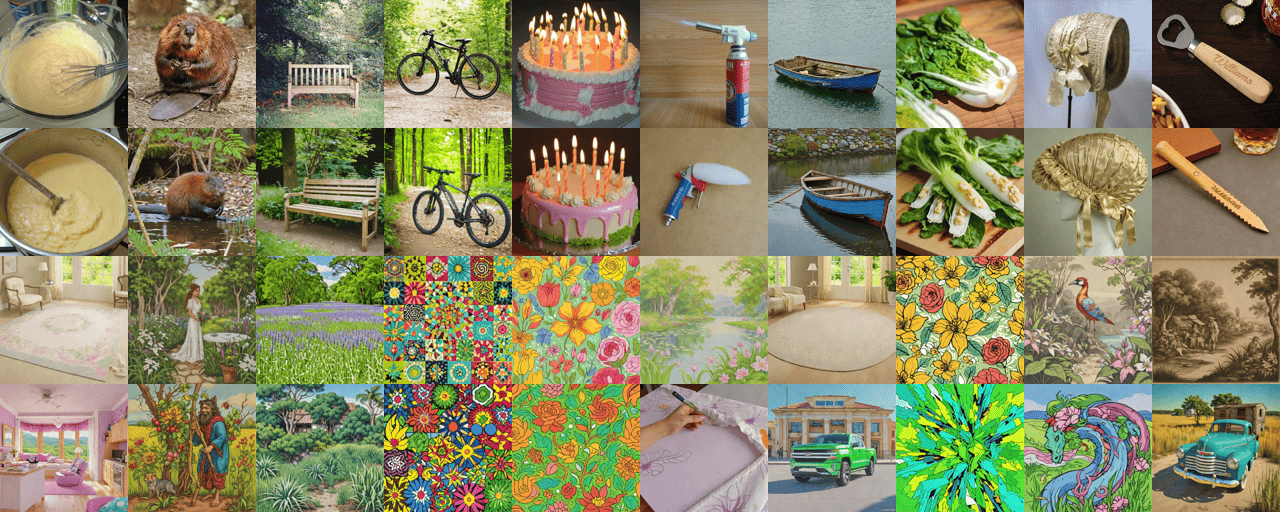}
    \caption{Random selected generated images in Subject 6 with NERV EEG encoder.}
    \label{fig:enter-label}
\end{figure}
\begin{figure}
    \centering
    \includegraphics[width=1\linewidth]{ 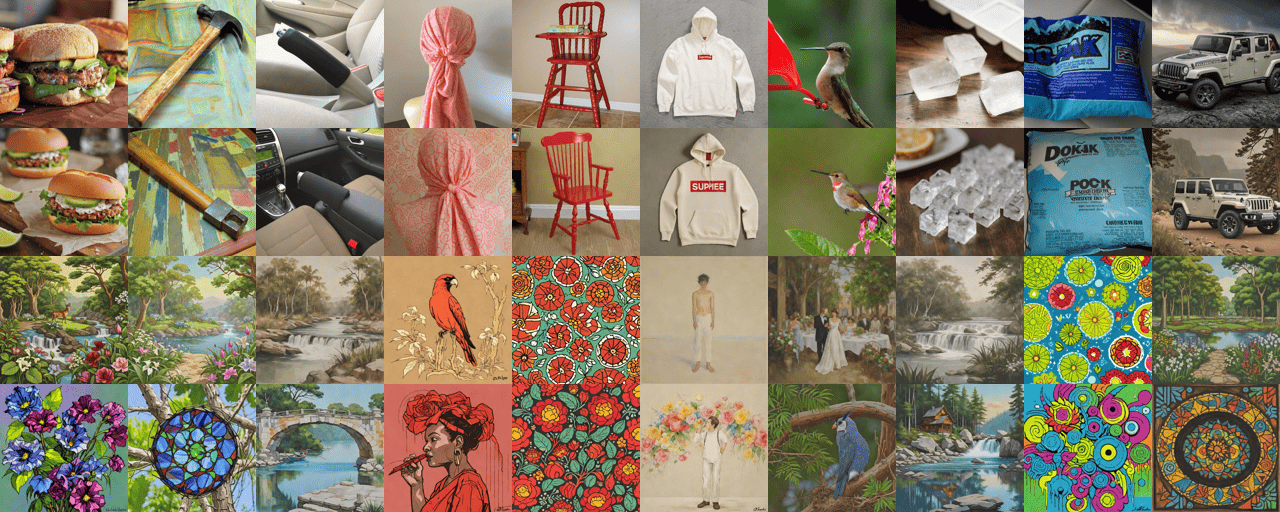}
    \caption{Random selected generated images in Subject 6 with NERV EEG encoder.}
    \label{fig:enter-label}
\end{figure}

\begin{figure}
    \centering
    \includegraphics[width=1\linewidth]{ 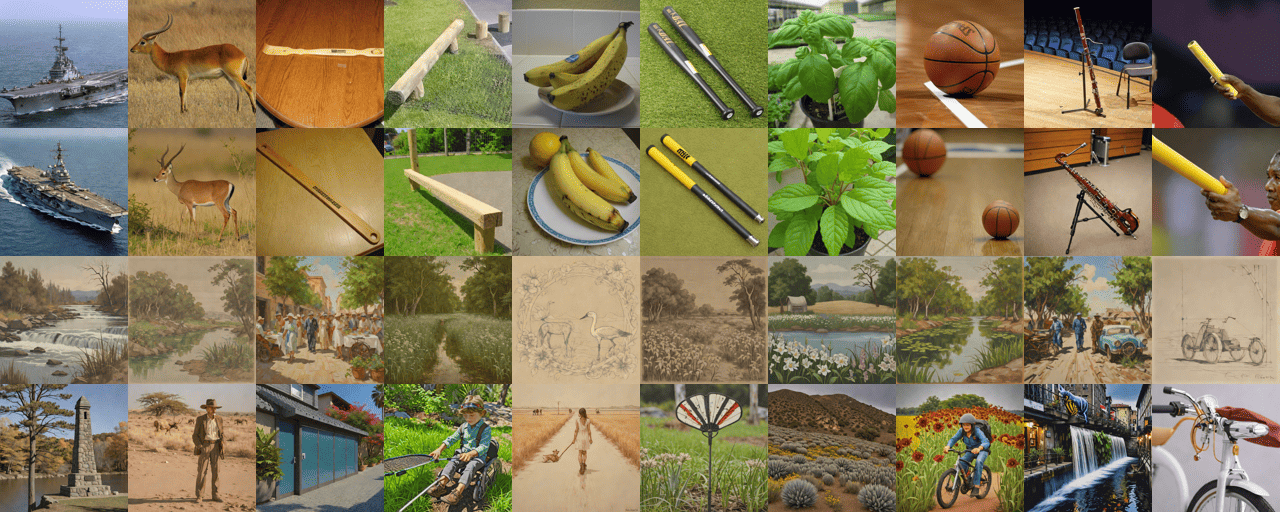}
    \caption{Random selected generated images in Subject 7 with NERV EEG encoder.}
    \label{fig:enter-label}
\end{figure}
\begin{figure}
    \centering
    \includegraphics[width=1\linewidth]{ 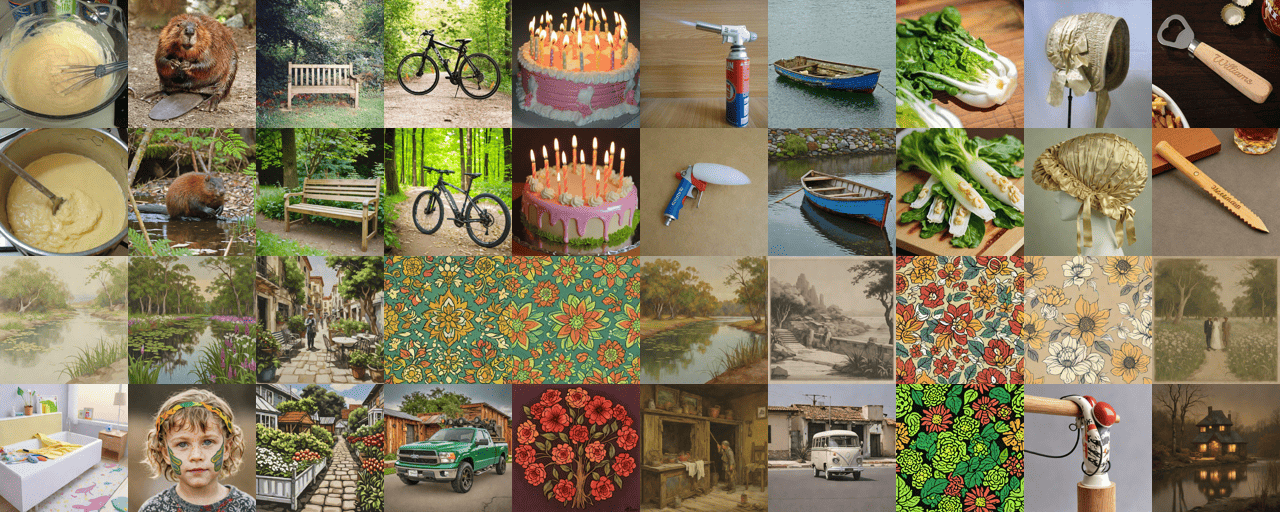}
    \caption{Random selected generated images in Subject 7 with NERV EEG encoder.}
    \label{fig:enter-label}
\end{figure}
\begin{figure}
    \centering
    \includegraphics[width=1\linewidth]{ 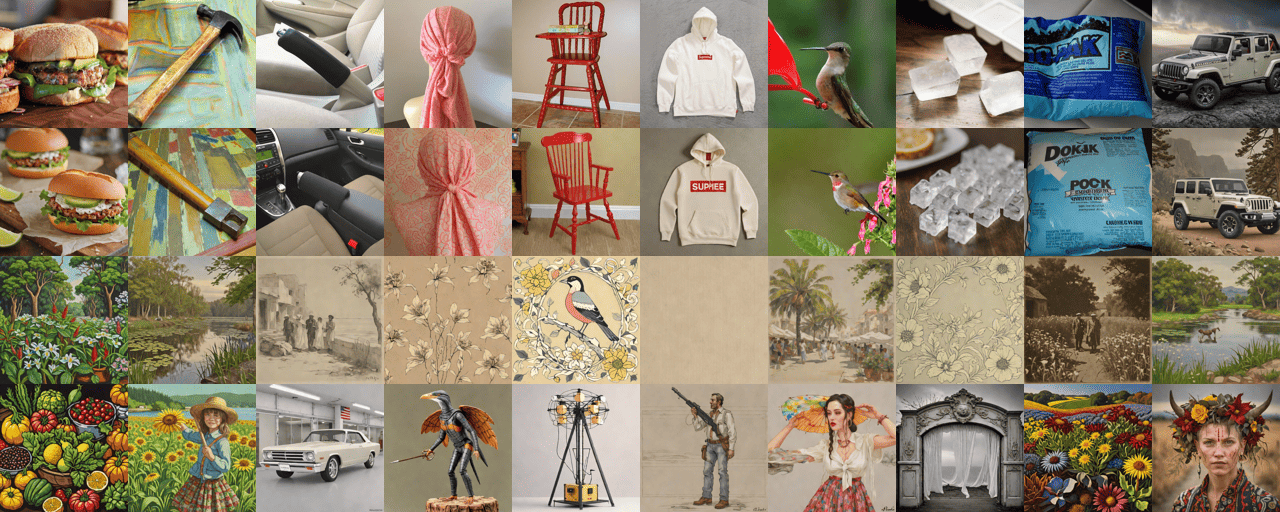}
    \caption{Random selected generated images in Subject 7 with NERV EEG encoder.}
    \label{fig:enter-label}
\end{figure}

\begin{figure}
    \centering
    \includegraphics[width=1\linewidth]{ 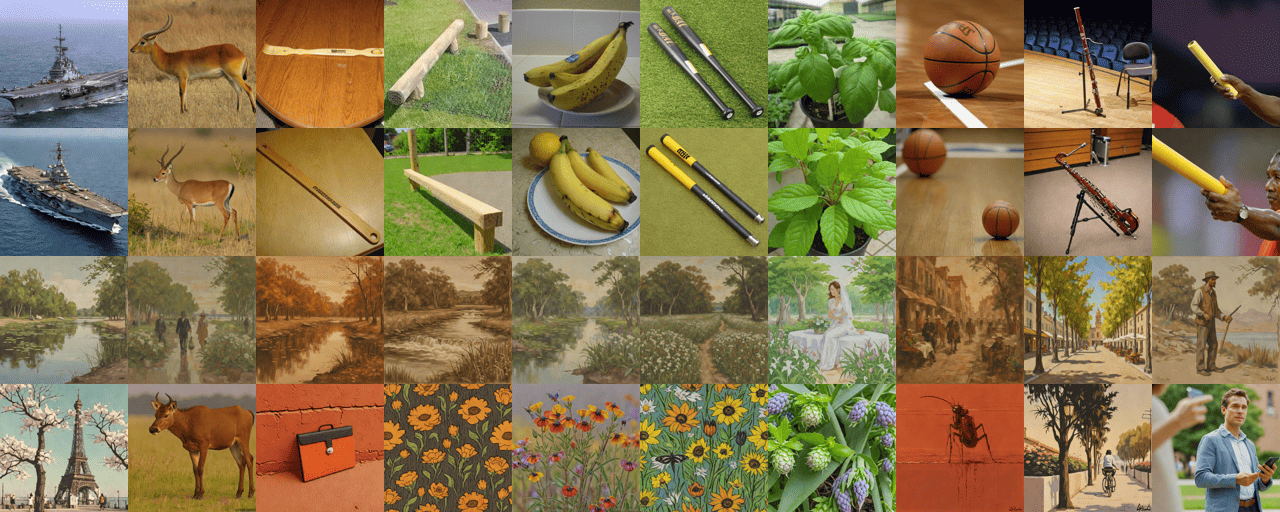}
    \caption{Random selected generated images in Subject 8 with NERV EEG encoder.}
    \label{fig:enter-label}
\end{figure}
\begin{figure}
    \centering
    \includegraphics[width=1\linewidth]{ 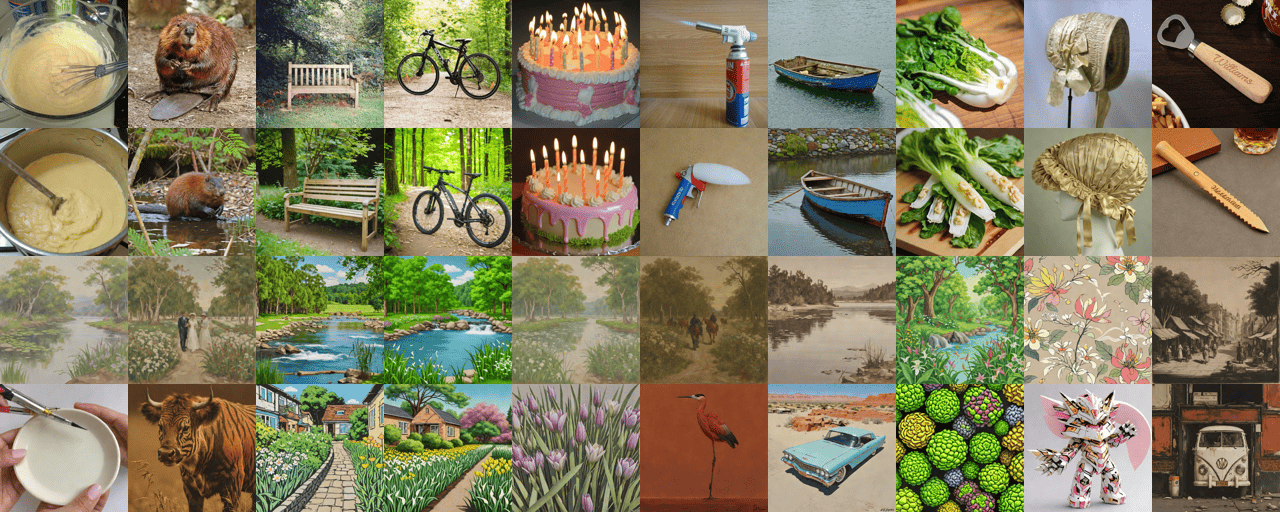}
    \caption{Random selected generated images in Subject 8 with NERV EEG encoder.}
    \label{fig:enter-label}
\end{figure}
\begin{figure}
    \centering
    \includegraphics[width=1\linewidth]{ 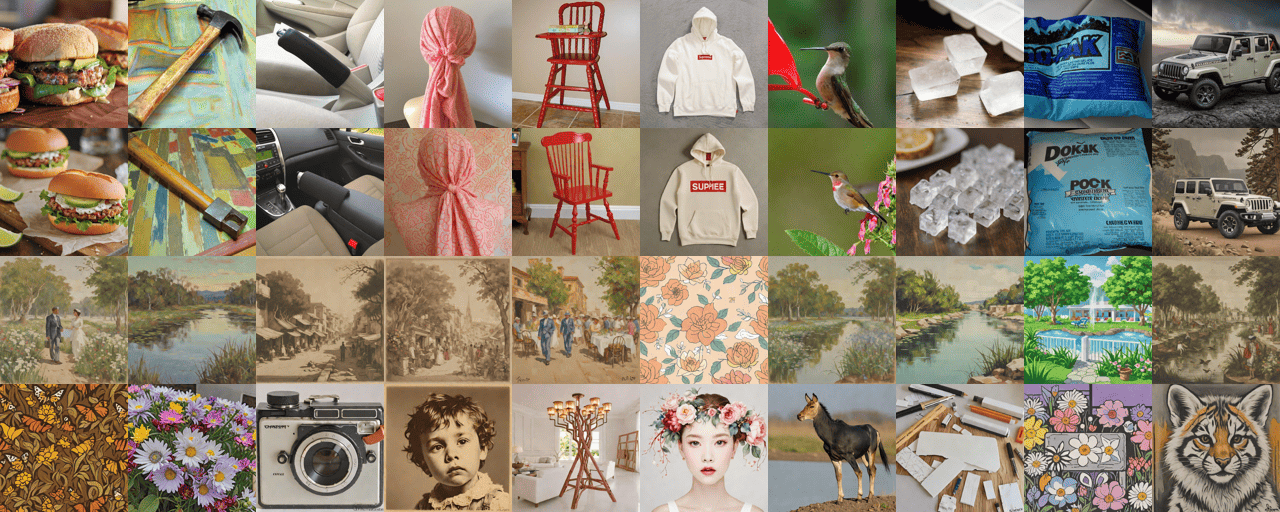}
    \caption{Random selected generated images in Subject 8 with NERV EEG encoder.}
    \label{fig:enter-label}
\end{figure}

\end{document}